\documentclass[rmp, aps,reprint,amsmath,amssymb,graphicx,superscriptaddress]{revtex4-1}

\usepackage{qcircuit}
\usepackage[dvips]{graphicx}
\usepackage{siunitx}
\usepackage{amsmath,amssymb,amsthm,mathrsfs,amsfonts,dsfont}
\usepackage{subfigure, epsfig}
\usepackage{braket}
\usepackage{bm}
\usepackage{enumerate}
\usepackage[dvipsnames]{xcolor}
\usepackage{color}
\usepackage{fancybox, graphicx}
\usepackage[skins]{tcolorbox}
\usepackage{multirow}
\usepackage{mathtools}
\DeclarePairedDelimiter{\ceil}{\lceil}{\rceil}

\usepackage[colorlinks]{hyperref}
\hypersetup{
	colorlinks=true,
	citecolor = {blue},
    linkcolor=blue,
    filecolor=magenta,      
    urlcolor=cyan,
}

\newtcolorbox[auto counter]{tbox}[2][]{%
	enhanced, float=hbt, drop fuzzy shadow southeast,
	colback=white!5!white, colframe=white!30!black,
	width= .97\columnwidth,sharp corners,boxrule=0.8pt,
	title={#2}, #1
}

\newcommand{\tr}{\mathrm{Tr}}

\newtcolorbox{codebox}{enhanced, width=.95\columnwidth, halign = flush left, drop fuzzy shadow southeast, boxrule=0.4pt, sharp corners, colframe=black, colback=white}
\begin{document}

\title{Quantum computational chemistry}
\begin{abstract}
One of the most promising suggested applications of quantum computing is solving classically intractable chemistry problems. \textcolor{black}{This may help to answer unresolved questions about phenomena like: high temperature superconductivity, \textcolor{black}{solid-state physics,} transition metal catalysis, or certain biochemical reactions. In turn, this increased understanding may help us to refine, and perhaps even one day design, new compounds of scientific and industrial importance. However, building a sufficiently large quantum computer will be a difficult scientific challenge. As a result, developments that enable these problems to be tackled with fewer quantum resources should be considered very important.} Driven by this potential utility, quantum computational chemistry is rapidly emerging as an interdisciplinary field requiring knowledge of both quantum computing and computational chemistry. This review provides a comprehensive introduction to \textcolor{black}{both computational chemistry and quantum computing}, bridging the current knowledge gap. We review the major developments in this area, with a particular focus on near-term quantum computation. Illustrations of key methods are provided, explicitly demonstrating how to map chemical problems onto a quantum computer, and solve them. We conclude with an outlook for this nascent field.
\end{abstract}

\date{\today}
\author{Sam McArdle}
\email{sam.mcardle.science@gmail.com}
\affiliation{Department of Materials, University of Oxford, Parks Road, Oxford OX1 3PH, United Kingdom}

\author{Suguru Endo}
\affiliation{Department of Materials, University of Oxford, Parks Road, Oxford OX1 3PH, United Kingdom}

\author{Al\'an Aspuru-Guzik}
\affiliation{Department of Chemistry and Department of Computer Science, University of Toronto, Toronto, Ontario M5S 3H6, Canada}
\affiliation{Vector Institute for Artificial Intelligence, Toronto, Ontario M5S 1M1, Canada}
\affiliation{Canadian Institute for Advanced Research (CIFAR) Senior Fellow, Toronto, Ontario M5S 1M1, Canada}

\author{Simon C. Benjamin}
\affiliation{Department of Materials, University of Oxford, Parks Road, Oxford OX1 3PH, United Kingdom}

\author{Xiao Yuan}
\email{xiao.yuan.ph@gmail.com}
\affiliation{Department of Materials, University of Oxford, Parks Road, Oxford OX1 3PH, United Kingdom}

\maketitle

\tableofcontents

\section{Introduction}\label{Sec:Introduction}
Quantum mechanics underpins all of modern chemistry. One might therefore imagine that we could use this theory to predict the behaviour of any chemical compound. This is not the case. As Dirac noted; ``\textit{The exact application of these laws leads to equations much too complicated to be soluble.}''~\cite{Dirac}. The problem described by Dirac is that the complexity of the wavefunction of a quantum system grows exponentially with the number of particles. This leaves classical computers unable to exactly simulate quantum systems in an efficient way. Feynman proposed a solution to this problem; using quantum hardware as the simulation platform, remarking that \textit{``If you want to make a simulation of nature, you'd better make it quantum mechanical, and by golly it's a wonderful problem, because it doesn't look so easy."}~\cite{Feynman1982}. 

\textcolor{black}{Although developing small quantum computers has taken over 30 years, we may soon be in a position to test Feynman's proposal}, following recent developments in quantum hardware including ion traps~\cite{Monz14QubitIon, harty2014high,HighFideity16,gaebler2016high}, superconducting systems~\cite{barends2014superconducting,song201710, chow2012superconducting,song2019observation,google2019supremacy}, and photonic systems~\cite{WangTenPhoton16,chen2017observation,12photon}. \textcolor{black}{It is believed that using quantum systems as our simulation platform will enable us to tackle classically intractable problems in chemistry, physics, and materials science. Classical computational methods have become an important investigative tool in areas like transition metal catalysis~\cite{vogiatzis2019catalysis} and high temperature superconductivity~\cite{dagotto1994correlatedelectrons}. Classical simulations enable us to rationalise experimental results, test physical models, and understand system properties. However, their ability to guide design is often precluded by the computational complexity of realistic models. As quantum computers are able to efficiently simulate quantum systems, it is believed that they will enable a more accurate understanding of the models in use today, as well as the ability to simulate more complex (and therefore, more realistic) models. This may lead to an increased understanding which we can leverage to make advances in areas as diverse as chemistry~\cite{Revolution}, biology~\cite{Reiher201619152}, medicine ~\cite{cao2018drugdiscovery}, and materials science~\cite{Babbush2017low}. It has even been speculated that as quantum hardware develops, quantum simulation may one day \textcolor{black}{progress from being accurate enough to confirm the results of experiments, to being more accurate than the experiments themselves. Quantum simulations of such high precision may in turn enable the design of new, useful compounds.} However, we stress that to achieve this ultimate goal we would need considerable further developments -- both in the technology required to build such as powerful quantum computer, and the theory behind an appropriate algorithm and model. This can be likened to the aerospace industry, where computational fluid dynamics calculations on classical computers have replaced wind tunnel testing in many stages of wing design~\cite{jameson1999aerospace}. However, for the most demanding parts of aerospace design, neither our largest classical computers, nor the physical models considered, are yet powerful enough to completely replace experimental testing~\cite{malik2012aerospace}. Instead, the two methods work together in synergy, to enable increased understanding, with greater efficiency.} \\

To date, several efficient quantum algorithms have been proposed to solve problems in chemistry. The runtime and physical resources required by these algorithms are expected to scale polynomially with both the size of the system simulated and the accuracy required. Experimental developments have accompanied these theoretical milestones, with many groups demonstrating proof of principle chemistry calculations. However, limited by hardware capabilities, these experiments focus only on small chemical systems that we are already able to simulate classically. Moreover, the gate counts currently estimated for transformative chemistry simulations likely signal the need for quantum error correction, which requires orders of magnitude more qubits, and lower error rates, than are currently available~\cite{mueck2015reform,babbush2018encoding,kivlichan2019condensedtrotter}. Despite ongoing experimental efforts, no group has yet demonstrated a single fully error corrected qubit. Even if the significant hardware challenges to build an error corrected quantum computer can be overcome~\cite{gambetta2017errorcorrected,Ladd2010quantumcomputingreview,monroe2013iontrapreview}, new theoretical developments may be needed to solve classically intractable chemistry problems on a quantum computer that we could realistically imagine building in the next few decades, \textcolor{black}{such as probing biological nitrogen fixation, or investigating new metal ion battery designs}. These breakthroughs may be achieved by connecting researchers working in quantum computing with those working in computational chemistry. We seek to aid this connection with this succinct, yet comprehensive, review of quantum computational chemistry (using quantum algorithms, run on quantum computers, to solve problems in computational chemistry) and its foundational fields. \\

Although quantum algorithms can solve a range of problems in chemistry, we focus predominantly on the problem of finding the low lying energy levels of chemical systems. This is known as `the electronic structure problem'. There are three reasons for this restriction of scope. Primarily, this problem is a fundamental one in classical computational chemistry. Knowledge of the energy eigenstates enables the prediction of reaction rates, location of stable structures, and determination of optical properties~\cite{helgaker2012wavefunctionadvances}. Secondly, the machinery developed to solve this problem on quantum computers is easily applied to other types of problems, such as finding transition states, or understanding the vibrational structure of molecules. Finally, most of the prior work in quantum computational chemistry has focused on this problem. As such, it provides an ideal context in which to explain the most important details of quantum computational chemistry.\\

This review is organised as follows. We first provide a brief overview of quantum computing and simulation in Sec.~\ref{Sec:QCS}. We then introduce the key methods and terminology used in classical computational chemistry in Sec.~\ref{Sec:classicalChemistry}. The methods developed to merge these two fields, including mapping chemistry problems onto a quantum computer, are described in Sec.~\ref{Sec:Encoding}. We continue our discussion of quantum computational chemistry in Sec.~\ref{Sec:QuantumChemistryAlgorithms} by describing algorithms for finding the ground and excited states of chemical systems. Sec.~\ref{Sec:errorMitigation} highlights the techniques developed to mitigate the effects of noise in non-error corrected quantum computers, which will be crucial for achieving accurate simulations in the near-future.  

In Sec.~\ref{Sec:illustrate} we provide several examples of how to map chemistry problems onto a quantum computer. We discuss techniques that can be used to reduce the simulation resources required, and the quantum circuits that can be used. This section seeks to illustrate the techniques described throughout the rest of the review, providing worked examples for the reader. We conclude this review in Sec.~\ref{Sec:conclude} with a comparison between classical and quantum techniques, and resource estimations for the different quantum methods. This section aims to help the reader to understand when, and how, quantum computational chemistry may surpass its classical counterpart.\\

A handful of related reviews on this topic exist in the literature. Summaries of early theoretical and experimental work in quantum computational chemistry were carried out by \textcite{kassal2011simulating,C2CP23700H}. More focused discussions of quantum algorithms introduced for chemistry simulation before 2015, and the computational complexity of problems in chemistry can be found in works by \textcite{kais2014quantum,yung2014introduction,veis2012review}. A comprehensive review was recently released by \textcite{cao2019review}. Said review, and our own, are complementary; \textcolor{black}{the review of \textcite{cao2019review} is well suited to experienced practitioners of classical electronic structure theory, and provides excellent detail on the computational complexity of quantum simulations, and how they asymptotically compare to cutting edge methods in classical chemistry. Our review provides a more practical guide to the field (especially for those new to electronic structure theory), showing explicitly how the workhorse methods of classical chemistry have been translated to work on quantum computers, and describing techniques to facilitate experimental demonstrations of quantum chemistry algorithms, such as resource reduction and error mitigation.} Together, these reviews provide a complete overview of the progress to date in quantum computational chemistry. \\

Despite being a relatively young field, quantum computational chemistry has grown extremely rapidly, and has already evolved beyond the stage that it can be fully described by a single review. As such, there are approaches to solving chemistry problems with a quantum computer which we are not able to describe fully in this review. \textcolor{black}{As stated above, we have chosen to prioritise the canonical topics in the field: using either near-term, or further-future digital quantum computers to solve the electronic structure problem \textit{ab initio}. We have focused on the most promising methods for solving this problem: variational algorithms with error mitigation, and the quantum phase estimation algorithm with quantum error correction. Extended discussion is reserved for methods which are key to understanding how quantum computers can be used to solve \textit{general} chemistry problems, or articles which have made important observations on ways to make these simulations more tractable. It is beyond the scope of this review to summarise work in directions complementary to these,} such as: \textcolor{black}{quantum machine learning based approaches to the electronic structure problem~\cite{xia2018quantummachinelearning,xia2018ising}}, using quantum computers as part of a problem decomposition approach to simulation~\cite{dallaire2016hubbard,dallaire2016hubbardarchitecture,kreula2016dmft,bauer2016dmft,rubin2016dmet, keen2019dmftexperiment,rungger2019dmft}, hybrid quantum algorithms for density functional theory~\cite{whitfield2014tddft, hatcher2019dftvqe}, relativistic quantum chemistry~\cite{veis2012relativistic,senjean2019dirac}, gate based methods for simulating molecular vibrations~\cite{mcardle2018vibrations,sawaya2018vibronic,sawaya2019resourceefficient}, analog simulators of molecular vibrations~\cite{joshi2014francknmr,huh2015boson,huh2017vibronic,clements2017experimental,shen2018vibronic, sparrow2018simulating,chin2018quantum,hu2018simulation, wang2019vibronicsuperconducting}, fermionic quantum computation for chemistry simulation~\cite{obrien2018majorana}, quantum methods for electron-phonon systems~\cite{macridin2018fermionboson,macridin2018electronphonon,wu2002BCSsim}, protein folding and molecular docking~\cite{perdomo2012finding,perdomo2008protein,babbush2012protein,babej2018proteinannealer,fingerhuth2018proteingate,lu2019protein,banchi2019moleculardocking,robert2019proteinfolding}, solving problems in chemistry using a quantum annealer~\cite{babbush2014adiabatic,genin2019annealer, teplukhin2018vibrationalannnealer,xia2018ising}, and \textcolor{black}{quantum algorithms for finding the eigenvalues of non-hermitian Hamiltonians~\cite{daskin2014nonunitary,wang2010nonunitary}}. \\

\section{Quantum computing and simulation}\label{Sec:QCS}
In this section, we introduce the basic elements of quantum computing and quantum simulation. We refer the reader to the works of \textcite{nielsen2002quantum,RevModPhys.86.153} for more detailed introductions.

\subsection{Quantum computing}\label{Subsec:QuantumComputing}
In this review, we focus on the qubit-based circuit model of quantum computation~\cite{nielsen2002quantum}. Other paradigms that vary to a greater or lesser extent include: adiabatic quantum computing~\cite{farhi2000adiabatic,aharonov2008adiabatic,albash2018rmpadiabatic}, one-way or measurement based quantum computing~\cite{Raussendorf2001oneway,Raussendorf2003measurement,jozsa2005introduction}, and continuous-variable quantum computing~\cite{lloyd1999continuous,braunstein2005continuous}. 

The canonical circuit model of quantum computation is so-named because of its resemblance to the logic circuits used in classical computing. In the classical circuit model, logic gates (such as AND, OR and NOT) act on bits of information. In the quantum case, quantum gates are acted upon the basic unit of information, the qubit. The qubit lives in a two-dimensional Hilbert space. The basis vectors of the space are denoted as $\{\ket{0},\ket{1}\}$, which are referred to as the computational basis states, 
\begin{equation}
\ket{0} = 
\begin{pmatrix}
1 \\
0\\
\end{pmatrix},~~
\ket{1} =
\begin{pmatrix}
0 \\
1 \\
\end{pmatrix}.
\end{equation}

A general single qubit state is described by 
\begin{gather}\label{EqQubit}
\ket{\varphi} = \alpha \ket{0}+\beta \ket{1} = 
\begin{pmatrix}
\alpha \\
\beta \\
\end{pmatrix}, \\
\alpha, \beta \in \mathbb{C}, \nonumber\\
|\alpha|^2 + |\beta|^2 = 1 .\nonumber
\end{gather}
When quantum logic gates act on the qubits, they manipulate both basis state vectors at the same time, providing (measurement limited) parallelism. Although the qubit is in a quantum superposition during the algorithm, when it is measured in the computational basis, it will be found in state $\ket{0}$ or state $\ket{1}$, not in a superposition. These measurement outcomes occur with probability $|\alpha|^2$ and $|\beta|^2$, respectively. For now, we will treat the qubit as an abstract two level system, before later elaborating on how they can be physically realised.

If there are $n$ qubits in the system, the state is described by a vector in the 2$^n$ dimensional Hilbert space formed by taking the tensor product of the Hilbert spaces of the individual qubits. States can be classified as either `product' or `entangled'. Product states can be decomposed into tensor products of fewer qubit wavefunctions, such as
\begin{equation}
\frac{1}{\sqrt{2}} (\ket{00} + \ket{01}) = \ket{0} \otimes \frac{1}{\sqrt{2}}( \ket{0} + \ket{1}).
\end{equation}
Entangled states cannot be decomposed into tensor products, such as the state
\begin{equation}\label{Eq:bell1}
\frac{1}{\sqrt{2}} (\ket{00} + \ket{11}) .
\end{equation}
In this review, we refer to the leftmost qubit in a vector as the $(n-1)$\textsuperscript{th} qubit, and the rightmost qubit as the zeroth qubit. \textcolor{black}{This choice enables us to write numbers in binary using computational basis states. For example, we can write $\ket{7} = \ket{1}\ket{1}\ket{1} = \ket{111}$. We can then place a quantum register of $n$ qubits in a superposition of the $2^n$ possible numbers that can be represented by $n$ bits. This is typically written as $\sum_{x=0}^{2^n -1} c_x  \ket{x}$.}

\begin{figure}[hbt]
{\includegraphics{{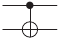}}}
\caption{The controlled-NOT (CNOT) gate. `$\bullet$' denotes the control qubit and `$\oplus$' denotes the target qubit.}\label{Fig:CNOT}
\end{figure}

A quantum circuit consists of a number of single and two qubit gates \textcolor{black}{acting} on the qubits. The qubits are initialised in a well defined state, such as the $\ket{\bar{0}}$ state ($\ket{\bar{0}} = \ket{0}^{\otimes n} = \ket{0} \otimes \ket{0} \otimes \dots \otimes \ket{0} $). A quantum circuit generally concludes with measurements to extract information. It may also employ additional intermediate measurements, for example, to check for errors. From a mathematical perspective, the gates are unitary matrices. 
Typical gates include the Pauli gates
\begin{equation}
X = 
\begin{pmatrix}
0 & 1 \\
1 & 0 \\
\end{pmatrix},~~
Y = 
\begin{pmatrix}
0 & -i \\
i & 0 \\
\end{pmatrix},~~
Z = 
\begin{pmatrix}
1 & 0 \\
0 & -1 \\
\end{pmatrix},
\end{equation}
the single qubit rotation gates
\textcolor{black}{\begin{align}
R_{x}(\theta) = e^{\left (\frac{{-i \theta X}\,}{2} \right)},~ R_{y}(\theta) = e^{\left (\frac{{-i \theta Y}\,}{2} \right)},~
R_{z}(\theta) = e^{\left (\frac{{-i \theta Z}\,}{2} \right)}
\end{align}}
the Hadamard and T gates
\begin{equation}
	\mathrm{Had} = \frac{1}{\sqrt{2}} 
\begin{pmatrix}
1& 1\\
1& -1
\end{pmatrix},
\quad \mathrm{T} =
\begin{pmatrix}
1& 0\\
0& e^{i\pi / 4}
\end{pmatrix},
\end{equation}
and multi-qubit entangling gates, such as the two qubit controlled-NOT (CNOT) gate shown in Fig.~\ref{Fig:CNOT}.
The action of the CNOT gate can be written mathematically as 
\begin{equation}
\ket{0}\bra{0}_C \otimes I_T + \ket{1}\bra{1}_C \otimes X_T ,
\end{equation}
where $T$ denotes the target qubit, and $C$ denotes the control qubit. \textcolor{black}{The matrix form of this operation is given by
\begin{equation}
\begin{pmatrix}
1& 0 & 0 & 0\\
0& 1 & 0 & 0\\
0& 0 & 0 & 1\\
0& 0 & 1 & 0
\end{pmatrix}.
\end{equation}}

\begin{figure}[hbt]
{\includegraphics{{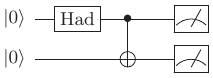}}}
\caption{A quantum circuit that generates the entangled state $(\ket{00}+\ket{11})/\sqrt{2}$ and measures each qubit in the computational basis. Time runs from left to right. Here, $\mathrm{Had}$ is the Hadamard gate, defined in the main text. \textcolor{black}{When measured, the qubits will either both be in the state 0 ($\ket{00}$), or both 1 ($\ket{11}$). Each of these two outcomes occurs with 50~\% probability.}}\label{Fig:circuit}
\end{figure}

These gates are used to create an example quantum circuit in Fig.~\ref{Fig:circuit}. This circuit generates the entangled state of 2 qubits given by Eq.~\eqref{Eq:bell1}, and then measures both of the qubits. 

With only single qubit operations and \textcolor{black}{CNOT gates}, it is possible to approximate an arbitrary multi-qubit gate to any desired accuracy~\cite{PhysRevA.51.1015}. As a result, the circuit model of quantum computing typically decomposes all algorithms into single and two qubit gates. We denote each gate by a unitary operator $U^{i,j}(\vec{\theta})$, where $i,j$ are the indices of the qubits the gates act on ($i=j$ for single qubit operations), and $\vec{\theta}$ are gate parameters (although the gates do not have to be parametrized, such as the Pauli gates). We can then mathematically describe a quantum circuit by
\begin{equation}
	\ket{\varphi} = \prod_k U^{i_k,j_k}_k(\vec{\theta}_k)\ket{\bar{0}},
\end{equation} 
where $k$ denotes the $k$\textsuperscript{th} gate in the circuit. The gates are ordered right to left. For example, the circuit in Fig.~\ref{Fig:circuit} would be written as

\begin{equation}
	\frac{1}{\sqrt{2}}(\ket{00}+\ket{11}) = \mathrm{CNOT}^{0,1} \mathrm{Had}^0 \ket{00}.
\end{equation} 

We extract information from the circuits by performing measurements of observables, $O$, which are represented by Hermitian matrices. Typically, we seek the average value over many measurements, $\bar{O}$, given by
\begin{equation}
	\bar{O} = \bra{\varphi}O\ket{\varphi},
\end{equation}
referred to as the expectation value of the operator $O$. Measuring the expectation value of qubit $i$ in the computational basis corresponds to $\bra{\varphi}Z_i\ket{\varphi}$. In practice, this means that we repeatedly prepare the state $\ket{\varphi}$, and measure the state of qubit $i$, labelling the outcomes $+1$ (for $\ket{0}$) and $-1$ (for $\ket{1}$). We then take the mean of these measurement outcomes. In order to measure qubits in the $X$ or $Y$ basis, single qubit rotations are first applied to change the basis of the relevant qubits, which are then measured in the $Z$ basis. \textcolor{black}{We can obtain the outcomes of measuring multi-qubit operators by taking the product of the measurement outcomes of single qubit operators measured in the same circuit iteration. For example, the expectation value of $Z_i Z_j$ could be obtained by preparing state $\ket{\varphi}$, measuring $Z_i$ on qubit $i$ and $Z_j$ on qubit $j$, multiplying these two measurement outcomes together, and averaging the results over many repetitions of this process.} These outcomes are typically correlated for entangled states. For example, the measurement outcome for $Z_1 Z_0$ on the state in Eq.~\eqref{Eq:bell1} is always $+1$ ($+1 \times +1$ for $\ket{00}$ and $-1 \times -1$ for $\ket{11}$).

The Pauli operators and identity matrix, multiplied by real coefficients, form a complete basis for any Hermitian single or \textcolor{black}{multi-qubit} operator. Therefore any \textcolor{black}{multi-qubit} observable can be expanded into strings of Pauli operators, the expectation values of which we can measure efficiently with a quantum computer. \\

It is important to distinguish between the number of \textit{physical} and \textit{logical} qubits in a quantum computer. \textcolor{Black}{Physical qubits are approximate two-level systems, which can be created in a range of different systems, including, but not limited to: energy levels in trapped ions~\cite{cirac1995trappedions,leibfried2003trappedionrmp}, polarization states of photons~\cite{Knill2001linearoptics}, spins in quantum dots~\cite{loss1998dots,hanson2007dotsrmp} or silicon~\cite{Kane1998silicon}, and energy levels of superconducting circuit resonators~\cite{shnirman1997superconducting,nakamura1999superconducting,wendin2017superconducting}.} In order to protect our qubits from decoherence caused by coupling to the environment~\cite{Landauer367, PhysRevA.51.992}, we can encode $m$ logical qubits into $n > m$ physical qubits. These logical qubits can simply be thought of as the abstract two level system described by Eq.~(\ref{EqQubit}). The codes used to construct logical qubits are analogous to classical error correcting codes, but are in general more complex, due to the delicate nature of quantum information and the `no-cloning theorem' of quantum mechanics. Depending on the code used, we can either detect, or detect \textit{and} correct the errors which occur. \textcolor{black}{The number of errors we can detect and/or correct depends on the code used (it is related to the \textit{distance} of the code)}. We must also account for the fact that the error checking measurements and correction procedure can cause additional errors to occur~\cite{knill1996accuracy}. We seek to build circuits in a `fault tolerant'~\cite{aharonov1997fault, gottesman1998theory,shor1996fault} manner, which limits the spread of errors during logical blocks of the computation. If this is achieved, then it is possible to scale up computations arbitrarily. If the physical error rate in the gates is below a certain (code-dependent) threshold value, the error rate in the logical operations can be made arbitrarily low, either by concatenation, or for certain codes, by growing the code. A more detailed discussion of error correction is given by \textcite{terhal2015error,devitt2013quantum, Raussendorf2012error,lidar2013qec}.

One of the most widely studied error correction codes is the surface code~\cite{kitaev1997quantum}, which is particularly suitable for 2D grids of qubits with nearest-neighbour connectivity. Physical error rates below the surface code threshold of around 1~\%~\cite{wang2011surface, fowler2012towards, stephens2014thresholds} have been achieved for trapped ion~\cite{HighFideity16,gaebler2016high} and superconducting~\cite{barends2014superconducting, google2019supremacy} qubits. However, with these error rates, we would require around $10^3 - 10^4$ physical qubits per logical qubit to perform interesting tasks in a fault-tolerant manner~\cite{fowler2012surface, PhysRevA.95.032338, campbell2017roads}. \textcolor{black}{For example, if we consider using a quantum computer to factor numbers in polynomial time~\cite{shor1994algorithms}, current estimates (at time of writing) suggest that we would require around 20 million physical qubits to factor a number that is too large to tackle using known classical algorithms~\cite{gidney2019factoring}. Building a machine of this size will be extremely difficult, in terms of isolating the qubits from the environment, developing scalable control systems, and minimising qubit crosstalk, as discussed by \textcite{gambetta2017errorcorrected,Ladd2010quantumcomputingreview}. We will discuss the comparably lower resources required for error corrected chemistry calculations in Sec.~\ref{Subsec:LongTermQuantumResources}. }\\

In contrast to the large error corrected machines described above, current quantum computers possess only tens of error-prone physical qubits. Nevertheless, quantum computers with more than around 50 qubits are considered too large to exactly simulate classically, and may thus be capable of solving problems which are intractable on even the largest classical supercomputers~\cite{google2019supremacy}. However, these problems are typically artificially constructed examples, rather than real-world problems~\cite{boixo2018characterizing, harrow2017quantum}. \textcite{preskill2018quantum} dubbed these machines `noisy intermediate-scale quantum' (NISQ) devices, and observed that it is currently unclear whether they will be able to outperform classical computers on useful tasks. The dichotomy between the resources needed for tackling problems like factoring, and the `supremacy' of a machine with more than around 50 qubits poses the question; \textit{`What, if anything, will near-term quantum computers be useful for?'}. The answer may lie with Feynman's original proposal; using quantum systems to simulate quantum systems.

\subsection{Quantum simulation}\label{Subsec:QuantumSimulation}
 
In this review, we focus on the digital quantum simulation of many-body quantum systems. Digital quantum simulation maps the target problem onto a set of gates which can be implemented by a quantum computer. A universal quantum computer can be programmed to perform many different simulations. This can be contrasted with analog quantum simulation, where the simulator emulates a specific real system of interest. However, analog simulators are generally considered more robust to noise, and therefore easier to construct \cite{RevModPhys.86.153}. To date, there have been several proposals for the simulation of chemistry using analog simulators~\cite{0953-4075-44-19-195302,huh2015boson,huh2017vibronic, chin2018quantum,arguello2018analog}, some of which have been experimentally realised~\cite{0295-5075-80-6-67008,clements2017experimental,HU2018293,sparrow2018simulating,shen2018vibronic}. Nevertheless, to perform accurate simulations of large chemical systems, we will likely require digital quantum simulation, as it is not yet clear how to protect large analog simulations from errors. Digital quantum simulation is more vulnerable to noise and device imperfections than analog simulation. While such imperfections can be addressed via error correction, this requires additional qubits and places stringent requirements on gate fidelities. In this review we focus solely on digital quantum simulation of chemistry problems. We refer the reader to the works of \textcite{houck2012chip,Schneidertrappedion,blatt2012quantum,aspuru2012photonic, RevModPhys.86.153} for information about digital quantum simulation of other physical systems, and analog quantum simulation.\\

\begin{figure*}[t]
{\includegraphics[width=15cm]{{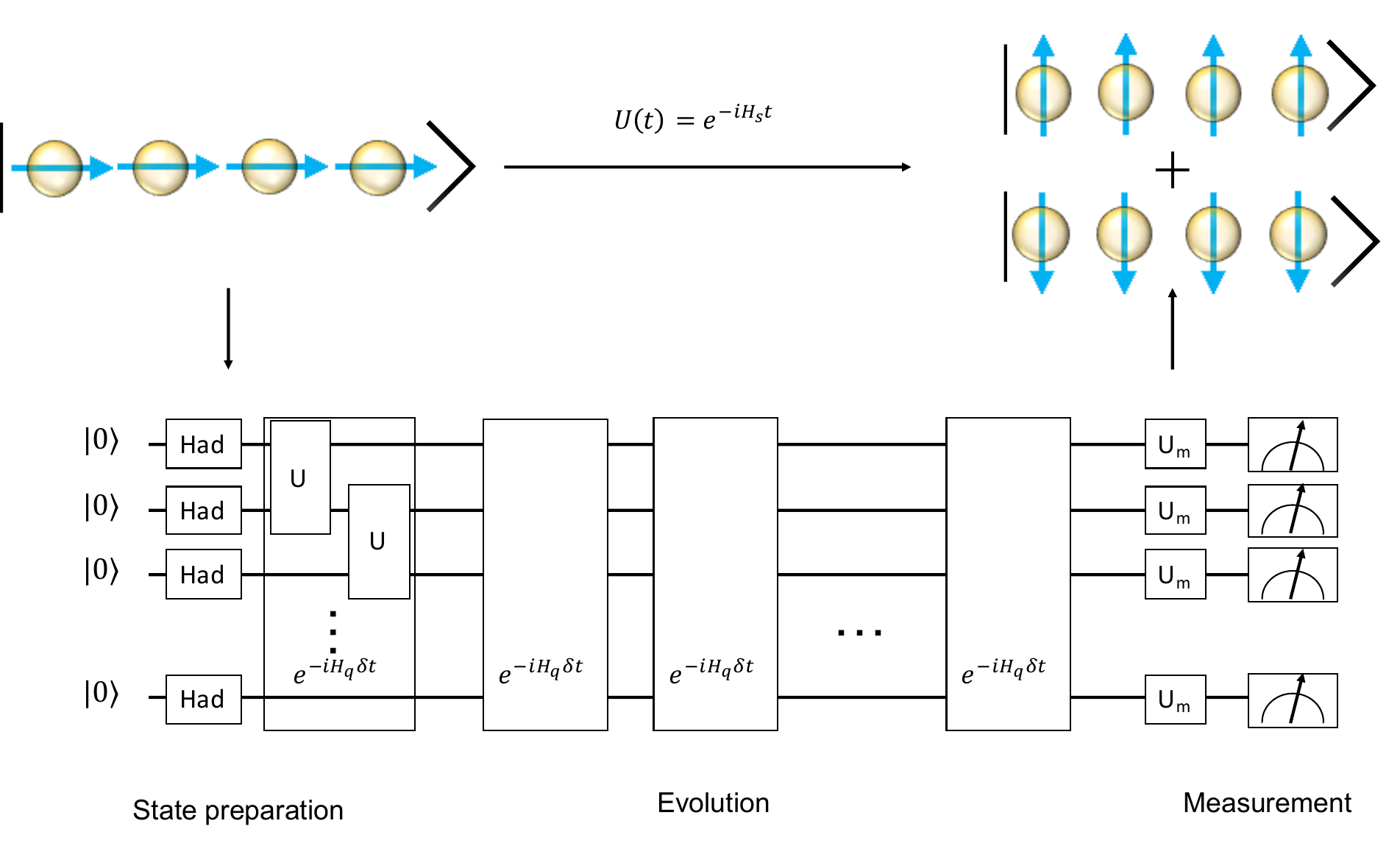}}}
\caption{Digital quantum simulation of time evolution of a spin chain, using a canonical Trotter-type method. We first map the system Hamiltonian, $H_{s}$, to a qubit Hamiltonian, $H_{q}$. Then the initial system wavefunction $\ket{\psi_{s}^i}$ is mapped to a qubit wavefunction $\ket{\psi_{q}^i}$. The time evolution of the system can be mapped to a Trotterized circuit that acts on the initial qubit wavefunction. Finally, well chosen measurements are applied to extract the desired information, such as particle correlation functions. \textcolor{black}{For a spin chain with an Ising Hamiltonian, $H = \sum_{\langle i,j \rangle} J_{ij} Z_i Z_j + \sum_i B_i X_i$, where the first sum is over nearest-neighbour spins $i$ and $j$, the unitaries $U_{ij}$ are given by $U_{ij} = \mathrm{CNOT}^{i,j}~R_z^j(2J_{ij})~ \mathrm{CNOT}^{i,j}~ R_x^i(2B_i)  $.}  }\label{Fig:dimulation}
\end{figure*}

The numerous problems in chemistry that can be simulated on a quantum computer can be divided into static and dynamics problems. Here, we use `dynamics' to mean evolving wavefunctions in time and seeing how certain observables vary, as opposed to chemical reaction dynamics, which are discussed separately below.\\

Methods for solving dynamics problems were formalised by \textcite{Lloyd1073} and further developed by \textcite{Abrams97}. As illustrated in Fig.~\ref{Fig:dimulation}, we can map the system Hamiltonian, $H_{s}$, to a qubit Hamiltonian, $H_{q}$. We similarly map the initial system wavefunction $\ket{\psi_{s}^i}$ to a qubit representation $\ket{\psi_{q}^i}$. We can then evolve the qubit wavefunction in time by mapping the system time evolution operator, $e^{-itH_{s}}$, to a series of gates. This can be achieved using a Lie-Trotter-Suzuki decomposition~\cite{trotter1959product}, commonly referred to as Trotterization or a Trotter circuit. This means that if the Hamiltonian of the system, $H_{s}$ can be written as 
\begin{equation}
H_s = \sum_j h_j,
\end{equation}
where $h_i$ are local terms which act on a small subset of the particles in the system, then a small time evolution under the Hamiltonian can be decomposed as
\begin{equation}\label{trotter}
e^{-iH_s \delta t} = e^{-i \sum_j h_j \delta t} \approx \prod_j e^{-i h_j \delta t} + \mathcal{O}(\delta t^2).
\end{equation}
The number of terms in the Hamiltonian scales polynomially with the number of particles for systems of interest in chemistry, \textcolor{black}{due to the two-body nature of the Coulomb interactions between particles~\cite{helgaker2014molecular}}. Each of the exponential terms in Eq.~(\ref{trotter}) can be realised efficiently using a quantum computer. As the dynamics of local Hamiltonians can be efficiently simulated on a quantum computer, but are generally thought to be inefficient to simulate on a classical computer, this problem belongs to the computational complexity class BQP (bounded-error quantum polynomial-time)~\cite{bernstein1997quantum,Lloyd1073}. It is important to note that, while it is widely believed to be true, it has not yet been mathematically proven that BPP $\neq$ BQP (where BPP is the complexity class containing problems that are solvable in polynomial time on a probabilistic classical computer). Further discussion of general Hamiltonian simulation methods is given in Sec.~\ref{Subsubsec:HamiltonianSimulation}.

It has also been shown that time evolution of open and closed quantum systems can be simulated using variational approaches~\cite{Li2017,endo2018variational,Fujirealtime19,chen2019demonstration}. A variational circuit is one with parametrized quantum gates, whose parameters are updated according to an algorithm specific update rule. They are discussed in more detail in Sec.~\ref{Subsec:VQE}. These techniques may enable simulation of time evolution using circuits with fewer gates than Trotterization. However, a variational circuit with set parameters is tailored to the time evolution of one or more specific initial states, in contrast to a Trotter circuit, which can be used to time evolve any valid initial state.

Once the system has been time evolved for the desired duration, we can extract useful dynamical quantities from these simulations. Examples of such quantities include the electronic charge density distribution, or particle correlation functions~\cite{Abrams97}. Further information on quantum dynamics simulation can be found in the review by \textcite{brown2010simulation}. \\

In chemistry, one is often concerned with determining whether two or more sub-systems will react with each other, when brought together with a certain energy. One might assume that this could be studied by simply initialising the reactants on the quantum computer, and time evolving under the system Hamiltonian, using the methods described above. However, it depends on the method used to map the chemical problem onto the quantum computer. We must take care to ensure that our model is able to accurately describe the system during all parts of the reaction. This is naturally taken care of using grid based methods, where the electrons and nuclei are treated on an equal footing, as discussed in Sec.~\ref{Subsubsec:firstqclassical} and Sec.~\ref{Subsubsec:FirstQRealSpaceMethods}. In contrast, if the problem is projected onto a finite basis set of electron spin-orbitals, we must be careful to ensure that: 1) The nuclear dynamics are accurately described, either through a precise classical treatment of the nuclear dynamics, or by using additional orbitals to describe the nuclear motion, and 2) The electron spin-orbitals used are able to accurately describe the positions of the electrons at all points in the reaction. This may require the orbitals to change in time. \textcolor{Black}{To the best of our knowledge, only the work by \textcite{berry2019timedep} considers a chemical reaction dynamics calculation on a quantum computer using basis set methods.} \\

We can obtain static properties by mapping the target wavefunction (such as the ground state wavefunction of the system) onto a qubit wavefunction. We can then use the quantum computer to calculate the expectation value of the desired observable, $\bra{\psi_{q}}O_{q}\ket{\psi_{q}}$. In particular, \textcite{Abrams99} showed that the phase estimation algorithm~\cite{kitaev1995phase} can be used to find the energy of a quantum system, and collapse it into the desired energy eigenstate. We will discuss this approach in Sec.~\ref{Subsec:qpe}. The ground state problem can also be tackled using variational algorithms~\cite{peruzzo2014variational}, which we will discuss in detail in Sec.~\ref{Subsec:VQE}.

Finding the low lying energy levels of a quantum Hamiltonian is in general an exponentially difficult problem for classical computers. Moreover, it is important to note that solving the ground state problem for a completely general local Hamiltonian is QMA-complete (quantum Merlin-Arthur complete, the quantum analogue of NP-complete)~\cite{kempe2006qma,cubitt2016qma}. Problems in this complexity class are not believed to be efficiently solvable with either a classical or quantum computer. \textcolor{black}{For such systems, it would appear that Nature itself cannot efficiently find the ground state.}

Despite this, the situation is not as bleak as it may initially seem. As stated in the introduction, we focus on finding the low lying energy levels of chemical systems (solving the electronic structure problem). It is widely believed that this problem should be efficiently solvable with a quantum computer, for physically relevant systems~\cite{whitfield2013complexity}. The electronic structure problem has received significant attention since it was first introduced in the context of quantum computational chemistry by \textcite{aspuru2005simulated}, and is widely considered to be one of the first applications of quantum computing. Solving the electronic structure problem is often a starting point for more complex calculations in chemistry, including the calculation of reaction rates, the determination of molecular geometries and thermodynamic phases, and calculations of optical properties~\cite{helgaker2012wavefunctionadvances}.

Before discussing how the electronic structure problem can be solved using a quantum computer in Sec.~\ref{Sec:Encoding} and Sec.~\ref{Sec:QuantumChemistryAlgorithms}, we first summarise the classical methods used to solve this problem in Sec.~\ref{Sec:classicalChemistry}. Many of these methods have formed the basis of the work done thus far in quantum computational chemistry.

\section{Classical computational chemistry}\label{Sec:classicalChemistry}
In this section, we introduce a selection of the most widely used techniques in \textit{ab initio} classical computational chemistry. As discussed in the introduction, we focus on tools developed to solve the electronic structure problem. The problem is formulated in Sec.~\ref{Subsec:problem}, and translated into the language of first and second quantisation in Sec.~\ref{Subsec:quantization}. In Sec.~\ref{Subsec:classicalchemistrymethods} we describe the different approximations that can be used to make this problem tractable for classical computers. In Sec.~\ref{Subsec:basis} we review some of the common spin-orbital basis functions used in basis set approaches to the molecular electronic structure problem. We discuss orbital basis changes, and their use in reducing the simulation resources in Sec.~\ref{Subsec:OrbitalReduction}. We have sought to produce a self-contained summary of the essential knowledge required for quantum computational chemistry, and we refer the reader to \textcite{helgaker2014molecular,szabo2012modern} for further information.

\subsection{The electronic structure problem}\label{Subsec:problem}
The Hamiltonian of a molecule consisting of $K$ nuclei and $N$ electrons is
\begin{equation}
\begin{aligned}
	H =& -\sum_i\frac{\hbar^2}{2m_e}\nabla^2_i  -\sum_I\frac{\hbar^2}{2M_I}\nabla^2_I - \sum_{i,I}\frac{e^2}{4\pi\epsilon_0}\frac{Z_I}{|\mathbf{r}_i-\mathbf{R}_I|}\\
	&+\frac{1}{2}\sum_{i\neq j}\frac{e^2}{4\pi\epsilon_0}\frac{1}{|\mathbf{r}_i-\mathbf{r}_j|}+\frac{1}{2}\sum_{I\neq J}\frac{e^2}{4\pi\epsilon_0}\frac{Z_IZ_J}{|\mathbf{R}_I-\mathbf{R}_J|}, 
\end{aligned}
\end{equation}
where $M_I$, $\mathbf{R}_I$, and $Z_I$ denote the mass, position, and atomic number of the $I$\textsuperscript{th} nucleus, and $\mathbf{r}_i$ is the position of the $i$\textsuperscript{th} electron. The first two sums in $H$ are the kinetic terms of the electrons and nuclei, respectively. The final three sums represent the Coulomb repulsion between: the electrons and nuclei, the electrons themselves, and the nuclei themselves, respectively. For conciseness, we work in atomic units, where the unit of length is $a_0 = 1$~Bohr ($0.529~\times~10^{-10}$~m), the unit of mass is the electron mass $m_e$, and the unit of energy is 1~Hartree ($1~\textrm{Hartree}~= ~{e^2}/{4\pi\epsilon_0a_0}~=~27.211~\textrm{eV}$). Denoting $M_I'=M_I/m_e$, the Hamiltonian in atomic units becomes
\begin{equation}
\begin{aligned}
	H =& -\sum_i\frac{\nabla^2_i}{2}  -\sum_I\frac{\nabla^2_I}{2M_I'} - \sum_{i,I}\frac{Z_I}{|\mathbf{r}_i-\mathbf{R}_I|}\\
	&+\frac{1}{2}\sum_{i\neq j}\frac{1}{|\mathbf{r}_i-\mathbf{r}_j|}+\frac{1}{2}\sum_{I\neq J}\frac{Z_IZ_J}{|\mathbf{R}_I-\mathbf{R}_J|}.
\end{aligned}
\end{equation}

We are predominantly interested in the \textit{electronic} structure of the molecule. As a nucleon is over one thousand times heavier than an electron, we apply the Born-Oppenheimer approximation, treating the nuclei as classical point charges. As a result, for a given nuclear configuration one only needs to solve the electronic Hamiltonian
\begin{equation}\label{ElectronicStructureH}
	H_e = -\sum_i\frac{\nabla^2_i}{2} -\sum_{i,I}\frac{Z_I}{|\mathbf{r}_i-\mathbf{R}_I|}+\frac{1}{2}\sum_{i\neq j}\frac{1}{|\mathbf{r}_i-\mathbf{r}_j|}.
\end{equation}

Our aim is to find energy eigenstates $\ket{E_i}$ and the corresponding energy eigenvalues $E_i$ of the Hamiltonian $H_e$. In the rest of this review, we drop the subscript $e$. In particular, we are interested in the ground state energy and the lowest excited state energies. We can solve this equation for a range of nuclear configurations to map out the potential energy surfaces of the molecule. Mapping out these potential energy curves explicitly is exponentially costly in the degrees of freedom of the molecule, and that there are a variety of methods being developed to solve this difficult problem more efficiently~\cite{christiansen2012vibrationalrev}. \\

We wish to measure the energy to an accuracy of at least $1.6 \times 10^{-3}$~Hartree, known as `chemical accuracy'. If the energy is known to chemical accuracy, then the chemical reaction rate at room temperature can be predicted to within an order of magnitude using the Eyring equation~\cite{eyring1935reaction,evans1935reaction}
\begin{equation}\label{Eyring}
\begin{aligned}
	\textrm{Rate} \propto e^{-\Delta E/ k_B T},
\end{aligned}
\end{equation}
where $T$ is the temperature of the system, and $\Delta E$ is the energy difference between the reactant and product states. In computational chemistry, we are typically more interested in the relative energies of two points on the potential energy surface than the absolute energy of a single point. Even if the individual energy values cannot be measured to within chemical accuracy, there is often a fortuitous cancellation of errors, which leads to the energy difference being found to chemical accuracy. However, in this review we consider chemical accuracy to mean an error of less than $1.6 \times 10^{-3}$~Hartree in the energy value at every point on the potential energy surface.

\textcolor{black}{\subsubsection{Chemical systems of interest}\label{Subsubsec:ProblemsOfInterest}
While classical computational chemistry has made tremendous progress in describing and predicting the properties of a multitude of systems, there are some systems that appear classically intractable to simulate with currently known techniques. Consequently, there has been significant interest in the possibility of using quantum computers to efficiently solve these problems. In particular, we are interested in solving so-called `strongly correlated systems' (we will explain this term more carefully in Sec.~\ref{Subsubsec:HFsec}. Here, it suffices to say that these are systems which possess wavefunctions with a high degree of entanglement. Many systems of commercial and scientific interest, such as catalysts and high temperature superconductors, are believed to be strongly correlated. In this section, we discuss two such strongly correlated systems that have been identified as interesting potential targets for a future quantum computer.}\\

\textcolor{black}{Many transition metals have found use as catalysts~\cite{vogiatzis2019catalysis}. However, many of these systems show strong correlation, often due to their open-shell nature, and spatially degenerate states~\cite{lyakh2012multireferenceCCreview}. This strong correlation often precludes their study \textit{in silico}, forcing us to carry out expensive, trial-and-error based discovery -- especially for transition metal based biological catalysts~\cite{podewitz2011bioinorganic}. One such system is the biological enzyme nitrogenase, which enables microorganisms which contain it to convert atmospheric dinitrogen (N$_2$) to ammonia (NH$_3$) under ambient conditions. This process is known as nitrogen fixation, and is one of the two main ways of producing ammonia for fertiliser -- the other is the energy intensive Haber-Bosch process. The Haber-Bosch process requires high temperatures and pressures, and is believed to consume up to 2~\% of the world's energy output~\cite{Reiher201619152}. While there has been significant experimental progress in understanding the structure of the nitrogenase enzyme over the last 100 years, the reaction mechanism is not yet fully understood~\cite{burgess1996nitrogenase}. The crux of understanding the enzyme appears to be several transition metal compounds found within it: the P cluster (containing iron and sulphur) and the iron molybdenum cofactor (FeMo-co, containing iron, molybdenum, carbon, hydrogen and oxygen)~\cite{hoffman2014nitrogenase}. Computational models of FeMo-co have been proposed which are beyond the reach of current classical methods for solving strongly correlated systems (see Sec.~\ref{Subsec:ClassicalLimits}) but would be accessible with a small error corrected quantum computer~\cite{Reiher201619152,berry2019qubitizationlowrank}. We discuss the resources required for these simulations in Sec.~\ref{Subsec:LongTermQuantumResources}.}\\

\textcolor{black}{High temperature superconductors are also believed to be strongly correlated systems. Since their discovery in the 1980's, there has been significant experimental and theoretical work to understand these compounds, which are not well described by the Bardeen-Cooper-Schrieffer theory of superconductivity. In the case of the so-called cuprate superconductors, it is widely believed that the mechanism of high temperature superconductivity is closely linked to the physics of the copper-oxygen planes that comprise them. A complete analytic or computational understanding of these layers is beyond the capabilities of current classical techniques. Several simplified models have been proposed, which have been found to capture some of the important behaviour of high temperature superconductors~\cite{dagotto1994correlatedelectrons}. In particular, models of fermions hopping on a 2D square lattice, such as the one-band Fermi-Hubbard model~\cite{hubbard1963hubbard}, and the related t-J model, appear to reproduce many experimental observations~\cite{anderson2002hubbard,lee2006hubbardmott}. While more complex models are likely required to fully understand the mechanism of cuprate superconductivity, a complete understanding of even the Fermi-Hubbard model is still elusive. Close to half filling, at intermediate interaction strengths, the system appears to show several competing orders in its phase diagram, which makes it difficult to reliably extract the ground state properties from classical numerical calculations~\cite{fradkin2015hubbardrmp}. Interestingly, this is the same regime that is believed to be most relevant to understanding cuprate superconductors~\cite{simonscollab2015hubbard}. Properties are typically obtained by performing ground state calculations on the Fermi-Hubbard model for a range of different system sizes, and then extrapolating to the thermodynamic limit~\cite{simonscollab2015hubbard}. We discuss the system sizes that can be tackled using modern classical techniques in Sec.~\ref{Subsec:ClassicalLimits}. We then examine the quantum resources required to surpass these calculations in Sec.~\ref{Subsec:LongTermQuantumResources}.}

\subsection{First and Second quantisation}\label{Subsec:quantization}

\textcolor{black}{As a consequence of the Pauli exclusion principle, the electronic wavefunction must be antisymmetric under the exchange of any two electrons. This antisymmetrisation can be accounted for in two ways, known as \textit{first} and \textit{second} quantisation. These names are largely historical; first quantisation was the approach initially taken by the pioneers of quantum mechanics, whereby variables like position and momentum are promoted to being operators (they are `quantised'). Second quantisation was developed afterwards, and quantises fields, rather than variables. There are key differences between these representations, which affect how simulations of physical systems are carried out using a quantum computer.}

As will be discussed in Sec.~\ref{Subsubsec:firstqclassical}, first quantised methods explicitly retain the antisymmetry in the wavefunction. In contrast, second quantisation maintains the correct exchange statistics through the properties of the operators which are applied to the wavefunction, as we will show in Sec.~\ref{Subsubsec:2ndqclassical}. These differences will become more apparent in the context of quantum computational chemistry mappings, which we will discuss in Sec.~\ref{Sec:Encoding}.

It is important to note that whether a simulation is carried out in first or second quantisation is distinct from whether a `basis set' or `grid based' method is used. This will be elaborated on in more detail in the following sections. \textcolor{black}{However, one key difference is that using} a basis set is known as a `Galerkin discretisation', which ensures that the energy converges to the correct value from above, as the number of basis functions tends to infinity. This property does not hold for grid based methods. A more detailed discussion on the differences between grid based and basis set methods can be found in the main text and Appendix~A of \textcite{Babbush2017low}.

\subsubsection{First quantisation}\label{Subsubsec:firstqclassical}

Here, we focus on classical first quantised simulation methods. Discussion of first quantised chemistry simulation on quantum computers is postponed until Sec.~\ref{Subsec:1stencoding}. \\

\paragraph{Grid based methods\\}
\addcontentsline{toc}{subsection}{\hspace{1cm}a. Grid based methods}
We consider the wavefunction in the position representation, $\{\ket{\mathbf{r}}\}$, which must be explicitly anti-symmetrised to enforce exchange symmetry. Mathematically, we describe the $N$ electron wavefunction as 
\begin{equation}
	\ket{\Psi} = \int_{\mathbf{x_1},\dots,\mathbf{x_N}} \psi(\mathbf{x_1},\dots,\mathbf{x_N}) \mathcal{A} \left( \ket{\mathbf{x_1},\dots,\mathbf{x_N}} \right) \mathrm{d}\mathbf{x_1},\dots,\mathrm{d}\mathbf{x_N} 
\end{equation}
where $\mathcal{A}$ denotes antisymmetrisation, $\mathbf{x_i} = (\mathbf{r}_i, \sigma_i) = (x_i, y_i, z_i,\sigma_i)$ gives the position and spin of the $i$\textsuperscript{th} electron and $\psi(\mathbf{x_1},\mathbf{x_2},\dots,\mathbf{x_N}) = \braket{\mathbf{x_1},\mathbf{x_2},\dots,\mathbf{x_N}|\Psi}$. We can simulate this system on a classical computer by evaluation of the wavefunction on a discretised spatial grid. However, the cost of storing the wavefunction scales exponentially with the number of electrons, $N$.
Suppose each axis of space is discretised into $P$ equidistant points. The discretised wavefunction is given by 
\begin{equation}\label{wave}
	\ket{\Psi} = \sum_{\mathbf{x_1},\dots,\mathbf{x_N}} \psi(\mathbf{x_1},\dots,\mathbf{x_N})\mathcal{A} \left( \ket{\mathbf{x_1},\dots,\mathbf{x_N}} \right),
\end{equation}
where $\ket{\mathbf{x_i}} = \ket{\mathbf{r_i}}\ket{\sigma_i}$ is a spatial and spin-coordinate, $\ket{\mathbf{r_i}} =  \ket{x_i}\ket{y_i}\ket{z_i},\forall i\in\{1,2,\dots,N\}$, $x_i, y_i, z_i\in\{0,1,\dots,P-1\}$, and $\sigma_i \in \{0,1\}$. In total, there are $P^{3N}\times 2^N$ complex amplitudes, showing that the memory required scales exponentially with the size of the simulated system. This makes it classically intractable to simulate more than a few particles on a grid using a classical computer. Consequently, we do not discuss classical methods that are specific to the grid based mapping in this review.

Grid based methods are useful when considering chemical reaction dynamics, or when simulating systems for which the Born-Oppenheimer approximation is not appropriate. In these scenarios, we must include the motion of the nuclei. If we consider the nuclear motion separately, we need to obtain the potential energy surfaces from electronic structure calculations. As mentioned in the previous section, mapping out these potential energy surfaces is exponentially costly. As such, it may be better to treat the nuclei and electrons on an equal footing, which is best achieved with grid based methods. This is discussed further by \textcite{kassal2008polynomial}.\\

\paragraph{Basis set methods \\}
\addcontentsline{toc}{subsection}{\hspace{1cm}b. Basis set methods}
We project the Hamiltonian onto $M$ basis wavefunctions, $\{{\phi_p(\mathbf{x_i})}\}$, where $\mathbf{x_i}$ is the spatial and spin coordinate of the $i$\textsuperscript{th} electron, $\mathbf{x_i} = (\mathbf{r_i}, \sigma_i)$. These basis functions approximate electron spin-orbitals. \textcolor{black}{The grid based method described above directly stores the wavefunction without exploiting any knowledge we may have about the general spatial form of the orbitals. In contrast, basis set methods exploit this knowledge to reduce the resources needed to simulate chemical systems.} We write the many electron wavefunction as a Slater determinant, which is an antisymmetrised product of the single electron basis functions. The wavefunction is given by
\begin{equation}\label{Slater2}
	\begin{aligned}
		&\psi(\mathbf{x_0} \dots \mathbf{x_{N-1}}) = \\
		  & \quad \frac{1}{\sqrt{N!}} \left.
		\begin{vmatrix}
		\phi_0(\mathbf{x_0}) & \phi_1(\mathbf{x_0}) &  ...  & \phi_{M-1}(\mathbf{x_0}) \\ 
		\phi_0(\mathbf{x_1}) & \phi_1(\mathbf{x_1}) &  ...  & \phi_{M-1}(\mathbf{x_1})\\
		. & . & . & . \\ 
		. & . & . & . \\ 
		. & . & . & . \\ 
		\phi_0(\mathbf{x_{N-1}}) & \phi_1(\mathbf{x_{N-1}}) &  ...  & \phi_{M-1}(\mathbf{x_{N-1}})  
		\end{vmatrix}
		 \right. .
	\end{aligned}
\end{equation}
Swapping the positions of any two electrons is equivalent to interchanging two rows of the Slater determinant, which changes the sign of the wavefunction. This provides the correct exchange symmetry for the fermionic wavefunction. While the number of spin-orbitals considered, $M$, is typically larger than the number of electrons in the system, $N$, the electrons can only occupy $N$ of the spin-orbitals in a given Slater determinant. As a result, the Slater determinant only contains the $N$ occupied spin-orbitals. 
\textcolor{black}{For example, imagine a fictitious system with two electrons, distributed among basis functions $A(\textbf{x})$ and $B(\textbf{x})$. Each of these basis functions could be occupied by an electron of either spin, so we effectively work with four basis functions $\{A_\uparrow(\textbf{x}), A_\downarrow(\textbf{x}), B_\uparrow(\textbf{x}), B_\downarrow(\textbf{x}) \}$. We consider the case where both electrons are in the $A$ orbitals (and therefore have opposite $s_z$ values, where $s_z$ is the $z$ component of the spin of the electron). As discussed above, the Slater determinant only contains the $N$ occupied spin-orbitals. We use Eq.~\eqref{Slater2} to obtain the wavefunction
\begin{equation}\label{SlaterExample}
	\begin{aligned}
		&\psi(\mathbf{x_0}, \mathbf{x_{1}}) =\\
		&  \quad \frac{1}{\sqrt{2}} \left.
		\begin{vmatrix}
		A_\uparrow(\mathbf{x_0}) & A_\downarrow(\mathbf{x_0})\\
		A_\uparrow(\mathbf{x_1}) & A_\downarrow(\mathbf{x_1})\\
		\end{vmatrix}
		 \right. = \\
		  & \frac{1}{\sqrt{2}}  \bigg{(} A_\uparrow(\mathbf{x_0}) A_\downarrow(\mathbf{x_1}) - A_\downarrow(\mathbf{x_0}) A_\uparrow(\mathbf{x_1})  \bigg{)}.
	\end{aligned}
\end{equation}
This wavefunction is antisymmetric under the exchange of the two electrons, as required.}

\textcolor{black}{As we will see in Sec.~\ref{Subsubsec:2ndqclassical}, the information encoded by a Slater determinant can be compressed by moving to the second quantised formalism. As a consequence, first quantised basis set methods are rarely, if ever, used in classical computational chemistry calculations. Nevertheless, it is important to be aware of what the wavefunction looks like in the first quantised basis set representation for two reasons. Firstly, as discussed above, the second quantised basis set method follows directly from the first quantised basis set approach. Secondly, first quantised basis set approaches have found use in quantum computational chemistry, as we will discuss in Sec.~\ref{Subsubsec:FirstQBasisSetMethods}.}

\subsubsection{Second quantisation}\label{Subsubsec:2ndqclassical}

\paragraph{Basis set methods\\}
\addcontentsline{toc}{subsection}{\hspace{1cm}a. Basis set methods}

The second quantised basis set approach follows naturally from the first quantised basis set method discussed in Sec.~\ref{Subsubsec:firstqclassical}. We again project the Hamiltonian onto $M$ basis wavefunctions, $\{{\phi_p(\mathbf{x_i})}\}$, and consider many-electron wavefunctions that must be antisymmetric under the exchange of any two electrons.  As we saw in Sec.~\ref{Subsubsec:firstqclassical}, to write down a Slater determinant we only need to indicate which spin-orbitals are occupied by electrons. This enables the introduction of a convenient shorthand for Slater determinants~\cite{szabo2012modern}
\begin{equation}\label{Slater}
    \begin{aligned}
    \psi(\mathbf{x_0} \dots \mathbf{x_{N-1}})   
		=\ket{f_{M-1}, \dots, f_p,\dots, f_0} = \ket{f},
    \end{aligned}
\end{equation}
where $f_p = 1$ when $\phi_p$ is occupied (and therefore present in the Slater determinant), and $f_p =0$ when $\phi_p$ is empty (and therefore not present in the determinant). The vector $\ket{f}$ is known as an occupation number vector, and the space of all such vectors is known as `Fock space'. The second quantised formalism is concerned with manipulating these occupation number vectors. As these occupation number vectors are a convenient short-hand for Slater determinants, we will refer to them throughout this review as Slater determinants. This is common practice in computational chemistry~\cite{szabo2012modern}.\\

Electrons are excited into the single electron spin-orbitals by fermionic creation operators, $a_p^\dag$. They are de-excited by annihilation operators, $a_p$. These operators obey fermionic anti-commutation relations
\begin{equation}\label{AntiComm}
\begin{aligned}
	\{a_p, a_q^\dag\} &= a_p a_q^\dag + a_q^\dag a_p = \delta_{pq}, \\
	\{a_p, a_q \} &= \{a_p^\dag, a_q^\dag\} = 0 .
\end{aligned}
\end{equation}

The determinants $\ket{f}$ form an orthonormal basis for the Fock space of the system. The actions of the fermionic operators on the determinants are given by
\begin{equation}\label{creationAnnOper}
	\begin{aligned}
		&a_p\ket{f_{M-1},f_{M-2},\dots, f_0} \\
		=& \delta_{f_p,1}(-1)^{\sum_{i = 0}^{p-1}f_i}\ket{f_{M-1},f_{M-2},\dots, f_p\oplus1,\dots, f_0},\\ \\
		&a_p^\dag\ket{f_{M-1},f_{M-2},\dots, f_0}\\
		 =& \delta_{f_p,0}(-1)^{\sum_{i = 0}^{p-1}f_i}\ket{f_{M-1},f_{M-2},\dots, f_p\oplus1,\dots, f_0},  
	\end{aligned}
\end{equation}
where $\oplus$ denotes addition modulo 2, \textcolor{black}{such that $ 0 \oplus 1 = 1$, $ 1 \oplus 1 = 0$}. The phase term $(-1)^{\sum_{i = 0}^{p-1}f_i}$ enforces the exchange
anti-symmetry of fermions. The spin-orbital occupation operator is given by 
\begin{equation}\label{numberOp}
	\begin{aligned}
		\hat{n}_i &= a^\dag_i a_i, \\
		\hat{n}_i \ket{f_{M-1},\dots, f_i, \dots, f_0} &= f_i \ket{f_{M-1},\dots, f_i, \dots, f_0}, \\
	\end{aligned}
\end{equation}
and counts the number of electrons in a given spin-orbital.\\

Observables must be independent of the representation used. Therefore, the expectation values of second quantised operators must be equivalent to the expectation values of the corresponding first quantised operators. As first quantised operators conserve the number of electrons, the second quantised operators must contain an equal number of creation and annihilation operators. We can use these requirements to obtain the second quantised form of the electronic Hamiltonian~\cite{helgaker2014molecular,szabo2012modern}. 
\begin{equation}\label{FH}
	H = \sum_{p,q}h_{pq}a^\dag_p a_q + \frac{1}{2}\sum_{p,q,r,s}h_{pqrs}a^\dag_p a^\dag_q a_ra_s,
\end{equation}
with 
\begin{equation}
	\begin{aligned}\label{Integrals}
		h_{pq}&=\int \mathrm{d}\textbf{x} \phi_p^*(\textbf{x}) \left(-\frac{\nabla^2}{2} -\sum_{I}\frac{Z_I}{|\mathbf{r}-\mathbf{R}_I|}\right) \phi_q(\mathbf{x}),\\
		 h_{pqrs}&=\int \mathrm{d}\mathbf{x}_1 \mathrm{d}\mathbf{x}_2\frac{\phi_p^*(\mathbf{x}_1) \phi_q^*(\mathbf{x}_2)\phi_r(\mathbf{x}_2) \phi_s(\mathbf{x}_1)}{|\mathbf{r}_1-\mathbf{r}_2|}.
	\end{aligned}
\end{equation}
\textcolor{black}{The first integral represents the kinetic energy terms of the electrons, and their Coulomb interaction with the nuclei. The second integral is due to the electron-electron Coulomb repulsion. The Hamiltonian only contains terms with up to four creation and annihilation operators (two creation, two annihilation) because the Coulomb interaction between the electrons is a two-body interaction.} As a result, the Hamiltonian contains up to $M^4$ terms, depending on the basis functions used. We examine the form of these basis functions, and how to select them in Sec.~\ref{Subsec:basis}. \textcolor{Black}{A special case of the electronic structure Hamiltonian is obtained for the Fermi-Hubbard model, introduced in Sec.~\ref{Subsubsec:ProblemsOfInterest}. The Fermi-Hubbard Hamiltonian considers fermions hopping between nearest-neighbour lattice sites with strength $t$. These fermions feel a repulsive (or attractive) force $U$ when they occupy the same lattice site, $i$. The Hamiltonian is given by
\begin{equation}
H=-t \sum_{\langle i, j\rangle, \sigma}\left(a_{i, \sigma}^{\dagger} a_{j, \sigma}+a_{j, \sigma}^{\dagger} a_{i, \sigma}\right)+U \sum_{i} n_{i, \uparrow} n_{i, \downarrow}
\end{equation}
where $\langle i, j\rangle$ denotes a sum over nearest-neighbour lattice sites, and $\sigma$ is a spin-coordinate. This Hamiltonian has only $\mathcal{O}(M)$ terms, where $M$ is the number of spin-sites. For convenience, throughout the rest of this review we will refer to both molecular spin-orbitals and lattice spin sites as spin-orbitals.} \\

Let us consider general and approximate solutions of the electronic structure Hamiltonian. If the electron-electron Coulomb interaction term in Eq.~\eqref{ElectronicStructureH} is neglected, we obtain a new Hamiltonian which describes the behaviour of $N$ independent electrons. We can define a suitable basis for this fictitious system as the set of molecular orbitals which diagonalise the non-interacting Hamiltonian. These molecular orbitals are typically linear combinations of orbitals localised around each of the atoms. We note that in practice, the molecular orbitals obtained by diagonalising the non-interacting part of the Hamiltonian will likely form a poor basis for the system. Instead, \textcolor{black}{a mean-field approximation (the Hartree-Fock procedure, to be described in Sec.~\ref{Subsubsec:HFsec})} can be used to obtain more suitable molecular orbitals. \textcolor{black}{As these single-particle molecular orbitals are chosen such that they diagonalise the non-interacting (or mean-field) Hamiltonian, energy eigenstates can be formed by taking tensor products of each electron in a different molecular spin-orbital. In order to obey the Pauli exclusion principle, these tensor products must be correctly anti-symmetrised. This can be achieved by creating Slater determinants from the molecular orbitals, as described by Eq.~\eqref{Slater2}}.

As they are eigenstates of a Hermitian operator, these Slater determinants form a complete basis of the problem Hilbert space. Consequently, the eigenstates of the true Hamiltonian can be expressed as linear combinations of these Slater determinants, written as
\begin{equation}\label{2ndQwavefunc}
	\begin{aligned}
		\ket{\Psi} = \sum_f \alpha_f \ket{f} ,
	\end{aligned}
\end{equation}
where $\alpha_f$ are complex coefficients which we refer to herein as the determinant amplitudes. These solutions are exact, provided that the molecular orbitals form a complete basis for the single particle states, and the $N$-electron wavefunction contains all of the determinants that these molecular orbitals can generate~\cite{helgaker2014molecular,szabo2012modern}. If all ${M} \choose {N}$ determinants are included, the wavefunction is known as the full configuration interaction (FCI) wavefunction. However, this wavefunction contains a number of determinants which scales exponentially with the number of electrons, making large calculations classically intractable. 

Second quantised basis set methods are the most widely used approach in classical computational chemistry, and have formed the basis of most of the work done so far in quantum computational chemistry. As a result, we discuss some of the approximate methods used in second quantised basis set simulations in Sec.~\ref{Subsec:classicalchemistrymethods} and Sec.~\ref{Subsec:basis}. In Sec.~\ref{Subsec:classicalchemistrymethods} we consider making ground state calculations classically tractable by approximating the exact ground state wavefunction with a restricted number of Slater determinants. In Sec.~\ref{Subsec:basis} we consider approximating the exact wavefunction by considering only the most important molecular orbitals. \textcolor{black}{However, first we briefly discuss the limited work that has been done on second quantised grid based methods, for the sake of completeness.}\\

\paragraph{Grid based methods\\}
\addcontentsline{toc}{subsection}{\hspace{1cm}b. Grid based methods}
\textcolor{black}{To the best of our knowledge, second quantised grid based methods have only been discussed by \textcite{Babbush2017low} in their Appendix A. Nevertheless, these methods follow naturally from the discussion of second quantised basis sets above. Our discussion of the topic closely follows the derivation of \textcite{Babbush2017low}. As a first step, we can consider our real space grid to be described by a set of basis functions that are delta functions, $\delta(r-r_i)$, each positioned at grid point $r_i = (x_i,y_i,z_i)$. The creation operators $a_{i,\sigma}^\dag$ then become $a_{x_i,y_i,z_i,\sigma}^\dag$~; rather than creating an electron in spin-orbital $i$, they now create an electron with spin $\sigma$ at the grid point $(x_i,y_i,z_i)$ in 3D space. As the basis functions do not overlap in space, the kinetic energy operator must be defined using a finite difference formula, rather than the integral in Eq.~(\ref{Integrals}). We must also define a suitable inner product between functions defined on the grid. These steps allow us to calculate the coefficients of each term in the Hamiltonian. As discussed above, this second quantised grid based method has not yet (to the best of our knowledge) been used in any classical or quantum computational chemistry algorithms.}

\subsection{Classical computation methods}\label{Subsec:classicalchemistrymethods}
In this section, we review four methods for approximating the ground state wavefunction with a restricted number of Slater determinants: the Hartree-Fock (HF), multiconfigurational self-consistent field (MCSCF), configuration interaction (CI), and coupled cluster (CC) methods. These methods create parametrized trial states, which can then be optimised to approach the ground state (to an accuracy determined by the approximations made). In order to isolate the errors arising from the method used, we assume that we are working in the full molecular orbital basis for our molecule, although in practice this would be classically intractable for large system sizes (with the size of the system dependent upon the accuracy of the method used). Restricting the size of the basis set will be discussed in Sec.~\ref{Subsec:basis}.

The methods discussed below are considered in the context of second quantised basis set calculations, as these translate most easily to the methods used in quantum computational chemistry. These methods are among the most straightforward and widely used in classical computational chemistry.

\subsubsection{Hartree--Fock}\label{Subsubsec:HFsec}
The Hartree--Fock (HF) method is a mean-field technique which aims to find the dominant Slater determinant in the system wavefunction. This is achieved by optimising the spatial form of the spin-orbitals to minimise the energy of the wavefunction. We generally consider a set of spin-orbitals, $M$, that is larger than the number of electrons in the molecule, $N$. As we only consider a single Slater determinant, we are essentially assuming that $N$ of the spin-orbitals are occupied, and $M-N$ are left unoccupied, or virtual. In the HF method, we first neglect the Coulomb repulsion term in the electronic structure Hamiltonian (Eq.~\eqref{ElectronicStructureH}), reducing the problem to one of $N$ independent electrons. We then assume that each electron moves in the average charge distribution of all of the other electrons, which introduces an effective potential. We can solve the $N$ coupled equations iteratively; first calculating the position of each electron, then updating the potential, and repeating this process until the orbitals converge. In the second quantised formalism, this procedure is carried out by using the orbitals to construct the `Fock operator', and diagonalising the Fock operator to obtain new orbitals. This process is repeated until the orbitals converge, and so HF is also referred to as the self-consistent field (SCF) method. The Fock operator, $\hat{f}$, is given by~\cite{helgaker2014molecular}
\begin{equation}\label{HartreeFock}
	\begin{aligned}
		\hat{f} &= \sum_{i,j} (h_{ij} + V_{ij}) a^\dag_i a_j, \\
		V_{ij} &= \sum_{k \in occ} (h_{ikkj} - h_{ikjk}),
	\end{aligned}
\end{equation}
where $V_{ij}$ describes the effective potential, and $occ$ is the set of occupied orbitals. We see that the Fock operator depends on the spatial form of the orbitals through $h_{ij}$, $h_{ikkj}$, and $h_{ikjk}$ which are obtained by calculating the integrals in Eq.~(\ref{Integrals}). When performing a HF calculation, we typically input a set of atomic orbitals, which are localised around each atom. These orbitals are used to calculate the Fock operator, which is then diagonalised to obtain new orbitals (which are linear combinations of the old orbitals). This process is repeated until the orbitals converge~\cite{szabo2012modern}. The new orbitals obtained are referred to as the canonical orbitals. This procedure generates single particle molecular orbitals from combinations of the single particle atomic orbitals.

The term $h_{ikkj}$ describes the Coulomb interaction of an electron with the charge distribution of the other electrons, while the term $h_{ikjk}$ describes exchange effects (also called Fermi correlation) arising from the required antisymmetrisation. However, as a mean-field solution, the HF method neglects the effects of dynamic and static correlation in the wavefunction. 

Dynamic correlation is a typically small correction~\cite{hattig2012correlatedreview}, arising from the Coulomb repulsion between electrons. Wavefunctions displaying dynamic correlation are often dominated by the Hartree-Fock determinant, and have small additional contributions from a (potentially large) number of excited state determinants. Static correlation occurs when more than one Slater determinant is equally dominant in the wavefunction~\cite{hattig2012correlatedreview}. In this case, the Hartree-Fock method provides a poor approximation to the ground state wavefunction. \textcolor{black}{The presence of strong static correlation can be evidenced by multiple near-degenerate solutions of the Hartree-Fock procedure.} Static correlation can arise because the wavefunction may require several determinants to be coupled in a proper spin configuration, or during bond breaking in order to account for the separation of the electrons onto the products (the latter case is often referred to as non-dynamic, or left-right correlation)~\cite{lyakh2012multireferenceCCreview}. The relative contribution of these effects depends on the orbital basis used. \textcolor{black}{\textcite{lyakh2012multireferenceCCreview} note that, while it is sometimes possible to reduce the level of static correlation by manually enforcing correct spin symmetries, and using appropriately localised orbitals, it is in general not possible to avoid strong static correlation when considering the whole of the potential energy surface.} As a result, static correlation often dominates in many systems of scientific interest, such as excited states~\cite{lischka2018excitedstatesmultiref}, or transition metals~\cite{vogiatzis2019catalysis}.\\

The Slater determinant generated from a HF calculation is typically taken as the reference state for post-HF methods, such as configuration interaction and coupled cluster, which seek to capture some of the dynamic correlation energy by including additional determinants, describing excitations above the HF state. \textcolor{black}{Although we will discuss orbital basis sets in more detail in Sec.~\ref{Subsec:basis}, we note here that the HF orbitals are not suitable for describing the virtual orbitals that electrons are excited into. This is because the HF method only optimises the occupied orbitals in the single Slater determinant wavefunction. In order to obtain suitable virtual orbitals, we can instead perform correlated calculations on individual atoms. The details of how to perform these correlated atomic calculations are beyond the scope of this review, and are not relevant for quantum computing applications. We refer the interested reader to Section 8.3 of the textbook by \textcite{helgaker2014molecular} for additional details. The importance of each of the virtual atomic orbitals is determined by their contribution to either the electron density (atomic natural orbitals) or the atomic correlation energy (correlation-consistent basis sets). These two metrics are closely related, and while either can be used, the latter produces more compact basis sets, and hence is used more frequently in practice~\cite{helgaker2014molecular}. We will discuss the use of correlation consistent basis sets further in Sec.~\ref{Subsubsec:correlatedbasissets}.}

\subsubsection{Multiconfigurational self-consistent field}\label{Subsubsec:MCSCF}
As discussed above, the Hartree-Fock method performs poorly for strongly correlated systems. Systems with strong static correlation are defined as those where multiple Slater determinants are equally dominant. These include excited states~\cite{lischka2018excitedstatesmultiref}, transition states~\cite{szalay2012mcscf}, systems at the dissociation limit~\cite{lyakh2012multireferenceCCreview}, and transition metals~\cite{vogiatzis2019catalysis}. In order to account for static correlation, we need to use a wavefunction which exhibits the required multireference nature. One such approach is the multiconfigurational self-consistent field (MCSCF) method. The MCSCF approach considers a wavefunction with several Slater determinants, and variationally optimises both the molecular orbitals, and the determinant amplitudes simultaneously~\cite{roos1980casscf}. \textcolor{black}{Mathematically, we write our MCSCF wavefunction as $\ket{\Psi} = e^{-\kappa} \sum_f \alpha_f \ket{f}$, where again $\alpha_f$ are determinant amplitudes, and $\ket{f}$ are Slater determinants. Here, $\kappa$ is an anti-Hermitian operator given by $\kappa = \sum_{i,j} k_{ij} a_i^\dag a_j$. Exponentiating $\kappa$ produces a unitary operator which rotates the orbital basis~\cite{helgaker2014molecular}. We then variationally minimise the energy by optimising both the amplitudes $\alpha_f$, and the entries $k_{ij}$ of $\kappa$.}

MCSCF can be considered the best approximation to the exact wavefunction for a given number of determinants~\cite{wang2008spectrum}. It is not possible to perform a complete MCSCF calculation on all possible determinants for systems with more than a few electrons, as the number of determinants scales exponentially with the number of electrons. We can attempt to use chemical intuition to select the most important Slater determinants, and perform an MCSCF calculation on this restricted number of determinants. Alternatively, we can use the complete active space self-consistent field (CASSCF) method~\cite{roos1980casscf}. This considers only the most important orbitals (an active space, see Sec.~\ref{Subsec:OrbitalReduction}) and performs an MCSCF calculation on all of the determinants that could be generated from distributing a certain number of electrons in these orbitals. MCSCF and CASSCF calculations are among the most effective classical methods at treating systems with strong static correlation~\cite{szalay2012mcscf}. \textcolor{Black}{Recent approaches, including replacing the CASSCF subroutine with tensor network methods, have enabled the treatment of even larger active spaces~\cite{knecht2016dmrgchemistry}. However, while the methods described above are often effective at recovering the static correlation energy, they struggle to recover the dynamic correlation energy. This requires additional techniques, such as multireference configuration interaction (see Sec.~\ref{Subsubsec:CI} and \textcite{szalay2012mcscf}) and multireference coupled cluster (see Sec.~\ref{Subsubsec:CC} and \textcite{lyakh2012multireferenceCCreview}) .}

\subsubsection{Configuration interaction}\label{Subsubsec:CI}
The configuration interaction (CI) method generates a correlated wavefunction by considering excitations above a reference state, typically the Hartree-Fock state. If all determinants are included, we recover the full configuration interaction (FCI) wavefunction, generated by considering all excitations above the Hartree-Fock wavefunction
\begin{equation}
	\begin{aligned}
	&|\Psi_{\mathrm{FCI}} \rangle \\
	=& \left(I + \sum_{i,\alpha} C_{i \alpha} a_i^\dag a_{\alpha} + \sum_{i > j,\alpha > \beta} C_{i j \alpha \beta} a_i^\dag a_j^\dag a_\alpha a_\beta + ...\right) |\Psi_{\mathrm{HF}} \rangle,
	\end{aligned}
\end{equation}
where $C$ are parameters to be optimised according to the Rayleigh-Ritz variational principle. \textcolor{black}{The Rayleigh-Ritz variational principle states that the energy expectation value of a parametrized wavefunction is greater than or equal to the lowest energy eigenvalue of the Hamiltonian being measured.} As considering all determinants is classically intractable, the CI method is typically limited to including a small number of excitations above the reference state: single excitations (CIS), double excitations (CISD), and occasionally triple excitations (CISDT). However, as low energy excitations dominate the ground state wavefunction in many physical systems, these truncations can still produce good approximations to the ground state energy~\cite{helgaker2014molecular,szabo2012modern}. The CI method is effective at recovering dynamic correlation, but less effective at recovering static correlation~\cite{helgaker2014molecular}. If the reference state is a MCSCF state, the method is known as multireference configuration interaction, which seeks to recover dynamic correlation on top of the static correlation described by the MCSCF state. 

The CI method suffers from two major limitations. The method converges slowly to the full configuration interaction wavefunction, as a result of its linear parametrization. In addition, the energy obtained from a truncated CI calculation does not scale correctly with the system size. A calculation is said to produce `size extensive' results for an observable if the correct scaling behaviour of that observable is obtained as the system is scaled to the thermodynamic limit~\cite{lyakh2012multireferenceCCreview}. The energy obtained from a truncated CI calculation is not size extensive. Truncated CI calculations also fail to satisfy a related property, known as `size consistency'. Size consistency arises from multiplicative separability of wavefunctions describing two non-interacting systems; i.e. $\ket{AB} = \ket{A}\ket{B}$ for $H_{AB} = H_A \oplus H_B$. One can easily verify that this does not hold for truncated CI calculations. Size consistency is important for obtaining the correct behaviour from calculations to determine reaction rates, as reactants at the beginning of the process are considered non-interacting.

\subsubsection{Coupled cluster}\label{Subsubsec:CC}
The coupled cluster (CC) method also includes additional determinants to recover the correlation energy, but uses a product parametrization. The CC wavefunction is given by
\begin{equation}
	\begin{aligned}
	|\Psi_\mathrm{CC} \rangle =& \prod_{i,\alpha} \left(I + C_{i \alpha} a_i^\dag a_\alpha\right) \times\\
	&\prod_{i > j,\alpha > \beta} \left(I + C_{i j \alpha \beta} a_i^\dag a_j^\dag a_\alpha a_\beta\right) \times ... |\Psi_\mathrm{HF} \rangle.
	\end{aligned}
\end{equation}
This formula can be recast in an exponential form, written as
\begin{equation}
	\begin{aligned}
	&|\Psi_\mathrm{CC}\rangle =  e^{T} |\Psi_\mathrm{HF} \rangle, \\
	\end{aligned}
\end{equation}
where $T = \sum_i T_i$, 
\begin{equation}
	\begin{aligned}
	&T_1 =  \sum_{i \in virt, \alpha \in occ} t_{i \alpha} a^\dag_i a_\alpha, \\ 
	&T_2 = \sum_{i, j \in virt, \alpha, \beta \in occ} t_{i j \alpha \beta}  a^\dag_i a^\dag_j a_\alpha a_\beta, \\
	&...,
	\end{aligned}
\end{equation}
where $occ$ denotes orbitals that are occupied in the Hartree-Fock state, $virt$ denotes orbitals that are unoccupied (virtual) in the Hartree-Fock state, and $t$ are excitation amplitudes. When all of the excitation operators $T_i$ are included, the CC method recovers the full configuration interaction wavefunction -- however, performing this calculation would be exponentially costly. As a result, the method is normally truncated at a lower excitation level, often single and double excitations (CCSD). \textcolor{Black}{The canonical implementation of CCSD does not store the wavefunction, as this would be exponentially costly (as the CCSD wavefunction has support on all possible Slater determinants). Instead, coupled non-linear equations can be derived. The solution to these equations is the CCSD approximation to the ground state~\cite{helgaker2014molecular, purvis1982coupled}. The time taken to solve these equations scales as $\mathcal{O}((M-N)^4 N^2)$~\cite{purvis1982coupled}, while the memory needed to store the molecular integrals needed scales as $\mathcal{O}(M^4)$. For more accurate results, the CCSD(T) method can be used, which treats the triple excitations pertubatively, and scales in time approximately as $\mathcal{O}(M^7)$. There has been significant work to reduce these high computational costs, often introducing approximations which exploit the locality of dynamical electron correlation in certain systems~\cite{schutz2000localCC,schutz2000localCC2}. This has reduced the scaling to be, in some cases, linear~\cite{schutz2011localCC3}.}

Because of its product parametrization, the CC method generates a trial wavefunction which includes all possible determinants, albeit with an incorrect parametrization. It therefore provides faster convergence than the configuration interaction method. The product parametrization also ensures size extensivity and size consistency. However, the CC method is not without its own shortcomings. Most notably, the wavefunction generated by the canonical CC method does not obey the Rayleigh-Ritz variational principle~\cite{helgaker2014molecular}. While it is possible to formulate alternative variants of the coupled cluster method, which are variational~\cite{voorhis2000variationalCC}, these are not as widely used by the computational chemistry community. \textcolor{Black}{Moreover, the conventional CC method described above is a single determinant reference state method. Consequently, it does not tend to perform well when applied to multireference states, which are required to treat systems with strong static correlation~\cite{lyakh2012multireferenceCCreview, lischka2018excitedstatesmultiref}. While there have been efforts to develop multireference coupled cluster approaches, these have their own limitations, and are not in widespread use, as discussed by \textcite{lyakh2012multireferenceCCreview}.} In Sec.~\ref{Subsec:VQE} we will describe a modified form of the CC method, known as unitary coupled cluster (UCC). This method is both variational and suitable for multireference states. While it is exponentially costly to implement with a classical computer, this method is efficient to implement using a quantum computer. \\

This section has treated the inaccuracies which result from approximating the full configuration interaction wavefunction, while including all molecular orbitals. The following section will discuss the converse case; we consider only a limited number of molecular orbitals, but assume that we include all possible determinants that they can generate, unless explicitly stated.

\subsection{Chemical basis sets}\label{Subsec:basis}
In this section, we describe some of the conventional orbital basis sets used in classical computational chemistry. Throughout this section, we will refer to the `true' orbitals of the system. These can be obtained by numerically solving the Schr\"odinger equation using grid based methods with a very fine grid spacing, which is only possible for small atoms or simple molecules. The orbital functions introduced in this section are approximations of these true orbitals.

Although the Schr\"odinger equation can be solved exactly for one electron atoms, the orbitals obtained become diffuse too rapidly to accurately describe many-electron atoms, especially close to the nuclei~\cite{helgaker2014molecular}. A better basis can be obtained by considering parametrized functions known as Slater-type orbitals (STO)
\begin{equation}\label{STO}
	\begin{aligned}
		R^{\mathrm{STO}}_n(r) \propto (\zeta r)^{n-1} e^{-\zeta r}, 
	\end{aligned}
\end{equation}
where $n$ is the energy level and $\zeta$ is a fitting parameter. By using different values of $\zeta$ for each orbital, we can generate a good basis~\cite{helgaker2014molecular}. Unlike the true atomic orbitals, these functions do not display oscillatory behaviour. Consequently, linear combinations of STOs are required to approximate the true orbitals. It is possible to only introduce a single basis function for each considered orbital in the molecule, and give each basis function a different $\zeta$ value. This is known as a single-zeta representation. Alternatively, we can introduce $n$ basis functions (where $n$ is not the energy level of the orbital, but a number defining the number of basis functions we wish to include), each with a different $\zeta$ value, for each orbital. This is known as an $n$-zeta representation. Introducing additional basis functions in this way increases the radial flexibility of the wavefunction. While the STO functions exhibit many desirable features, they make evaluating the two-electron integrals in Eq.~\eqref{Integrals} computationally expensive.

To simplify the two-electron integrals, we can instead use Gaussian basis functions. The Gaussian basis functions are obtained by considering the Schr\"odinger equation with a three dimensional Harmonic oscillator potential. The form of a Gaussian-type orbital (GTO) is given by
\begin{equation}\label{GTO}
	\begin{aligned}
		R^{\mathrm{GTO}}_{nl}(r) \propto (\sqrt{\alpha_{nl}} r)^l e^{-\alpha_{nl} r^2}, 
	\end{aligned}
\end{equation}
where $\alpha_{nl}$ is a fitting parameter and $l$ denotes the angular momentum quantum number of the orbital. Because of the dependence on $r^2$ in the exponent, GTOs are more localised than STOs. As a result, GTOs do not approximate the atomic charge distribution as well as STOs, so more are required to describe a given orbital. However, this limitation is compensated by the ease of integral evaluation. Furthermore, the disadvantages of GTOs are less prominent in molecular calculations~\cite{helgaker2014molecular}.

The most common basis sets construct approximate STOs from linear combinations of GTOs. These approximate STOs are used as the basis functions for our atomic orbitals. 
The number and type of orbitals defines the basis set. There is a compromise between the accuracy obtained and the number of basis functions used. The number of orbitals considered determines the runtime and memory requirements of classical chemistry algorithms. In the case of quantum computational chemistry, the number of basis functions determines the number of qubits and gate operations required to solve the problem, which we will discuss explicitly in Sec.~\ref{Sec:Encoding} and Sec.~\ref{Sec:QuantumChemistryAlgorithms}, respectively.

\subsubsection{STO-$n$G and split-valence basis sets}\label{Subsubsec:stoorbs}
Some of the most simple bases are the STO-$n$G basis sets (Slater Type Orbital-$n$ Gaussians)~\cite{hehre1969stong}. In an STO-$n$G basis, each atomic orbital is considered to be an approximate STO. The STOs are approximated using $n$ GTOs. STO-$n$G basis sets are often called minimal basis sets, as they contain only the orbitals required to write the Hartree--Fock (HF) state (and those orbitals of similar energy). Calculations using minimal basis sets are of limited accuracy, giving only a qualitative description of the system. It is important to note that when carrying out a HF calculation in an STO-$n$G basis, the true HF energy (i.e. the energy obtained by performing a HF calculation using a grid based method, on an infinitely precise grid) will not be obtained, as the STO-$n$G basis sets only approximate the true HF orbitals. As an example of an STO-$n$G basis set we consider lithium, which has 3 electrons, of which 2 can reside in the $1s$ orbital, leaving 1 in the second energy level. We include in the minimal basis set $\{1s, 2s, 2p_x, 2p_y, 2p_z\}$ orbitals. We include both the $2s$ and $2p$ orbitals because they are of the same energy level.

More accurate basis sets can be formed by adding increased radial flexibility to the valence orbitals (the orbitals of the highest occupied energy level), by considering a double-zeta representation of the valence orbitals. This can be achieved using split-valence~\cite{Pople} basis sets, such as the 6-31G basis. These basis sets can be further improved by adding additional orbitals with higher angular momenta, which make the angular part of the wavefunction more flexible. These orbitals are called `polarisation functions', as they describe the polarisation of atomic charge caused by bonding (for example, the 6-31G* basis).

\textcolor{black}{Although the aforementioned basis sets are too small for calculations of such high accuracy that quantum computers are warranted, it is important to discuss them here, as they have been used extensively in the small experimental demonstrations possible on today's quantum hardware.}

\subsubsection{Correlation-consistent basis sets}\label{Subsubsec:correlatedbasissets}
Additional accuracy can be obtained by using \mbox{cc-PV$n$Z} basis sets (correlation consistent polarised valence $n$ zeta), introduced by \textcite{Dunning}. These include additional unoccupied (`virtual') orbitals to recover the correlation energy. The virtual orbitals are generated from correlated calculations on atoms. The core orbitals have a single-zeta representation, while the valence orbitals have an $n$-zeta representation. The virtual orbitals considered are polarisation functions, with higher angular momenta than the valence orbitals. The polarisation functions are selected by the size of their contribution to the correlation energy.

For atomic hydrogen in the cc-PVDZ (D $=$ double, so $n=2$) the highest occupied energy level (the valence level) is the first level, and so we take a double-zeta representation of the $1s$ state, considering \{$1s$, $1s'$\} orbitals. The $1s'$ orbital is often referred to as a $2s$ orbital. This is because the additional function chosen to describe the valence orbital has the same angular momentum as the ordinary $1s$ orbital, but is more diffuse -- so it resembles a $2s$ orbital. We then include polarisation functions, which have a higher angular momentum value than the valence functions. In total, there are five basis functions for cc-PVDZ hydrogen: \{$1s$, $1s'$, $2p_x$, $2p_y$, $2p_z$\}. For lithium in the cc-PVDZ basis, the core orbital is \{$1s$\}. The valence orbitals (which have a double-zeta representation) are \{$2s$, $2p_x$, $2p_y$, $2p_z$, $2s'$, $2p_x'$, $2p_y'$, $2p_z'$\}, and the polarisation functions are \{$3d_{zz}$, $3d_{xz}$, $3d_{yz}$, $3d_{xy}$, $3d_{x^2 - y^2}$ \}, which we write as \{$5\times 3d$\}. This yields 14 basis functions. For lithium in the cc-PVTZ basis (T $=$ triple so $n=3$), we first include the 14 orbitals above. As we consider a triple-zeta representation of the valence orbitals, we need additional \{$2s'', 2p_x'', 2p_y'', 2p_z''$\} orbitals. We then include additional polarisation functions; \{$5 \times 3d', 7 \times 4f$\}. This leads to a total of 30 orbitals. \\

cc-PV$n$Z basis sets with higher values of $n$ contain orbitals that better approximate the true atomic orbitals than those with lower $n$ values. However, even large ($n=5$) basis sets struggle to exactly represent the true HF orbitals of simple molecules such as N$_2$~\cite{helgaker2014molecular}. This limitation can be overcome by measuring the ground state energy in several different cc-PV$n$Z bases, and then extrapolating to the basis set limit. \\

\begin{figure}[t]
{\includegraphics[width=9cm]{{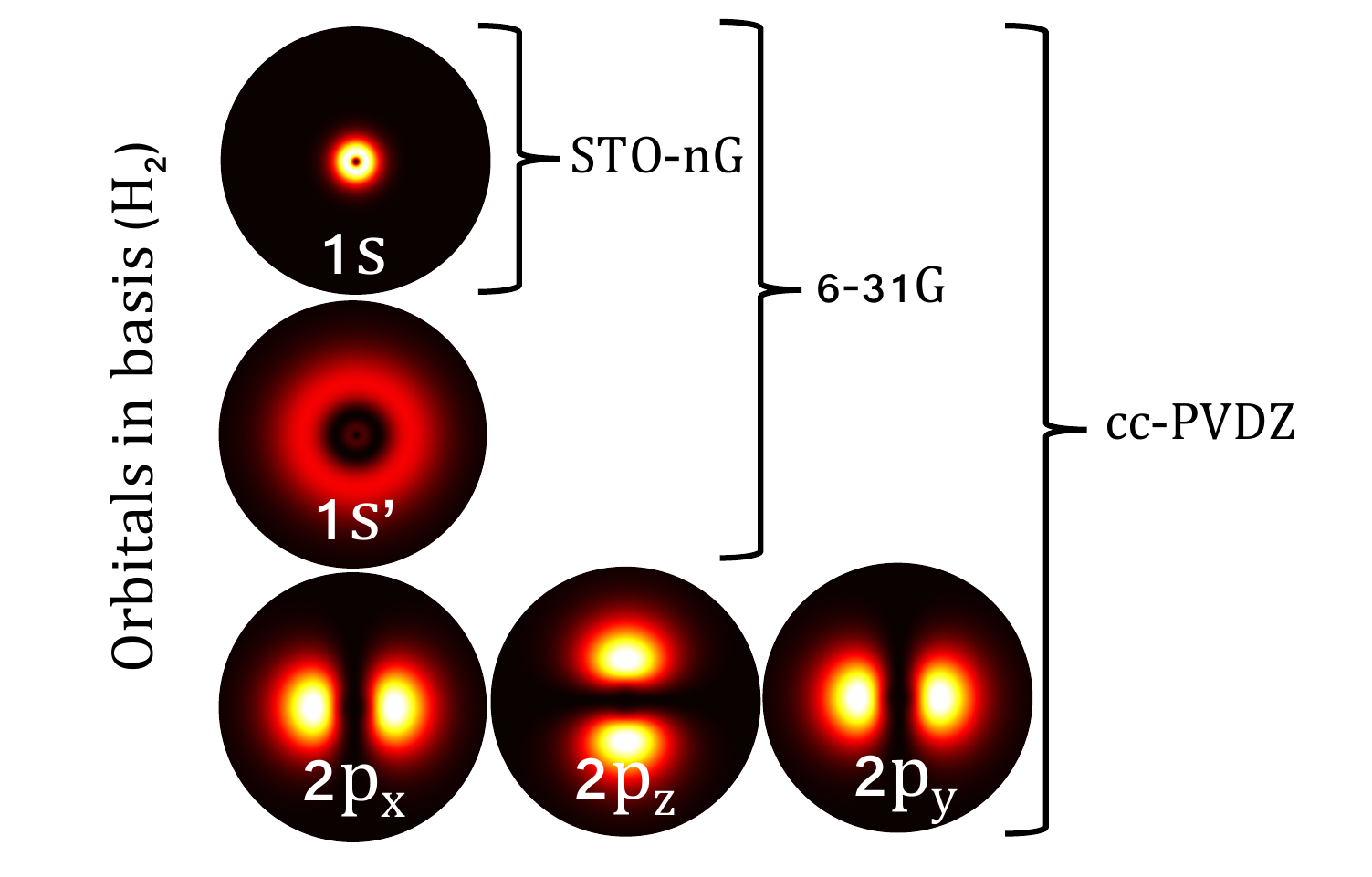}}}
\caption{The orbitals included in different basis sets for the di-hydrogen molecule. The $1s'$ orbital is often written as $2s$. The plots show the radial probability distributions for the true Hydrogenic orbitals, which the basis orbitals approximate.}\label{Fig:Basis}
\end{figure}

\subsubsection{Plane wave basis sets}\label{Subsubsec:PlaneWave}
While the aforementioned basis sets have a long history of use in classical computational chemistry (and as a result, early work in quantum computational chemistry), they are not necessarily optimal basis sets for calculations performed on quantum computers. While these basis sets result in an accurate description of the system with relatively few basis functions, they also lead to Hamiltonians containing up to $\mathcal{O}(M^4)$ terms. As we will see in Sec.~\ref{Sec:QuantumChemistryAlgorithms}, the number of terms in the Hamiltonian plays a key role in the cost of some quantum chemistry algorithms. It is therefore interesting to question whether there are basis sets that are more useful for quantum computational chemistry. Two examples of such bases are the plane wave and plane wave dual basis sets introduced for quantum computing by \textcite{Babbush2017low}. The plane wave basis functions, $\phi_{\nu}(r)$, are given by
\begin{equation}\label{PlaneWave}
	\begin{aligned}
		\phi_{\nu} = \sqrt{\frac{1}{V}} \exp\bigg{(}\frac{2 \pi i \nu r}{L}\bigg{)},
	\end{aligned}
\end{equation}
for a plane wave with wavevector corresponding to the $\nu$\textsuperscript{th} harmonic of the computational cell with length $L$ and volume $V$. The plane wave dual basis is obtained by taking the discrete Fourier transform of the plane wave basis states, so are like a smooth approximation to a grid~\cite{Babbush2017low}. These basis sets diagonalise the kinetic and potential operators, respectively. This reduces the number of Hamiltonian terms from $\mathcal{O}(M^4)$ to $\mathcal{O}(M^3)$ in the plane wave basis, and $\mathcal{O}(M^2)$ in the plane wave dual basis. \textcolor{black}{This in turn leads to a reduction in the asymptotic scaling of quantum chemistry algorithms to find the ground state energy of molecules and solid state systems (we will discuss the magnitude of this potential speedup in Sec.~\ref{Sec:QuantumChemistryAlgorithms})}. These plane wave basis sets are well suited to periodic systems, and have a long history of use in classical density functional theory calculations. However, to describe molecular systems, approximately 10 to 100 times as many plane wave basis functions are required as GTOs~\cite{Babbush2017low}. 

A similar reduction in the number of Hamiltonian terms can be obtained using gausslet basis sets~\cite{white2017gausslet}, or `discontinuous Galerkin' sets~\cite{mcclean2019galerkin}, which can both require fewer functions to accurately describe individual molecules than plane wave basis sets. Creating efficient basis sets for quantum computational chemistry remains an open and fundamental area of research.

\subsection{Reduction of orbitals}\label{Subsec:OrbitalReduction}
It is sometimes the case that certain orbitals are very likely to be either occupied or virtual in all Slater determinants in the wavefunction. As calculating the ground state energy is essentially a question of distributing electrons among orbitals, we can simplify our calculation by using this information. Specifically, we are able to remove spin-orbitals from the calculation if their expected occupation number is close to 0 or 1. Our calculation is reduced to including only the most important (ambiguously occupied) orbitals. This is known as performing the calculation in a reduced active space.

In order to determine the occupation of orbitals, we can use the reduced density matrices (RDMs) of the system. The expectation value of any 1- or 2-electron Hermitian operator, $O$, with a state $\ket{\Psi} = \sum_f \alpha_f \ket{f}$, is given by~\cite{helgaker2014molecular}
\begin{equation}\label{RDMS}
	\begin{aligned}
	\bra{\Psi} O \ket{\Psi} = \sum_{i,j} O_{ij} \rho^1_{ij} + \sum_{i,j,k,l} V_{ijkl} \rho^2_{ijkl}, \\ 
	\rho^1_{ij} = \bra{\Psi} a^\dag_i a_j \ket{\Psi}, \quad \rho^2_{ijkl} = \bra{\Psi} a^\dag_i a^\dag_k a_l a_j \ket{\Psi},
	\end{aligned}
\end{equation}
where $\rho^1$ is the single-particle reduced density matrix (1-RDM), $\rho^2$ is the two-particle reduced density matrix (2-RDM), and $O_{ij}$ and $V_{ijkl}$ are defined in a similar way to the coefficients in Eq.~(\ref{Integrals}). When eliminating orbitals in this way, the RDMs are defined with respect to a state which is an approximation of the ground state, which could be the results of a classically tractable configuration interaction or coupled cluster calculation. These RDMs contain all of the information required to evaluate $\langle O \rangle_\Psi$. From the definition above, we can see that the diagonal elements of $\rho^1$ are the expectation values of the number operator for the corresponding orbitals. As $\rho^1$ is a Hermitian operator, we can diagonalise it with a unitary transform. This is a basis change from the canonical orbitals to the `natural molecular orbitals'. The diagonal elements of the basis transformed $\rho^1$ are called the natural orbital occupation numbers (NOONs).

Spin-orbitals with a NOON close to 0 or 1 (compared to the other NOONs) can be assumed to be empty or occupied, respectively. Occupied orbitals are typically referred to as `core' orbitals, while the empty orbitals are known as `virtual' orbitals. As a result, we can reduce our problem by considering only the ambiguously occupied orbitals. In Sec.~\ref{Sec:illustrate} we provide an explicit example of how this method can be used to reduce the number of orbitals required to simulate lithium hydride in an STO-3G basis set.

\textcolor{black}{
\textcite{takeshita2019virtualorbs} showed how to re-integrate some of the energy contribution of the virtual orbitals into a quantum simulation, without requiring additional qubits to represent the virtual orbitals. Their method utilises an increased number of measurements, and classical post-processing. We will describe this method in more detail in Sec.~\ref{Subsubsec:SubspaceExpansionExcited}. Those authors also provide a technique to improve the energy estimate from a calculation by optimising the active space using orbital rotations, as is done in MCSCF calculations (Sec.~\ref{Subsubsec:MCSCF}).}\\

This section has introduced the concepts in classical computational chemistry necessary to understand how quantum computers can be used for chemical simulation. The following sections introduce methods developed to solve chemistry problems using quantum computers. We return to classical computational chemistry methods in Sec.~\ref{Subsec:ClassicalLimits}, where we assess the strengths, weaknesses, and computational limits of the methods introduced here.

\section{Quantum computational chemistry mappings}\label{Sec:Encoding}
In this section, we describe the techniques developed to enable quantum computers to represent problems in chemistry. In Sec.~\ref{Subsec:1stencoding} and Sec.~\ref{Subsec:2ndencoding} we introduce methods for encoding fermions into qubits (both in first and second quantisation). We then describe methods which utilise knowledge of the structure of chemistry problems to reduce the resources required, in Sec.~\ref{Subsec:reduction}. As discussed in Sec.~\ref{Subsec:quantization} the distinguishing feature between first and second quantised methods is whether antisymmetry is enforced in the wavefunction directly (first quantised), or in the behaviour of the operators which act on the wavefunction (second quantised). As in the previous section, we consider a system with $M$ spin-orbitals (when discussing basis set approaches) and $N$ electrons. \textcolor{black}{We summarise the number of qubits required to store the wavefunction in each of the representations in Table~\ref{Quantizationtable}.}

\begin{table}[!h]
\caption{\textcolor{black}{The number of qubits required to store the wavefunction using the different encoding methods. $M$ is the number of spin-orbitals used in a basis set simulation, $N$ is the number of simulated particles in the problem, and $m=\mathrm{log_2}(P)$, where $P$ is the number of grid points per axis for a grid based simulation. The number of qubits can often be further reduced, as discussed in Sec.~\ref{Subsubsec:HamiltonianReduction}.}} 
\begin{tabular}{c|c} 
\hline 
Mapping & Number of qubits \\
\hline 
First quantised, basis sets & $N\ceil{\mathrm{log_2}(M)}$\\
\hline 
First quantised, grid based & $(3m+1)N$\\
\hline 
Second quantised, basis sets & $M$ \\
\hline 
\end{tabular} 
\label{Quantizationtable} 
\end{table}

In order to make the mappings more clear, we show how the wavefunction would look under each mapping for a fictitious system. When considering a basis set mapping of this system, we consider spin-orbitals $\ket{A_\uparrow}$, $\ket{A_\downarrow}$, $\ket{B_\uparrow}$, $\ket{B_\downarrow}$. \textcolor{black}{We are free to arbitrarily define the Hartree-Fock state of our fictitious system, and choose it to be both electrons in the $\ket{A}$ orbital. We are interested in the wavefunction when the $z$ component of the spin is zero.} When considering a grid based approach, for the sake of simplicity we consider spinless electrons on a 1D grid. The two single particle wavefunctions considered are approximately `n' shaped (see Eq.~\eqref{gridbasedex1}) and `u' shaped (see Eq.~\eqref{gridbasedex2}).

\subsection{First quantised encoding methods}\label{Subsec:1stencoding}
Here we give an overview of first quantised quantum simulation, which can be carried out using either a discrete single-particle basis, or grid based methods.

\subsubsection{Grid based methods}\label{Subsubsec:FirstQRealSpaceMethods}
As discussed in Sec.~\ref{Subsubsec:firstqclassical}, the wavefunction of an $N$-particle system can be represented in real space on a discretised grid of $P$ points per axis, and is given by
\begin{equation}\label{wave2}
	\ket{\Psi} = \sum_{\mathbf{x_1},\dots,\mathbf{x_N}} \psi(\mathbf{x_1},\dots,\mathbf{x_N})\mathcal{A} \left( \ket{\mathbf{x_1},\dots,\mathbf{x_N}} \right),
\end{equation}
where $\ket{\mathbf{x_i}} = \ket{\mathbf{r_i}}\ket{\sigma_i}$ is a spatial and spin-coordinate, $\ket{\mathbf{r_i}} =  \ket{x_i}\ket{y_i}\ket{z_i},\forall i\in\{1,2,\dots,N\}$, $x_i, y_i, z_i\in\{0,1,\dots,P-1\}$, and $\sigma \in \{0,1\}$. We consider the case where $P=2^m$, where $m$ is an arbitrary number which determines the precision of our simulation. While it is classically intractable to store the required $P^{3N} \times 2^N =2^{(3m+1)N}$ complex amplitudes for large quantum systems, it is possible using a quantum computer. If we write the basis vector $\ket{x=2^m-1}$ in binary as $\ket{11....11}$, we can see that it only requires $m$ bits. An $m$ qubit register can be in a superposition of $2^m$ possible states. As a result, it only requires $(3m+1)N$ qubits to store the $N$ electron wavefunction described by Eq.~(\ref{wave2}). Using a grid based method, rather than basis sets, means that the Born-Oppenheimer approximation is not required. As a result, we are able to treat the electrons and nuclei on an equal footing using grid based methods, which can be important for systems undergoing reactions. 

\textcolor{black}{In order to make the first quantised grid based mapping more understandable, we consider three example wavefunctions. Without loss of generality, we neglect the spin coordinate of the electrons, and consider only a single spatial dimension for each electron. The first example considers a single spinless electron, on a four point grid. The `n'-shaped wavefunction considered is given by
\begin{equation}\label{gridbasedex1}
\begin{aligned}
	\ket{\varphi} &= \frac{1}{\sqrt{6}} \ket{00} + \frac{1}{\sqrt{3}} \ket{01} + \frac{1}{\sqrt{3}} \ket{10} + \frac{1}{\sqrt{6}} \ket{11}\\
	&= \frac{1}{\sqrt{6}} \ket{\textbf{0}} + \frac{1}{\sqrt{3}} \ket{\textbf{1}} + \frac{1}{\sqrt{3}} \ket{\textbf{2}} + \frac{1}{\sqrt{6}} \ket{\textbf{3}}
\end{aligned}
\end{equation}
and can be stored using two qubits. The second example considers an antisymmetrised product state of two electrons, where one electron is in the state $\ket{\varphi}$, and the other has a `u'-shaped wavefunction
\begin{equation}\label{gridbasedex2}
	\ket{\phi} = \frac{1}{\sqrt{3}} \ket{\textbf{0}} + \frac{1}{\sqrt{6}} \ket{\textbf{1}} + \frac{1}{\sqrt{6}} \ket{\textbf{2}} + \frac{1}{\sqrt{3}} \ket{\textbf{3}}.
\end{equation}
The pair wavefunction is then given by
\begin{equation}\label{gridbasedex3}
\begin{aligned}
	\ket{\Phi} =& \frac{1}{\sqrt{2}}\left(\ket{\varphi}_1\ket{\phi}_2 - \ket{\phi}_1\ket{\varphi}_2 \right), \\
	=& \frac{1}{6\sqrt{2}}( \ket{\textbf{1}}_1\ket{\textbf{0}}_2 - \ket{\textbf{0}}_1\ket{\textbf{1}}_2 + \ket{\textbf{1}}_1\ket{\textbf{3}}_2 - \ket{\textbf{3}}_1\ket{\textbf{1}}_2  \\ 
	&+ \ket{\textbf{2}}_1\ket{\textbf{0}}_2 - \ket{\textbf{0}}_1\ket{\textbf{2}}_2 + \ket{\textbf{2}}_1\ket{\textbf{3}}_2 - \ket{\textbf{3}}_1\ket{\textbf{2}}_2),
\end{aligned}
\end{equation}
where the subscripts label the electrons. In general, an entangled state of two spinless electrons on a $P$ point 1D grid can be written in this way as
\begin{equation}\label{gridbasedex4}
	\ket{\Psi} = \sum^{P-1}_{i=0} \sum^{P-1}_{j=0} \psi_{ij} \left( \ket{\textbf{i}}_1\ket{\textbf{j}}_2 - \ket{\textbf{j}}_1\ket{\textbf{i}}_2 \right),
\end{equation}
where again, the subscripts label the electrons.}

Grid based methods were first introduced for the quantum simulation of general quantum systems by \textcite{wiesner1996simulations,zalka1998simulating}. They were then adapted for simulating problems in chemistry by \textcite{lidar1999thermal} and \textcite{kassal2008polynomial}. Physically relevant states can then be prepared using the algorithms outlined by \textcite{doi:10.1063/1.3115177}. \textcite{kassal2008polynomial} showed how to time evolve the wavefunction under the electronic stucture Hamiltonian. As we will discuss in Sec.~\ref{Subsec:qpe}, time evolution is a key subroutine of algorithms that find the ground states of chemical systems. Finally, the relevant observables can be measured~\cite{kassal2008polynomial, whitfield2015unified}. A thorough investigation of the resources required to perform these simulations in a fault-tolerant manner was carried out by \textcite{jones2012faster}. The time evolution algorithm of \textcite{kassal2008polynomial} was subsequently made more efficient by \textcite{KivReal}, who also performed a more thorough analysis of both gate counts and errors. We will discuss the method used by \textcite{KivReal} in more detail in Sec.~\ref{Subsubsec:ChemistryTimeEvolution}.

Although the spatial resolution of grid based methods increases exponentially with the number of qubits used, it is not possible to use this to exponentially improve the accuracy of the calculation. This is because all known grid based algorithms have gate counts which scale polynomially with the inverse grid spacing. As a result, any attempt to exponentially increase the simulation accuracy by exponentially reducing the grid spacing causes the gate count to increase exponentially. \textcite{KivReal} also showed that there exist systems where the grid spacing must decrease exponentially with the number of particles in the system to maintain constant accuracy. Consequently, these systems are not efficient to simulate using this method. However, those authors noted that such pathological cases can also exist for basis set methods, but are typically dealt with efficiently using a clever choice of basis function.

The simulation of chemical systems using a grid based method can require considerably more qubits than in basis set approaches. For example, it would require 96 logical qubits to store the position of a single spinless particle to 32 bits of accuracy using a grid based approach. This can be contrasted with basis set approaches, where interesting molecules or Fermi-Hubbard models could be simulated with around 100 logical qubits (as we will discuss in Sec.~\ref{Subsec:LongTermQuantumResources}). Consequently, grid based approaches are typically considered unsuitable for near-term quantum computers, which will have relatively few qubits.

\subsubsection{Basis set methods}\label{Subsubsec:FirstQBasisSetMethods}
The original algorithm for simulating quantum systems in the first quantisation using a discrete basis was given by \textcite{Abrams97}. If we consider $M$ single-particle basis functions (such as the molecular orbitals or lattice spin sites described in Sec.~\ref{Subsubsec:2ndqclassical}), we can enumerate these from $0$ to $M-1$. We can store these spin-orbitals using $\ceil{\mathrm{log_2}(M)}$ qubits, denoting spin-orbital 0 as $\ket{0...00}$, spin-orbital 1 as $\ket{0...01}$ and so on, such that spin-orbital $M-1$ is represented as $\ket{1...11}$. We then use $N$ registers of these $\ceil{\mathrm{log_2}(M)}$ qubits (one register for each electron) to describe the states of all of the electrons in the system. As a result, it requires $N\ceil{\mathrm{log_2}(M)}$ qubits to store the wavefunction.

If we consider a product state generated by each electron being in a single orbital, we observe that the wavefunction does not have the correct antisymmetry. As such, it must be antisymmetrised. The original approach, by \textcite{Abrams97}, accomplished this using $\mathcal{O}(N^2 \mathrm{log_2^2}(M))$ gates and $\mathcal{O}(N \mathrm{log_2}(M))$ ancilla qubits. This was improved by \textcite{berry2018improved}, who used a circuit with $\mathcal{O}(N\mathrm{log_2^c}(N)\mathrm{log_2}(M))$ gates, with a depth of $\mathcal{O}(\mathrm{log_2^c}(N) \mathrm{log_2log_2}(M))$, where $c \geqslant 1$ and depends on the choice of sorting network used, and $\mathcal{O}(N\mathrm{log_2}(N))$ ancilla qubits.

\textcolor{black}{We can apply the first quantised basis set mapping to the fictitious system described above. We first label each of the orbitals: $\ket{A_\uparrow} = \ket{00} = \ket{\textbf{0}}$, $\ket{A_\downarrow} = \ket{01} = \ket{\textbf{1}}$, $\ket{B_\uparrow} = \ket{10} = \ket{\textbf{2}}$, $\ket{B_\downarrow} = \ket{11} = \ket{\textbf{3}}$. The Hartree-Fock state has both electrons in the $\ket{A}$ orbitals. An incorrectly symmetrised HF state would therefore be \textcolor{black}{$\ket{A_\uparrow}_1 \ket{A_\downarrow}_2 = \ket{\textbf{0}}_1\ket{\textbf{1}}_2$}, where the subscripts denote which electron each orbital describes. The correctly antisymmetrised HF wavefunction would be
\begin{equation}\label{Eq:1stQExampleMappingHF}
\ket{\Psi_{\mathrm{HF}}} = \frac{1}{\sqrt{2}} (\ket{\textbf{0}}_1\ket{\textbf{1}}_2 - \ket{\textbf{1}}_1\ket{\textbf{0}}_2 ).
\end{equation}
\textcolor{black}{If we now consider excitations above the HF state, then a general wavefunction with $s_z=0$ that has been correctly antisymmetrised is given by}
\begin{equation}\label{Eq:1stQExampleMappingFull}
\begin{aligned}
\ket{\Psi} =& \frac{\alpha}{\sqrt{2}}\left(\ket{\textbf{0}}_1\ket{\textbf{1}}_2 - \ket{\textbf{1}}_1\ket{\textbf{0}}_2 \right) \\
+&\frac{\beta}{\sqrt{2}} \left(\ket{\textbf{2}}_1\ket{\textbf{3}}_2 - \ket{\textbf{3}}_1\ket{\textbf{2}}_2 \right) \\
+& \frac{\gamma}{\sqrt{2}} \left(\ket{\textbf{0}}_1\ket{\textbf{3}}_2 - \ket{\textbf{3}}_1\ket{\textbf{0}}_2 \right) \\
+& \frac{\delta}{\sqrt{2}}
\left(\ket{\textbf{1}}_1\ket{\textbf{2}}_2 - \ket{\textbf{2}}_1\ket{\textbf{1}}_2 \right).
\end{aligned}
\end{equation}
As we have $N=2$ electrons, and $M=4$ spin-orbitals, we can see that we only require $N \ceil{\mathrm{log_2}(M)} = 2 \times \ceil{\mathrm{log_2}(4)} = 4$ qubits to store the wavefunction.}

The Hamiltonian can be obtained by projecting it onto the single-particle basis functions
\textcolor{black}{
\begin{equation}\label{Eq:FirstQHamil}
\begin{aligned}
H &= \sum_{i=1}^N \sum_{\alpha,\beta = 0}^{M-1} h_{\alpha \beta} \ket{\phi_\beta}_i \bra{\phi_\alpha}_i \\
&+ \frac{1}{2}\sum_{i \neq j}^N \sum_{\alpha, \beta, \gamma, \delta}^{M-1} h_{\alpha \beta \gamma \delta} \ket{\phi_\alpha}_i \ket{\phi_\beta}_j  \bra{\phi_\gamma}_j \bra{\phi_\delta}_i , 
\end{aligned}
\end{equation}
where
\begin{equation}\label{firstqints}
	\begin{aligned}
		h_{\alpha \beta}&=\int \mathrm{d}\textbf{x} \phi_\beta^*(\textbf{x}) \left(-\frac{\nabla^2}{2} -\sum_{I}\frac{Z_I}{|\mathbf{r}-\mathbf{R}_I|}\right) \phi_\alpha(\mathbf{x}),\\
		 h_{\alpha \beta \gamma \delta}&=\int \mathrm{d}\mathbf{x}_1 \mathrm{d}\mathbf{x}_2\frac{\phi_\alpha^*(\mathbf{x}_1) \phi_\beta^*(\mathbf{x}_2)\phi_\gamma(\mathbf{x}_2)\phi_\delta(\mathbf{x}_1)}{|\mathbf{r}_1-\mathbf{r}_2|}.
	\end{aligned}
\end{equation}
For example, if we denote terms in the first sum of Eq.~(\ref{Eq:FirstQHamil}) as $H^i_{\phi_\alpha \phi_\beta}$, and consider our model system with spin-orbitals $\phi_\alpha = \{\ket{A_\uparrow}$, $\ket{A_\downarrow}$, $\ket{B_\uparrow}$, $\ket{B_\downarrow} \}$, then the term $H^1_{A_\uparrow B_\uparrow}$ (which acts on electron 1) is given by
\begin{equation}
	\begin{aligned}
		h_{A_\uparrow B_\uparrow} \ket{B_\uparrow}_{e_1} \bra{A_\uparrow}_{e_1} = \\
		h_{A_\uparrow B_\uparrow} \ket{10}_{e_1} \bra{00}_{e_1} = \\
		h_{A_\uparrow B_\uparrow} (\ket{1}_{q_3}\bra{0}_{q_3}) \otimes (\ket{0}_{q_2}\bra{0}_{q_2}) = \\
		h_{A_\uparrow B_\uparrow} \left ( \frac{1}{2}(X_{q_3} - iY_{q_3})\right ) \otimes \left ( \frac{1}{2}(I_{q_2} - Z_{q_2}) \right ) =\\
		\frac{h_{A_\uparrow B_\uparrow}}{4} (X_{q_3}I_{q_2} - iY_{q_3}I_{q_2} - X_{q_3}Z_{q_2} + iY_{q_3}Z_{q_2}),
	\end{aligned}
\end{equation}
where $e_i$ denotes electron $i$, $q_i$ denotes the $i$\textsuperscript{th} qubit (counting from the right), $X, Y, Z, I$ are the Pauli operators introduced in Sec.~\ref{Subsec:QuantumComputing}, and $h_{A_\uparrow B_\uparrow}$ is given by the first integral in Eq.~(\ref{firstqints}). There are up to $\mathcal{O}(N^2 M^4)$ possible 2-body terms, each leading to up to $\mathcal{O}(2^{2\mathrm{log_2}(M)})$ Pauli terms -- meaning the Hamiltonian can contain up to $\mathcal{O}(N^2 M^6)$ Pauli terms, which are up to $2\mathrm{log_2}(M)$-local.}

Once the Hamiltonian has been obtained, we can use it to time evolve the wavefunction, which maintains the correct antisymmetry~\cite{Abrams97}. As mentioned previously, and shown in Sec.~\ref{Subsec:qpe}, time evolution is a key subroutine of algorithms to find the ground state of chemical systems.

\subsection{Second quantised basis set encoding methods}\label{Subsec:2ndencoding}
To simulate chemical systems in the second quantised representation on a quantum computer, we need to map from operators which act on indistinguishable fermions to operators acting on distinguishable qubits. An encoding method is a map from the fermionic Fock space to the  Hilbert space of qubits, such that every fermionic state can be represented by a qubit state. There are multiple methods of encoding, which we describe below. In the following section, we only discuss second quantised basis set methods, as second quantised grid based methods have only briefly been discussed in the context of quantum computational chemistry (see \textcite{Babbush2017low} Appendix A). \\

\subsubsection{Jordan-Wigner encoding}\label{Subsubsec:JWencoding}
In the Jordan--Wigner (JW) encoding~\cite{jordan1928jwtransform}, we store the occupation number of a spin-orbital in the $\ket{0}$ or $\ket{1}$ state of a qubit (unoccupied and occupied, respectively). More formally,
\begin{equation}
\begin{aligned}
		\ket{f_{M-1},f_{M-2},\dots, f_0} & \rightarrow \ket{q_{M-1},q_{M-2},\dots, q_0}, \\
		q_p = f_p & \in \{0, 1 \}.
\end{aligned}
\end{equation}
The fermionic creation and annhilation operators increase or decrease the occupation number of a spin-orbital by 1, and also introduce a multiplicative phase factor (see Eq.~(\ref{creationAnnOper})). The qubit mappings of the operators preserve these features, and are given by,
\begin{equation}
	\begin{aligned}
		a_p &= Q_p\otimes Z_{p-1}\otimes \dots\otimes  Z_{0},\\
		a_p^\dag &= Q_p^\dag \otimes Z_{p-1}\otimes \dots\otimes  Z_{0},
	\end{aligned}
\end{equation}
where $Q = \ket{0}\bra{1} = \frac{1}{2}(X + iY)$ and $Q^\dag = \ket{1}\bra{0} = \frac{1}{2}(X - iY)$. The $Q$ or $Q^\dag$ operator changes the occupation number of the target spin-orbital, while the string of $Z$ operators recovers the exchange phase factor $(-1)^{\sum_{i = 0}^{p-1}f_i}$. We refer to the action of the $Z$ operators as `computing the parity of the state'. \textcolor{black}{Using the JW encoding, the second quantised fermionic Hamiltonian is mapped to a linear combination of products of single-qubit Pauli operators
\begin{equation}
	H = \sum_j h_j P_j = \sum_j h_j\prod_i \sigma_i^j,
\end{equation}
where $h_j$ is a real scalar coefficient, $\sigma_i^j$ represents one of $I$, $X$, $Y$  or $Z$, $i$ denotes which qubit the operator acts on, and $j$ denotes the term in the Hamiltonian. Each term $P_j$ in the Hamiltonian is typically referred to as a `Pauli string', and the number of non-identity single-qubit Pauli operators in a given string is called its `Pauli weight'. All of the second quantised encoding methods discussed in this section produce Hamiltonians of this form.} An example JW mapping is shown in Table.~\ref{JWtable}. \textcolor{black}{In order to further clarify the second quantised JW encoding, we apply it to the fictitious system described earlier. \textcolor{black}{As stated previously, we assume that the Hartree-Fock state for this system has both electrons occupying the $\ket{A}$ orbitals.} We store the occupations of the spin-orbitals $\ket{A_\uparrow}$, $\ket{A_\downarrow}$, $\ket{B_\uparrow}$, $\ket{B_\downarrow}$, which we order as $\ket{f_{B_\downarrow}, f_{B_\uparrow}, f_{A_\downarrow}, f_{A_\uparrow}}$, with $f_i = 0,1$. The Hartree-Fock state is then given by
\begin{equation}
 \ket{\Psi_{\mathrm{HF}}} = \ket{0011}. 
 \end{equation}
This state corresponds to the antisymmetrised Slater determinant shown in Eq.~(\ref{Eq:1stQExampleMappingHF}). The $s_z=0$ wavefunction is then
\begin{equation}
 \ket{\Psi} = \alpha \ket{0011} + \beta \ket{1100} +\gamma \ket{1001} + \delta \ket{0110}.
\end{equation}
This state can be compared with the first quantised basis set mapping shown in Eq.~(\ref{Eq:1stQExampleMappingFull}).} Working in the JW basis, it is easy to see the advantage that quantum computers have over their classical counterparts for chemistry problems. As discussed in Sec.~\ref{Subsubsec:2ndqclassical}, the full configuration interaction wavefunction contains a number of determinants which scales exponentially with the number of electrons, as roughly $\mathcal{O}(M^N)$. As such, it requires an amount of memory that scales exponentially with the system size. 
However, using a quantum computer, we can instead store the full configuration interaction (FCI) wavefunction using only $M$ qubits~\cite{aspuru2005simulated}. A register of $M$ qubits can be in a superposition of $2^M$ computational basis states. In the JW basis, every Slater determinant required for the FCI wavefunction can be written as one of these computational basis states. As such, quantum computers can efficiently store the FCI wavefunction. This is also true for the other second quantised encodings.\\

\begin{table*}[t]
\caption{Example mappings of a fermionic Fock state and its fermionic operators onto the corresponding qubit state, and qubit operators. $\hat{n}_i$ is the fermionic number operator.}
\begin{center}
 \begin{tabular}{c|c|c|c} 
 \hline
Fermion & Jordan-Wigner & Parity & Bravyi-Kitaev \\ [0.5ex] 
 \hline
 $a\ket{0001}+b\ket{0010}$ & $a\ket{0001}+b\ket{0010}$ & $a\ket{1111}+b\ket{1110}$ & $a\ket{1011}+b\ket{1010}$ \\
  $+c\ket{0100}+d\ket{1000}$ & $+c\ket{0100}+d\ket{1000}$ & $+c\ket{1100}+d\ket{1000}$ & $+c\ket{1100}+d\ket{1000}$ \\
 $a_0$ & $Q_0$ & $X_3X_2X_1 Q_0$ & $X_3X_1Q_0$\\
 $a_1$ & $Q_1 Z_0$ & $X_3X_2\left(Q_1\ket{0}\bra{0}_0-Q_1^\dag\ket{1}\bra{1}_0\right)$&$X_3\left(Q_1\ket{0}\bra{0}_0-Q_1^\dag\ket{1}\bra{1}_0\right)$  \\
 $a_2$ & $Q_2Z_1Z_0$ & $X_3\left(Q_2\ket{0}\bra{0}_1-Q_2^\dag\ket{1}\bra{1}_1\right)$& $X_3Q_2Z_1$ \\
  $a_3$ & $Q_3Z_2Z_1Z_0$ & $Q_3\ket{0}\bra{0}_2-Q_3^\dag\ket{1}\bra{1}_2$& $\frac{1}{2}\left(Q_3(\mathbf{1}+Z_2Z_1)-Q_3^\dag(\mathbf{1}-Z_2Z_1)\right)$ \\
$\hat{n}_i = a_i^\dag a_i$ & $ \ket{1} \bra{1}_i$ & $\ket{1}\bra{1}_{i=0}$~,~$\frac{1}{2}\left(\mathbf{1}-Z_iZ_{i-1}\right)_{i=1,2,3}$ &$\ket{1}\bra{1}_{i=0,2}$~,~ $\frac{1}{2}(\mathbf{1}-Z_1Z_0)_{i=1}$~,~ $\frac{1}{2}(\mathbf{1}-Z_3Z_2Z_1)_{i=3}$\\
 \hline
\end{tabular}
\end{center}
\label{JWtable}
\end{table*}

The primary advantage of the JW encoding is its simplicity. However, while the occupation of a spin-orbital is stored locally, the parity is stored non-locally. The string of $Z$ operators means that a fermionic operator mapped to qubits generally has a weight of $\mathcal{O}(M)$ Pauli operators, each acting on a different qubit.

An alternative to the JW mapping that has not yet found particular use within the field, but is worth the reader being aware of, is the parity encoding. This approach stores the parity locally, and the occupation number non-locally. We use the $p$\textsuperscript{th} qubit to store the parity of the first $p$ modes,
\begin{equation}\label{parityencoding}
\begin{aligned}
	\ket{f_{M-1},f_{M-2},\dots, f_0} & \rightarrow \ket{q_{M-1},q_{M-2},\dots, q_0}, \\
		q_p = \bigg{[}\sum_{i=0}^p & f_i\bigg{]} ~~(\textrm{mod}~2) .
\end{aligned}
\end{equation}
The transformed creation and annihilation operators are described by \textcite{BravyiKitaev12}. An example of the parity mapping is shown in Table.~\ref{JWtable}.

\subsubsection{Bravyi--Kitaev encoding}\label{Subsubsec:BKencoding}
The Bravyi--Kitaev (BK) encoding~\cite{BRAVYI2002210} is a midway point between the JW and parity encoding methods, in that it compromises on the locality of occupation number and parity information. The orbitals store partial sums of occupation numbers. The occupation numbers included in each partial sum are defined by the BK matrix, $\beta_{pq}$.
\begin{equation}\label{BKencodingEq}
\begin{aligned}
	\ket{f_{M-1},f_{M-2},\dots, f_0} & \rightarrow \ket{q_{M-1},q_{M-2},\dots, q_0}, \\
	 q_p = & \left[\sum_{q = 0}^{p}\beta_{pq}f_q\right]~~(\textrm{mod}~2).
\end{aligned}
\end{equation}
 It is defined recursively~\cite{BRAVYI2002210,BravyiKitaev12} via
\begin{equation}
\begin{aligned}
\beta_1 & = [1], \\ 
\beta_{2^{x + 1}} = &
\begin{pmatrix}
    \beta_{2^x} & \textbf{0} \\
    \textbf{A} & \beta_{2^x} \\
 \end{pmatrix},
\end{aligned}
\end{equation}
where $\textbf{A}$ is an $(2^x \times 2^x)$ matrix of zeros, with the bottom row filled with ones, and $\textbf{0}$ is a $(2^x \times 2^x)$ matrix of zeros. As an example, when $M = 4$ ($x=1$), the matrix $\beta_{pq}$ is
\begin{equation}\label{Eq:bkmatrix}
\beta_4 = 
\begin{pmatrix}
    1 & 0 & 0 & 0 \\
    1 & 1 & 0 & 0 \\
    0 & 0 & 1 & 0 \\
    1 & 1 & 1 & 1 
 \end{pmatrix}.
\end{equation}

When the number of qubits is not a power of two, the BK encoding is carried out by creating the BK matrix for the next largest power of two, and only using the first $M$ rows.
The qubit operators for the BK encoding are considerably more complicated than those in the JW or parity encodings. We refer to Table~\ref{JWtable} for an example, and works by \textcite{BravyiKitaev12, tranter2015bravyikitaev} for a more detailed discussion. Applying the BK mapping to a fermionic operator results in a qubit operator with a Pauli weight of $\mathcal{O}(\log_2 M)$. A thorough comparison of the BK and JW mappings was performed by \textcite{tranter2018comparison} for 86 molecular systems. \textcolor{black}{They found that the BK transform was at least as efficient, in general, as the JW transform when finding the ground states of the molecular systems. In many cases using the BK transform made the calculations considerably more efficient. \\}

Another version of the BK encoding also exists in the literature. This is referred to as the BK-tree method, as it takes its inspiration from a classical data structure known as a Fenwick tree~\cite{havlicek2017operator}. We explicitly show how to use this mapping with molecules in Sec.~\ref{Sec:illustrate}. As with the standard BK mapping, the BK-tree encoding balances how it stores occupation and parity information. As a result, it too only requires $\mathcal{O}(\log_2 M)$ qubit operations to realise a fermionic operator, in general. However, there are subtle differences between the two mappings. It has been noted that the BK-tree mapping produces qubit operators with a greater weight than the standard BK mapping~\cite{openfermionforum}. This would suggest that it is less suitable for near-term quantum computation. However, the BK-tree mapping also possesses advantages over the standard BK encoding. The BK-tree mapping is uniquely defined even when the number of orbitals, $M$, is not a power of 2. As a result, when using the BK-tree mapping we are always able to use the qubit reduction by symmetry technique, which we will discuss in Sec.~\ref{Subsec:reduction}. We have observed that it is only possible to use this technique with the standard BK mapping when the number of orbitals is a power of two. As a result, it is important to consider the benefits of both mappings, before choosing which one to use. \textcolor{black}{A further generalisation of the BK-tree encoding is to consider ternary trees~\cite{jiang2019ternary}, which leads to asymptotic reductions in the Pauli weight of Hamiltonian terms.}

\subsubsection{Locality preserving mappings}\label{Subsubsec:OtherEncodings}

Mappings have been developed which endow the qubit operators with the same locality as the underlying fermionic Hamiltonian. These mappings typically require more qubits than the JW encoding. \textcite{VerCirac} developed a scheme to eliminate the strings of $Z$ operators introduced by the JW transform, resulting in qubit operators with the same locality as the fermionic operators. This is achieved by doubling the number of qubits. Similar ideas were introduced by \textcite{ball2005aux} and \textcite{farrelly2014causal}. These ideas were later generalised and expanded upon by \textcite{AuxFermion, steudtner2018codes}.\\

There is also another variant of the BK transform, known as the Bravyi-Kitaev superfast transform (BKSF)~\cite{BRAVYI2002210}. This mapping first represents each spin-orbital by a vertex on a graph, and each interaction term in the Hamiltonian as an edge on the graph. Qubits are then associated to the edges. In general, a graph will have more edges than vertices, so this increases the number of qubits required. However, the number of gates required to implement a fermionic operator will scale as $\mathcal{O}(d)$ where $d$ is the degree of the graph. Assuming fairly local interactions for a molecule, the degree of the graph will be less than the number of vertices. As a result, the BKSF transform may require fewer gates than the JW mapping. We refer the reader to the work of \textcite{Superfast,chien2019superfast} for a detailed discussion of the BKSF transform, and comparison to the JW transform. The BKSF transform has been generalised~\cite{setia2018superfast} to either: 1) Reduce the weight of each of the Pauli operators in the Hamiltonian to $\mathcal{O}(\log d)$, or 2) Provide some protection from errors. A related mapping, known as the majorana loop stabilizer code, was introduced by \textcite{jiang2018majorana}. It also preserves the locality of the underlying model, and offers some protection from errors. \textcolor{black}{We will discuss this error detecting/correcting property in more detail in Sec.~\ref{Subsec:StabiliserMitigation}.}

\subsection{Resource reduction}\label{Subsec:reduction}
In this section, we focus on general techniques that can be used to reduce the resources required for quantum chemistry simulation. In particular, we focus on methods to remove qubits from the simulation using symmetries, and low rank decomposition methods for reducing the cost of quantum circuits.

\subsubsection{Hamiltonian reduction}\label{Subsubsec:HamiltonianReduction}
We focus on techniques to reduce the number of qubits required for the second quantised approach, using $\mathbb{Z}_2$ symmetries. More general qubit reduction schemes have also been developed~\cite{bravyi2017tapering}, but these have yet to be numerically or experimentally investigated. 

In the JW, parity and BK encoding methods, the number of qubits is equal to the number of spin-orbitals considered, $M$. However, as the Hamiltonian possesses symmetries, the wavefunction can be stored in a smaller Hilbert space. Here, we will describe the method by \textcite{bravyi2017tapering}, which utilises two such symmetries: conservation of electron number and spin. This method enables the systematic reduction of two qubits when using the parity, BK (with the caveat that the number of orbitals is a power of two), or BK-tree encoding. For a system with $M$ spin-orbitals, we can arrange the orbitals such that the first $M/2$ spin-orbitals describe spin up states, and the last $M/2$ spin-orbitals describe spin down states. For non-relativistic molecules, the total number electrons and the total $s_z$ value are conserved. Examining the BK matrix presented in Eq.~(\ref{Eq:bkmatrix}), we see that every element in the final row is one, and the first half of the elements in the $M/2$\textsuperscript{th} row are also one. Consequently, the final element of the vector encoded by this matrix, $q_{M-1}$, is equal to the number of electrons (mod 2). Similarly, the $M/2$\textsuperscript{th} element in the encoded vector, $q_{\frac{M}{2} -1}$, is equal to the number of spin up electrons (mod 2). As the electron number and total $s_z$ value are conserved by the Hamiltonian, these qubits are only acted on by the identity or Pauli~$Z$ operators. We can replace these operators by their corresponding eigenvalues ($+1$ for the identity, $+1$ for $Z_{M-1}$ if the total number of electrons is even, $-1$ for $Z_{M-1}$ if the total number of electrons is odd, $+1$ for $Z_{\frac{M}{2} -1}$ if the number of spin up electrons is even, and $-1$ for $Z_{\frac{M}{2} -1}$ if the number of spin up electrons is odd). The Hamiltonian then only acts on $(M-2)$ qubits, so two qubits can be removed from the simulation. Exactly the same method can be used for the parity and BK-tree encodings. We will explicity show how this method can be used to remove two qubits from chemical Hamiltonians in Sec.~\ref{Sec:illustrate}. We remark that while this transformation leaves the ground state of the system unchanged, it does alter the excited states that can be found. In particular, we are restricted to finding those states with an electron number and total $s_z$ value equal to the values determined as described above. \textcolor{black}{These techniques have been extended by \textcite{setia2019pointgroups} to include molecular point group symmetries.}

\textcolor{black}{
\subsubsection{Low rank decomposition techniques}\label{Subsubsec:LowRank}
As discussed in Sec.~\ref{Subsec:problem}, the electronic structure Hamiltonian in the canonical molecular orbital basis contains $\mathcal{O}(M^4)$ terms, where $M$ is the number of spin-orbitals in the molecule. This means that many quantum circuits implementing time evolution under the molecular Hamiltonian naively scale in a similar way. As we shall discuss in Sec.~\ref{Subsec:qpe} and Sec.~\ref{Subsec:VQE}, such circuits are a key component of many quantum algorithms. \textcite{motta2018low} utilised a low rank decomposition of the Hamiltonian in a Gaussian orbital basis to reduce the cost of time evolution. This decomposition is made possible by the structure present in the molecular Hamiltonian, which arises from the pairwise nature of the electron-electron interactions. Those authors leveraged a doubly factorised Cholesky decomposition of the Hamiltonian, with an efficient quantum circuit implementation. By carefully truncating the number of terms in the decomposition, they were able to reduce the cost of time evolution, with a controllable error. Similar decompositions were also introduced for other quantum circuits of interest. We will discuss the impact of these cost reductions in more detail in Sec.~\ref{Subsec:qpe} and Sec.~\ref{Subsec:VQE}.} 

This low rank decomposition technique was also used by \textcite{huggins2019efficientmeasurements}, to reduce the number of circuit repetitions required for variational algorithms to find the ground state energy of chemical systems. We discuss this technique in more detail in Sec.~\ref{Subsubsec:Measurement}. \\

We note that most of the work to date in quantum computational chemistry has focused on second quantised basis set methods. While first quantised basis set simulations require asymptotically fewer qubits than second quantised basis set simulations, for the smallest simulable systems (such as small molecules in minimal basis sets), second quantised basis set methods require either fewer qubits and/or shorter gate sequences. This has caused second quantised basis set methods to become the \textit{de facto} option for experimental demonstrations of quantum computational chemistry algorithms, due to the limits of current quantum hardware.

\section{Quantum computational chemistry algorithms}\label{Sec:QuantumChemistryAlgorithms}
In this section, we focus on methods used to solve the electronic structure problem with a quantum computer. We describe the quantum phase estimation algorithm and related methods in Sec.~\ref{Subsec:qpe}. We then discuss the variational quantum eigensolver (VQE) in Sec.~\ref{Subsec:VQE}. Both of these sections are concerned with finding the ground state energies of chemical systems. We conclude this section with a discussion of methods that can be used to find excited states in Sec.~\ref{Subsec:excited}.

\textcolor{black}{As mentioned in Sec.~\ref{Sec:Introduction}, the techniques developed for solving the electronic structure problem can often be generalised to solve other problems in computational chemistry. For example, \textcite{obrien2019derivatives} showed how to calculate the energy derivatives of molecular Hamiltonians (which can be used for geometry optimisation and transition state discovery) by using algorithms based on phase estimation (Sec.~\ref{Subsec:qpe}), or based on a linear response quantum subspace expansion (Sec.~\ref{Subsubsec:SubspaceExpansionExcited}). The same properties can also be calculated using the method of \textcite{mitarai2019derivatives}, who leveraged the variational quantum eigensolver (Sec.~\ref{Subsec:VQE}), with analytic gradient measurements (Sec.~\ref{Subsubsec:VQEOptimisation}), and by \textcite{parrish2019derivatives} who used the contraction VQE method (Sec.~\ref{Subsubsec:ContractionVQE}). }

It can be argued that the VQE and phase estimation, as presented herein, represent near-term and long-term methods (respectively) for solving chemistry problems with a quantum computer. However, in reality, aspects of each algorithm can be incorporated into the other, creating new methods~\cite{yung2014transistor, wang2018generalisedvqe} which occupy the intermediate region in the quantum computational chemistry timeline. Algorithms to find the ground state using methods which differ from both phase estimation and the VQE have also been proposed such as techniques based on time series estimation~\cite{somma2019timeseries,somma2002timeseries,obrien2018phaseestimation}, or the method by \textcite{ge2017faster}. \\

\subsection{Quantum phase estimation}\label{Subsec:qpe}
\subsubsection{Implementation}\label{Subsubsec:QPEalgorithm}
Phase estimation~\cite{kitaev1995phase} can be used to find the lowest energy eigenstate, $\ket{E_0}$, and excited states, $\ket{E_{i>0}}$, of a physical Hamiltonian~\cite{Abrams99}. In the case of quantum computational chemistry, this qubit Hamiltonian can encode a fermionic Hamiltonian, obtained using the methods discussed in Sec.~\ref{Sec:Encoding}.

\begin{figure}[t]
{\includegraphics{{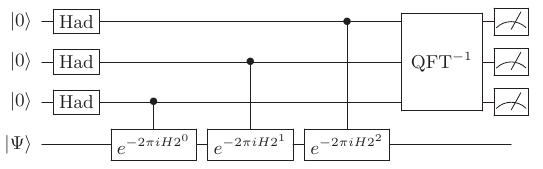}}}
\caption{The canonical quantum phase estimation circuit with three ancilla qubits. When the ancilla qubits are in state $\ket{x}$, a control rotation $e^{-2\pi iHx}$ is applied to the target state $\ket{\Psi}$. QFT denotes the quantum Fourier transform~\cite{shor1994algorithms,coppersmith1994qft}. By measuring the ancilla qubits in the computational basis, they collapse to an eigenvalue of $H$ and the register qubits collapse to an estimate of the corresponding energy eigenstate.
}\label{Fig:QPE}
\end{figure}

The canonical phase estimation algorithm is described as follows~\cite{nielsen2002quantum}, and shown in Fig.~\ref{Fig:QPE}.
  
\begin{enumerate} 
	\item We initialise the qubit register in state $\ket{\Psi}$, which has non-zero overlap with the true full configuration interaction (FCI) target eigenstate state of the system. We require an additional register of $\omega$ ancilla qubits. We can expand the state $\ket{\Psi}$ in terms of energy eigenstates of the Hamiltonian, writing that $\ket{\Psi} = \sum_i c_i\ket{E_i}$, where $c_i$ are complex coefficients.
    
    \item We apply a Hadamard gate to each ancilla qubit, placing the ancilla register in the superposition $\frac{1}{\sqrt{2^\omega}}\sum_x\ket{x}$, where $x$ are all possible bit-strings that can be constructed from $\omega$ bits. We then apply the controlled gates shown in Fig.~\ref{Fig:QPE}:
  
		\begin{equation}
			\frac{1}{\sqrt{2^\omega}}\sum_i \sum_x \ket{x} c_i \ket{E_i} \rightarrow \frac{1}{\sqrt{2^\omega}} \sum_i \sum_x e^{-2\pi iE_ix}c_i\ket{x}\ket{E_i}.
		\end{equation} 

		\item We apply the inverse quantum Fourier transform to the ancilla qubits in order to learn the phase, which encodes the information about the energy eigenvalue:
		\begin{equation}
			\frac{1}{\sqrt{2^\omega}}\sum_i \sum_x e^{-2\pi iE_ix}c_i\ket{x}\ket{E_i}\xrightarrow[]{\mathrm{QFT}^{-1}}\sum_i c_i\ket{\mathrm{bin}(E_i)}\ket{E_i}.
		\end{equation}

		\item We measure the ancilla qubits in the $Z$ basis, which gives an estimate of the energy eigenvalue as a binary bit-string, $\mathrm{bin}(E_i)$, with probability $|c_i|^2$. This procedure collapses the main register into the corresponding energy eigenstate, $\ket{E_i}$.
	\end{enumerate}

\textcolor{black}{The number of ancilla qubits, $\omega$, required for the method of phase estimation described above is determined by the desired success probability and precision in the energy estimate. \textcite{nielsen2002quantum} showed that to obtain a binary estimate of the energy, precise to $n$ bits, with success probability $p$, requires 
\begin{equation}\label{PhaseEstPrecision}
    \omega = n +  \left \lceil \mathrm{log_2}\bigg{(}2+\frac{1}{2p}\bigg{)} \right \rceil 
\end{equation}
ancilla qubits.} Phase estimation has been experimentally demonstrated in a variety of quantum architectures~\cite{lanyon2010towards, PhysRevLett.104.030502, wang2015quantum, PRXH2,paesani2017bayesian,Santagatieaap9646,li2011solving}

To realise the standard phase estimation algorithm given above, we sequentially need to time evolve the main register under the Hamiltonian $H$ for times $t_0=2\pi , t_1=4\pi, ... , t_{\omega-1} = 2^\omega \pi $. The total coherent time evolution, $T$, is then given by approximately $T = 2^{\omega+1} \pi$. Using Eq.~(\ref{PhaseEstPrecision}), for a success probability of $p=0.5$, we require $\omega = n + 2$ ancilla qubits. The total evolution time can be related to the binary precision $\epsilon_\mathrm{PE} = 1/2^n$, to show that $T = 8\pi / \epsilon_\mathrm{PE}$. Given that our success probability for this estimate is $p=0.5$, we expect to have to repeat the procedure twice to obtain a good estimate of the ground state. This is equivalent to a total of $16\pi/\epsilon_\mathrm{PE}$ calls to the unitary $e^{-iH}$~\cite{Reiher201619152}. In order to account for the fact that $c_0 < 1$, we must multiply the number of repetitions of phase estimation by $1/|c_0|^2$, on average, to obtain the ground state.

The basic phase estimation algorithm described above can be improved in many ways. It can be modified to use only a single ancilla qubit, which is used to measure each bit in the energy eigenvalue sequentially~\cite{aspuru2005simulated,kitaev1995phase}. It can also be made more efficient~\cite{svore2013phaseestimation, kivlichan2019randomizedphaseestimation}, parallelised~\cite{knill2007parallelphaseest,Reiher201619152}, or made more resilient to noise~\cite{obrien2018phaseestimation}. We can further improve upon the asymptotic scaling of phase estimation by using classically obtainable knowledge about the energy gap between the ground and first excited states~\cite{berry2018improved}. The ultimate limit for the number of calls required to $e^{-iH}$ is $\pi / \epsilon_\mathrm{PE}$ (for a completely general Hamiltonian, $H$), which is approximately obtained using Bayesian approaches~\cite{wiebe2016bayesian, paesani2017bayesian, berry2009phaseestimation, Higgins2007phaseest}, or entanglement based approaches~\cite{babbush2018encoding}. For the case of a molecular Hamiltonian \textcite{Reiher201619152} showed that a number of calls scaling as $\pi / 2\epsilon_\mathrm{PE}$ will suffice. \\

\textcolor{black}{The finite precision, $\epsilon_\mathrm{PE}$, obtained in $E_0$ is not the only source of error in the algorithm. There are also errors arising from imperfect implementation of the controlled unitary evolutions applied to the main register, which we denote as $\epsilon_\mathrm{U}$. This error can arise, for example, from a decomposition of $e^{-iH}$ into arbitrary single and two qubit gates, as occurs during a Trotter decomposition. There are also errors arising from constructing arbitrary gates from a discrete set of gates -- such as approximating single qubit rotations from multiple T and Hadamard gates. These are typically referred to as circuit synthesis errors, $\epsilon_\mathrm{CS}$, and can be quantified using techniques such as the Solovay-Kitaev theorem~\cite{dawson2005solovaykitaev}. For the specific case of a Trotter decomposition of $e^{-iH}$, \textcite{Reiher201619152} showed that the error in the energy eigenvalue obtained from phase estimation is upper bounded by $\epsilon_\mathrm{PE} + \epsilon_\mathrm{U} + \epsilon_\mathrm{CS}$. } \textcolor{black}{In general, it is difficult to optimally allocate resources between these error budgets in order to minimise the total error~\cite{Reiher201619152,kivlichan2019condensedtrotter}.\\}

Regardless of which version of phase estimation is used, there are two universal features. Firstly, it is necessary for the register to initially be in a state with a non-zero overlap with the target eigenstate. Secondly, we must have a way to coherently implement a unitary operator defined by an efficiently invertible function of the Hamiltonian. This unitary operator is often (but not always) chosen to be the time evolution operator, $e^{-iH}$, used above. We will discuss techniques to satisfy both of these requirements in Sec.~\ref{Subsubsec:QPEstateprep} and Sec.~\ref{Subsubsec:HamiltonianSimulation}, below.\\

\subsubsection{State preparation}\label{Subsubsec:QPEstateprep}
Initialising the qubit register in a state which has a sufficiently large overlap with the target eigenstate (typically the ground state), is a non-trivial problem. This is important, because a randomly chosen state would have an exponentially vanishing probability of collapsing to the desired ground state, as the system size increases. Even more problematically, \textcite{mcclean2014locality} showed that phase estimation can become exponentially costly, by considering the imperfect preparation of eigenstates of non-interacting subsystems. This highlights the necessity of developing state preparation routines which result in at worst a polynomially decreasing overlap with the FCI ground state, as the system size increases. Several techniques have been proposed for state preparation. One approach is to prepare reference states obtained from classically tractable calculations, such as: configuration interaction states~\cite{babbush2015chemical,wang2009preparing}, open-shell spin symmetry-adapted states~\cite{sugisaki2016constructing,sugisaki2018openshell}, multireference states~\cite{sugisaki2019multireference}, or states produced by adaptive sampling configuration interaction methods~\cite{tubman2018orthogonality}. Alternatively, we can use: the variational methods discussed in Sec.~\ref{Subsec:VQE}~\cite{yung2014transistor}, quantum algorithms for imaginary time evolution~\cite{motta2019imaginary}, or adiabatic state preparation~\cite{aspuru2005simulated}. We focus here on adiabatic state preparation, an approach inspired by the adiabatic model of quantum computation~\cite{farhi2000adiabatic}.

For any Hamiltonian $H_s$, we can prepare a state $\ket{\Psi}$ that is close to its ground state via adiabatic state preparation~\cite{albash2018rmpadiabatic}. To do so, we first start with a simple Hamiltonian $H_0$ and prepare its ground state. We then time evolve the system under a Hamiltonian that changes slowly from $H_0$ to $H_s$, thus preparing a state that is close to the ground state of $H_s$. The efficiency of adiabatic state preparation depends on the gap between the ground state and the first excited state along the path between $H_0$ and $H_s$. For chemical systems, ASP may be achieved by initialising the system in the ground state of the Hartree-Fock Hamiltonian ($H_0$), and interpolating between the initial and final Hamiltonians using an annealing schedule such as $H(t) = (1-t/T)H_0 + (t/T)H_s$, where $t$ is the time and $T$ is the maximum desired simulation time~\cite{aspuru2005simulated}. Alternative paths that may be more efficient for problems of chemical interest have also been investigated~\cite{veis2014adiabatic, wecker2015stronglycorrelated}.
The maximum annealing time, $T$, is given by
\begin{equation}\label{annealing}
    T \approx \mathcal{O}\left(\frac{M^4}{\mathrm{min_s}\Delta (t)}\right),
\end{equation}
where $\Delta(t) = E_1(t) - E_0(t)$ and $M$ is the number of spin-orbitals in the molecule. \textcite{Reiher201619152} noted that the true scaling may be closer to $\mathcal{O}(M^2 / \mathrm{min_t} \Delta(t))$. It is difficult to know the size of the gap along the entire adiabatic path \textit{a priori}, restricts our ability to perform ASP in the minimum amount of time. \textcolor{Black}{One possible method for reducing the annealing time required is to introduce additional `driving' Hamiltonians, as was numerically investigated by \textcite{matsuura2019vanqver}.} \textcolor{black}{Although Eq.~(\ref{annealing}) does not explicitly depend on the initial state used, it is intuitively preferable to start in a state that has good overlap with the target ground state. We would expect the anneal path to be shorter, and we may be more confident that the gap between the ground and excited state does not shrink exponentially. This view is supported by the numerical simulations of \textcite{veis2014adiabatic}, who found that annealing times for methylene (CH$_2$) could be reduced by up to four orders of magnitude by using an initial state with larger overlap with the true ground state. We note however that if an initial state with sufficiently large overlap with the ground state is available, we may be able to forgo adiabatic state preparation entirely, and instead carry out phase estimation directly on that initial state. As discussed above, phase estimation only requires a non-negligible overlap with the target ground state.}

There are a variety of methods that can be used to evolve the system under this time-dependent Hamiltonian, which are discussed below.

\subsubsection{Hamiltonian simulation}\label{Subsubsec:HamiltonianSimulation}

As discussed above, both the canonical phase estimation algorithm and adiabatic state preparation require implementation of the time evolution operator, $e^{-iHt}$, where $H$ may or may not be time dependent. There are several ways to do this, each with their own advantages and disadvantages.\\

\paragraph{Product formulae\\}
\addcontentsline{toc}{subsection}{\hspace{1cm}a. Product formulae}

The most simple method for time evolution has already been described in Sec.~\ref{Subsec:QuantumSimulation}; Trotterization. If a time-independent Hamiltonian $H$, can be decomposed as $H=\sum_i h_i$, where $h_i$ are local Hamiltonians, then a first order Lie-Trotter-Suzuki approximation~\cite{trotter1959product} of the time evolution is
\begin{equation}
	e^{-iHt} = \left(\prod_i e^{-ih_it/S}\right)^S + \mathcal{O}(t^2/S).
\label{trottereq}
\end{equation}
This approach is also referred to as the `product formula' method. In practice, to achieve accuracy $\varepsilon$, the number of Trotter steps $S=\mathcal{O}(t^2/\varepsilon) $ should be large in order to suppress the errors in the approximation. This is effectively a stroboscopic evolution under time evolution operators corresponding to each of the terms in the Hamiltonian. It is also possible to use higher order product formulae~\cite{suzuki1976trotter,Berry2007,PhysRevA.78.052325}, which scale better with respect to the simulation error than the first order method. Randomisation procedures (such as randomly ordering the terms in the Trotter sequence, or stochastically choosing which terms to include in the Hamiltonian), have been shown to improve the accuracy obtained using product formulae~\cite{childs2018faster,campbell2018random,ouyang2019stochastichamiltonian}.

Product formulae can also be used to simulate dynamics under a time dependent Hamiltonian, $H(t)$. \textcite{wiebe2011simulating} showed that the accuracy of such simulations depends on the derivatives of the Hamiltonian (although this dependence may be alleviated by incorporating randomisation procedures~\cite{PoulinTimeDependent}).

As discussed above, the error in the simulation is determined by the Trotter formula used, the number of Trotter steps, and the ordering of the terms. It is important to note that the gate counts of all product formula based methods scale as $\mathcal{O}(\mathrm{poly}(1/\epsilon))$. \textcite{algorithmic} investigated using extrapolation to suppress the error arising from using a finite number of Trotter steps, as is often done in classical computations. \\

\paragraph{Advanced Hamiltonian simulation methods \\}
\addcontentsline{toc}{subsection}{\hspace{1cm}b. Advanced Hamiltonian simulation methods}

Alternative methods have been introduced which may realise the time evolution operator more efficiently than Trotterization, including: quantum walk based methods~\cite{childs2011walks,berry2015hamiltonian}, multiproduct formulae~\cite{low2019multiproduct, childs2012multiproduct}, Taylor series expansions~\cite{berry2012black,PhysRevLett.114.090502,Berry15optimal} or Chebyshev polynomial approximations \cite{subramanian2018implementing}, and qubitization~\cite{low2016hamiltonian, low2018hamiltonian} in conjunction with quantum signal processing~\cite{LowQSPprx, PhysRevLett.118.010501}. \textcolor{black}{The cost of these methods depends on how the Hamiltonian is accessed or `queried' during the computation. One approach is the `linear combinations of unitaries' (LCU) query model, which decomposes the Hamiltonian or time evolution operator into a linear combination of unitary operators, which are then applied in superposition using oracle circuits (which must be explicitly constructed for a given problem). As a linear combination of unitaries is itself not necessarily unitary, these approaches may require additional techniques (such as amplitude amplification~\cite{berry2014oblivAA}) to maintain a high probability of success.} Alternatively, we can use oracles to access the non-zero entries of Hamiltonians which are $d$-sparse (they have at most $d$ non-zero elements in each row and column, where $d$ is a polylogarithmic function of the matrix dimension).

In addition the the product formula approach, the Taylor series and quantum signal processing + qubitization techniques have found the most use to date in quantum computational chemistry algorithms. Both of the aforementioned query models can be used for the Taylor series and quantum signal processing + qubitization approaches. The Taylor series and quantum signal processing + qubitization algorithms scale exponentially better with regards to the accuracy of the simulation than product formula based methods. \\

\textcolor{black}{The Taylor series method expands the time evolution operator as a truncated Taylor series, where each term in the expansion is unitary. We can then use the aforementioned LCU oracles to implement these unitary operators in superposition, thus realising time evolution under the Hamiltonian.} Variants of the Taylor series method for simulating time dependent Hamiltonians have also been developed~\cite{kieferova2018simulating,berry2019timedep}. \textcolor{black}{Qubitization provides another way of accessing the Hamiltonian during quantum computation, using a `block encoding' of the Hamiltonian. Qubitization naturally implements a quantum walk operator with eigenvalues $e^{-i\mathrm{arcsin}(E_k/\alpha)}$ where $E_k$ is the $k$\textsuperscript{th} eigenvalue of the Hamiltonian $H$ and $\alpha$ is a normalisation factor. We can then use quantum signal processing to invert the $\mathrm{arcsin}$ function, recovering time evolution with the desired eigenvalues $e^{-iE_kt}$.} We refer the readers to work by \textcite{childs2017toward} and the review of \textcite{cao2019review} for a summary of recent progress in Hamiltonian simulation and a comparison of the different methods.

\subsubsection{Phase estimation for chemistry simulation}\label{Subsubsec:ChemistryTimeEvolution}

Both product formulae and more advanced methods of Hamiltonian simulation have been applied to solve problems in quantum computational chemistry. We discuss the asymptotic results obtained using these different methods for four classes of problems: molecules in Gaussian orbital basis sets, systems treated with plane wave basis sets, the Fermi-Hubbard model, and first quantised grid based simulations. We will discuss the explicit gate and qubit counts required to implement some of these methods in Sec.~\ref{Subsec:LongTermQuantumResources}. Once again, the number of spin-orbitals in the molecule, or spin-sites in a lattice is given by $M$, and the number of electrons is given by $N$. \textcolor{black}{We also refer the reader to the tables produced by \textcite{Babbush2017low, babbush2018encoding}, which chart developments in the asymptotic scaling of phase estimation based approaches to quantum computational chemistry.} \\

\paragraph{Gaussian orbital basis sets \\}\label{Para:HamSimGaussians}
\addcontentsline{toc}{subsection}{\hspace{1cm}a. Gaussian orbital basis sets}

As discussed in Sec.~\ref{Subsec:basis}, Gaussian orbitals can be used to compactly describe single molecules. They result in a second quantised Hamiltonian with $\mathcal{O}(M^4)$ terms. The overall cost of phase estimation using product formulae depends on the cost of implementing a single Trotter step (which can be influenced by the number of terms in the Hamiltonian), and the number of Trotter steps required to achieve the desired accuracy, and the accuracy desired (typically taken as a constant value, such as to within chemical accuracy). 

Early works on finding the ground state of molecular systems in Gaussian basis sets used first and second order product formalae~\cite{aspuru2005simulated,whitfield2011simulation,wecker2014gates, hastings2014improving, poulin2014trotter,Reiher201619152,babbush2015chemical}. A series of improvements throughout these papers reduced the scaling of phase estimation for molecules from $\mathcal{O}(M^{11})$~\cite{wecker2014gates} to (empirically) $\mathcal{O}(M^5)$~\cite{babbush2015chemical}. A notable feature of these Trotter-based simulations is that rigorous bounds on the Trotter error are believed to be loose by several orders of magnitude, which may impact the scaling of these approaches. We discuss this in more detail in Sec.~\ref{Subsubsec:OutstandingQPEproblems}.\\

As would be expected, the introduction of more advanced Hamiltonian simulation algorithms led to a reduction in the asymptotic scaling of chemistry algorithms. Using the Taylor series method, algorithms were developed which scale as $\mathcal{O}(M^5)$~\cite{babbush2016exponential} and $\mathcal{O}(N^2 M^3)$~\cite{babbush2017exponential} for molecules in Gaussian basis sets. The first result uses the second quantised representation of the Hamiltonian. The second, more efficient result is obtained by constructing oracle circuits which access the non-zero elements of the Hamiltonian in a basis of Slater determinants (known as the CI matrix~\cite{toloui2013cimatrix}). The Hamiltonian has a sparsity of $\mathcal{O}(N^2 M^2)$ in this basis. These algorithms calculated the molecular integrals on the quantum computer, exploiting an analogy between the discretisation of space in Riemannian integration and the discretisation of time in time dependent Hamiltonian simulation methods~\cite{babbush2016exponential}. Techniques to further increase the sparsity of the problem, such as using symmetries present in the electronic structure Hamiltonian, have also been proposed; however these have yet to be thoroughly investigated in the context of Hamiltonian simulation~\cite{gulania2019young, whitfield2013spinfree}.\\

\textcolor{black}{As discussed above, qubitization~\cite{low2016hamiltonian,low2018hamiltonian} is a technique originally introduced in conjunction with quantum signal processing~\cite{LowQSPprx, PhysRevLett.118.010501} to approximate the unitary $e^{-iHt}$, in order to perform time evolution of a given state.} However, it was later noted that for certain functions $f(H)$, qubitization could be used on its own to directly implement the unitary $e^{-if(H)t}$ without approximation errors~\cite{berry2018improved,poulin2018spectral}. One can then perform phase estimation on this unitary to obtain the energy eigenvalues of the Hamiltonian, provided $f(H)$ is efficiently invertible. This technique was used by \textcite{berry2019qubitizationlowrank}, in conjunction with a low rank decomposition of the Hamiltonian~\cite{motta2018low} (described in Sec.~\ref{Subsubsec:LowRank}). This produced an algorithm with a scaling of $\mathcal{O}(M^{1.5} \lambda)$, where $\lambda$ is the 1-norm of the Hamiltonian. While rigorous bounds on $\lambda$ in a Gaussian basis are not known, \textcite{berry2019qubitizationlowrank} noted that it usually scales as at least $\mathcal{O}(M^{1.5})$.

\paragraph{Plane wave basis sets \\}\label{Para:HamSimPlane}
\addcontentsline{toc}{subsection}{\hspace{1cm}b. Plane wave basis sets}
As discussed in Sec.~\ref{Subsubsec:PlaneWave}, plane wave basis sets are particularly suitable for periodic systems, such as materials. While they can also be used to simulate single molecules, we require approximately 10 to 100 times as many plane waves as Gaussian orbitals for accurate simulations. The plane wave and plane wave dual basis sets reduce the number of terms in the Hamiltonian to $\mathcal{O}(M^3)$ and $\mathcal{O}(M^2)$, respectively. As a result, using the plane wave basis sets can reduce the asymptotic scaling of several approaches to chemistry simulation. 

The best product based formulae approaches described above scale as $\mathcal{O}(M^{1.67}N^{1.83})$ in a plane wave basis~\cite{Babbush2017low}. This can be further improved for systems where it is appropriate to target an extensive error in the energy~\cite{kivlichan2019condensedtrotter}. This means that the error in the energy scales with the size of the system, and therefore with $M$. As the scaling of a Trotterized simulation is inversely proportional to the energy error desired, factors of $M$ can be cancelled from both the numerator and denominator of the algorithm scaling. For 3D materials, such as diamond, the scaling of Trotter based approaches to phase estimation in a plane wave dual basis was reduced to around $\mathcal{O}(M^{2})$.\\

The plane wave basis also benefited the algorithms described above which utilise the Taylor series method for Hamiltonian simulation. The scaling of the CI matrix based algorithm of \textcite{babbush2017exponential} was reduced from $\mathcal{O}(N^2 M^3)$ in a Gaussian basis to $\mathcal{O}(M^{2.67})$ in a plane wave basis~\cite{Babbush2017low}. 

A time dependent form of the Taylor series approach was used for simulation in the `interaction picture', which enables more efficient time evolution in a plane wave dual basis, scaling as $\mathcal{O}(M^2)$~\cite{low2018interaction} (although this algorithm uses $\mathcal{O}(M \mathrm{log_2}(M))$ qubits). A similar interaction picture method was introduced by \textcite{babbush2018sublinear}, using plane waves and first quantisation. This algorithm has a gate scaling of $\mathcal{O}(N^{8/3} M^{1/3})$ and requires $\mathcal{O}(N \mathrm{log_2}(M))$ qubits, plus ancillas~\cite{babbush2018sublinear}. \textcolor{black}{This was the first quantum computational chemistry algorithm to achieve sublinear scaling in $M$;} a feature that can be used to mitigate the inability of plane waves to compactly describe molecular systems. \\

Qubitization has also been used to produce highly efficient algorithms for the plane wave and related basis sets. \textcite{babbush2018encoding} used qubitization and a form of phase estimation which saturates the Heisenberg limit to construct algorithms for simulations of matter in a plane wave dual basis. Their algorithm has a gate scaling of $\mathcal{O}(M^3)$, for condensed phase electronic structure problems.

\textcolor{black}{Continuing the trend of exploiting different chemical representations to improve the efficiency of qubitization, \textcite{mcclean2019galerkin} investigated the use of discontinuous Galerkin basis sets. As discussed in Sec.~\ref{Subsubsec:PlaneWave}), these basis sets provide a more compact description of molecular systems than plane waves, but also reduce the number of terms in the Hamiltonian to $\mathcal{O}(M^{2})$. \textcite{mcclean2019galerkin} estimated from numerics on hydrogen chains that these basis sets could reduce the scaling of qubitization to around $\mathcal{O}(M^{2.6})$. Further work is required to better ascertain and optimise the scaling of this technique.}

Qubitization can also be used in first quantization with the plane wave basis to achieve a sublinear scaling with the number of plane waves. \textcite{babbush2018sublinear} obtained a scaling of $\mathcal{O}(N^{4/3} M^{2/3} + N^{8/3} M^{1/3})$, similar to the scaling of their interaction picture algorithm discussed above.

\paragraph{Fermi-Hubbard model \\}\label{Para:HamSimFH}

\addcontentsline{toc}{subsection}{\hspace{1cm}c. Fermi-Hubbard model}
As described in Sec.~\ref{Subsubsec:2ndqclassical}, the Fermi-Hubbard Hamiltonian contains $\mathcal{O}(M)$ terms, many of which commute with each other. This can improve the efficiency of many of the algorithms discussed above.

\textcite{wecker2015stronglycorrelated} investigated using adiabatic state preparation and a Trotter based approach to phase estimation to probe the phase diagram of the Fermi-Hubbard model. Their approach to finding the ground state required $\mathcal{O}(M^3)$ gates. This was subsequently improved by \textcite{kivlichan2019condensedtrotter}, who considered the case where an extensive error in the energy is targeted, and found that the cost of Fermi-Hubbard model simulations was effectively between $\mathcal{O}(1)$ and $\mathcal{O}(M^{1/2})$ using Trotterisation.

The qubitization algorithm of \textcite{babbush2018encoding}, discussed above, was also applied to the Fermi-Hubbard model. Those authors obtained a gate scaling of $\mathcal{O}(M)$ for the Fermi-Hubbard model. When considering an intensive error in the energy, this approach outperforms the Trotter-based method of \textcite{kivlichan2019condensedtrotter}. \\

New procedures have also been developed specifically for time evolution under lattice Hamiltonians. These Hamiltonians have geometrically local interactions between qubits that are laid out on a lattice -- such as the Fermi-Hubbard model under a locality preserving mapping (see Sec.~\ref{Subsubsec:OtherEncodings}). \textcite{haah2018quantum} made use of arguments about the speed of information propagation in lattice systems to obtain a simulation algorithm requiring $\mathcal{O}(Mt)$ gates to simulate time evolution for time $t$ under lattice Hamiltonians. Similar results were obtained by \textcite{childs2019lattice}, who proved that a $k$\textsuperscript{th} order product formula can simulate time evolution of an $M$ qubit lattice Hamiltonian using $\mathcal{O}((Mt)^{1+\frac{1}{k}}) $ elementary operations. These algorithms are almost optimal in terms of asymptotic gate complexity. \\

\paragraph{Grid based methods \\}\label{Para:HamSimGrid}
\addcontentsline{toc}{subsection}{\hspace{1cm}d. Grid based methods}
As discussed in Sec.~\ref{Subsubsec:firstqclassical}, grid based methods are particularly suitable for high accuracy calculations -- in particular, calculations which do not make the Born-Oppenheimer approximation, and so treat both the electrons and nuclei on an equal footing.

The first quantised grid based algorithm algorithm of \textcite{kassal2008polynomial} proceeds as follows. The qubits are used to create a discretised grid, as described in Sec.~\ref{Subsec:1stencoding}. Physically relevant states can then be prepared using the algorithms outlined by \textcite{kassal2008polynomial, doi:10.1063/1.3115177}. The state can be propagated in time by repeatedly using the quantum Fourier transform to move between the position and momentum bases, so that the potential and kinetic terms are diagonal (respectively) and therefore simple to apply. \textcite{kassal2008polynomial} did not calculate an explicit scaling for their algorithm.\\

\textcite{KivReal} improved upon the method described above, using the Taylor series method for time evolution. Their method discretised the kinetic and potential terms of the Hamiltonian, separated them into linear combinations of unitary operators, and applied the Taylor expansion method to simulate time evolution. Those authors obtained a scaling of $\mathcal{O}((N/h^2 + N^2)t)$ (excluding logarithmic factors), where $h$ is the grid spacing and $t$ is the simulation time.

\subsubsection{Outstanding problems}\label{Subsubsec:OutstandingQPEproblems}

While the advanced Hamiltonian simulation methods described above are asymptotically more efficient than Trotterization, the Trotter error bounds appear to be loose by several orders of magnitude~\cite{poulin2014trotter,babbush2015chemical}. A study of spin Hamiltonians~\cite{childs2017toward} found that the asymptotic scaling of Trotter methods was much worse than the qubitization + quantum signal processing, and Taylor series methods. However, when numerical simulations were performed, Trotter methods required lower gate counts than the other methods. \textcolor{black}{There are three main factors that may make the Trotter error bound loose in chemical simulations. Firstly, it is difficult to obtain a tight error bound for a Trotterized time evolution under a physical Hamiltonian, as analytic proofs typically utilise multiple triangle inequalities, each of which may be loose. Secondly, these error bounds can be understood as the worst error that could be induced in any state which the unitary acts on. However, for chemistry simulation, we are often only interested in a small subset of the possible states, which again may reduce the Trotter error. Finally, we are often interested not in the error in the state, but in the error of some observable. This fact was exploited by \textcite{heyl2019quantum} who considered the error in local observables of spin chains, and obtained an improved simulation complexity.}

A challenge facing the Taylor series method is that it may require many elementary logic operations, resulting in a large T gate count when considering fault-tolerant approaches (although work by \textcite{sanders2019stateprep} may help to alleviate this problem). However, while the Taylor series method is asymptotically more expensive than qubitization + quantum signal processing, there has not yet been a variant of the latter technique formulated for time dependent Hamiltonians. Time dependent techniques are used in the interaction picture method that underpins the algorithms of \textcite{babbush2018sublinear,low2018interaction}. \textcolor{black}{In addition, it was noted by \textcite{childs2017toward} that implementing quantum signal processing required intensive classical pre-computation to obtain the parameters used in the quantum circuit. As such, it is not yet possible to conclusively state which method will perform best for chemical systems.}

Despite this progress, all of the methods discussed above require circuits with a large number of gates. As a result, these methods are typically assumed to require fault-tolerance. As near-term quantum computers will not have enough physical qubits for error correction, the long gate sequences required by these algorithms make them unsuitable for near-term hardware. Consequently, alternative methods are required for near-future chemistry simulation. We discuss one such approach in the following section.

\subsection{Variational algorithms}\label{Subsec:VQE}

A promising algorithm for near-term quantum hardware is the variational quantum eigensolver (VQE), first proposed and experimentally realised by \textcite{peruzzo2014variational}, and elaborated on by \textcite{VQETheoryNJP}. The VQE aims to find the lowest eigenvalue of a given Hamiltonian, such as that of a chemical system. The VQE is a hybrid quantum-classical algorithm. This means that it uses the quantum computer for a state preparation and measurement subroutine, and the classical computer to process the measurement results and update the quantum computer according to an update rule. This exchanges the long coherence times needed for phase estimation for a polynomial overhead due to measurement repetitions and classical processing. To date, the VQE has only been applied to second quantised basis set simulations, and so our discussion of it will only be concerned with that scenario.

\begin{figure}[t]
{\includegraphics[height=6cm]{{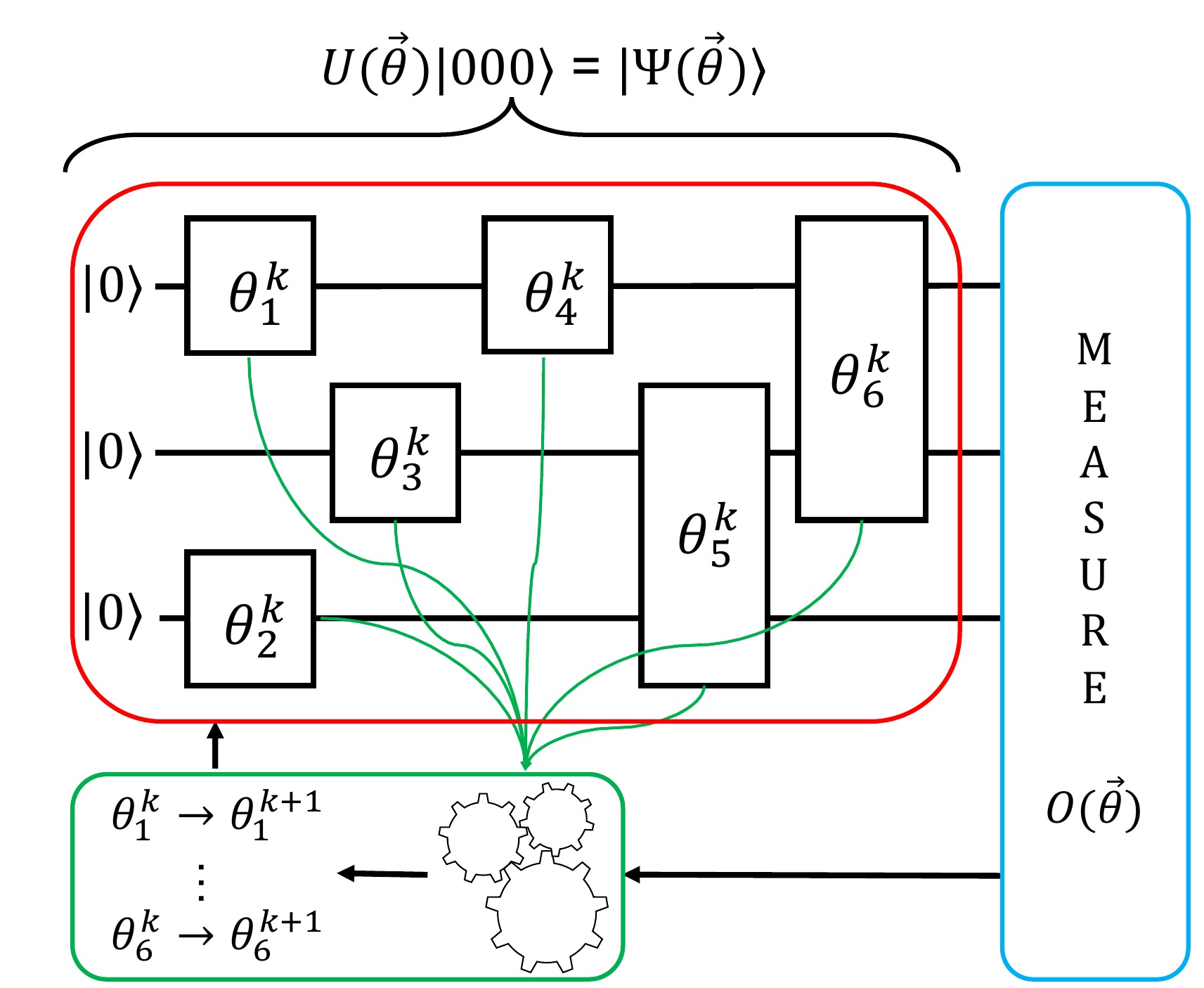}}}
\caption{A schematic of the variational quantum eigensolver (VQE). \textcolor{black}{The VQE attempts to find the ground state of a given problem Hamiltonian, by classically searching for the optimal parameters $\vec{\theta}$ which minimise $\bra{\Psi(\vec{\theta})} H \ket{\Psi(\vec{\theta})}$.} The state preparation and measurement subroutines (red, upper left, and blue, right) are performed on the quantum computer. The measured observable $O(\vec{\theta})$ and parameter values are fed into a classical optimisation routine (green, lower), which outputs new values of the parameters. The new parameters are then fed back into the quantum circuit. The gates acting on the qubits can be any parametrized gates, e.g. single qubit rotations or controlled rotations. Non-parametrized gates (e.g. X, Y, Z, CNOT) are also allowed. The circuit $U(\vec{\theta})$ and trial wavefunction it produces $|\Psi(\vec{\theta})\rangle$ are both known as the VQE ansatz. The process is repeated until the energy converges.}\label{Fig:VQEDiagram}
\end{figure}

The VQE relies upon the Rayleigh-Ritz variational principle. This states that for a parametrized trial wavefunction $\ket{\Psi (\vec{\theta})}$
\begin{equation}\label{Vary}
	\begin{aligned}
		\bra{\Psi (\vec{\theta})}H\ket{\Psi (\vec{\theta})} \geq E_0,
	\end{aligned}
\end{equation}
where $E_0$ is the lowest energy eigenvalue of the Hamiltonian $H$, and $\vec{\theta}$ is a vector of independent parameters, $\vec{\theta}=(\theta_1, ... , \theta_n)^\mathrm{T}$. This implies that we can find the ground state wavefunction and energy by finding the values of the parameters which minimise the energy expectation value. As classical computers are unable to efficiently prepare, store and measure the wavefunction, we use the quantum computer for this subroutine. We then use the classical computer to update the parameters using an optimisation algorithm. This sequence is shown in Fig.~\ref{Fig:VQEDiagram}. The qubit register is initialised in the zero state. We can optionally apply a non-parametrized set of gates to generate a mean-field~\cite{wecker2015stronglycorrelated,KivLinearDepth,jiang2018hubbard} or multi-reference state~\cite{dallaire2018low,babbush2015chemical,sugisaki2016constructing,tubman2018orthogonality,sugisaki2018openshell,sugisaki2019multireference} describing the chemical system of interest
\begin{equation}\label{StartingState}
	\begin{aligned}
		|\Psi_{\mathrm{ref}} \rangle = U_{\mathrm{prep}}|\bar{0} \rangle.
	\end{aligned}
\end{equation}
A series of parametrized gates $U(\vec{\theta}) = U_{N}(\theta_N)\dots U_{k}(\theta_k)\dots U_{1}(\theta_1)$ are then applied to the qubits. Here, $U_{k}(\theta_k)$ is the $k^{\textrm{th}}$ single or two qubit unitary gate, controlled by parameter $\theta_k$. This circuit generates the trial wavefunction
\begin{equation}\label{VQEprep}
	\begin{aligned}
		\ket{\Psi(\vec{\theta})}= U(\vec{\theta})|\Psi_{\mathrm{ref}} \rangle.
	\end{aligned}
\end{equation}
We refer to $\ket{\Psi(\vec{\theta})}$ as the ansatz state, and $U(\vec{\theta})$ as the ansatz circuit. However, the reader will find that in the literature the word ansatz is typically used interchangeably to describe both. The set of all possible states that can be created by the circuit $U$ is known as the `ansatz space'. We must select an ansatz appropriate for both the available hardware, and chemical system being simulated. The merits, drawbacks and implementation of common ans\"atze will be discussed in Sec.~\ref{Subsubsec:VQEAnsatz}.  

Once the wavefunction has been generated, we need to measure the expectation value of the Hamiltonian. Chemical Hamiltonians in the second quantised basis set approach can be mapped to a linear combination of products of local Pauli operators, using the transformations introduced in Sec.~\ref{Subsec:2ndencoding}. We write that
\begin{equation}
	H = \sum_j h_j P_j = \sum_j h_j\prod_i \sigma_i^j,
\end{equation}
where $h_j$ is a real scalar coefficient, $\sigma_i^j$ represents one of $I$ , $X$ , $Y$  or $Z$, $i$ denotes which qubit the operator acts on, and $j$ denotes the term in the Hamiltonian. We can then use the linearity of expectation values to write that 
\begin{equation}\label{VQEenergy}
	\begin{aligned}
		E(\vec{\theta}_k) = \sum_j^N h_j \langle \Psi (\vec{\theta}_k) | \prod_i \sigma_i^j | \Psi (\vec{\theta}_k) \rangle.
	\end{aligned}
\end{equation}

These state preparation and measurement steps should be repeated many times in order to measure the expectation value of every term in the Hamiltonian to the required precision. As the quantum computer is reinitialised for each repetition, the required coherence time is reduced compared to quantum phase estimation. This is known as the Hamiltonian averaging method of calculating the energy~\cite{mcclean2014locality}, and requires $\mathcal{O}(1 / \epsilon^2)$ measurements to determine the energy to a precision $\epsilon$. We will discuss the measurement aspect of the VQE in more detail in Sec.~\ref{Subsubsec:Measurement}.

Once the energy has been measured, it is passed to a classical optimisation routine, together with the current values of $\vec{\theta}_k$ (other observables, such as the energy gradient, could also be supplied to the optimisation routine). The optimisation routine outputs new values of the circuit parameters, $\vec{\theta}_{k+1}$. These are used to prepare a new trial state, $| \Psi (\vec{\theta}_{k+1}) \rangle$, which is ideally lower in energy. These steps are repeated until the energy converges to a minimum. We will summarise previous investigations into classical optimisation routines in Sec.~\ref{Subsubsec:VQEOptimisation}. 

The VQE has been experimentally demonstrated on most leading quantum architectures including: superconducting qubits~\cite{PRXH2,kandala2017hardware, PhysRevX.8.011021}, trapped ions~\cite{kokail2019latticeschwinger, TrappedIon, PhysRevA.95.020501, nam2019watervqe}, and photonic systems~\cite{peruzzo2014variational}, and shows many desirable features. It appears to be robust against some errors~\cite{VQETheoryNJP,PRXH2}, and capable of finding the ground state energies of small chemical systems using low depth circuits~\cite{kandala2017hardware}. While the VQE appears promising for near-term chemistry simulation, several considerable challenges remain. Firstly, the VQE is a heuristic method; it is currently unclear whether the quantum circuits proposed will yield better approximations of the ground state than the classical methods already available. This challenge is exacerbated by the difficulty of optimising the wavefunction, as optimisation routines could fail to find the global minimum. Secondly, the number of measurements required to obtain the energy to the desired accuracy can be large. Finally, the VQE is typically considered in the context of near-term quantum computers, without error correction. While techniques have been proposed to protect these calculations from the effects of noise (which we will discuss in Sec.~\ref{Sec:errorMitigation}), it is still possible that noise may prevent us from implementing sufficiently long circuits to surpass classical methods.

\subsubsection{Ans\"atze}\label{Subsubsec:VQEAnsatz}
The parametrized circuits, or `ans\"atze', for the VQE lie between two extremes: hardware-efficient and chemically inspired. \textcolor{Black}{There has been relatively little work comparing the effectiveness of different ans\"atze for anything but the smallest chemistry problems. As quantum computers with tens of qubits become more widely available, and quantum simulators become more powerful, there will be greater scope to test different ans\"atze for larger problem sizes.}

\paragraph{Hardware efficient ans\"atze \\}
\addcontentsline{toc}{subsection}{\hspace{1cm}a. Hardware efficient ans\"atze}
Hardware efficient ans\"atze have been in use since the first VQE experiment by \textcite{peruzzo2014variational}. These ans\"atze are comprised of repeated, dense blocks of a limited selection of parametrized gates that are easy to implement with the available hardware. They seek to build a flexible trial state using as few gates as possible. As such, they are well suited to the quantum computers currently available, which have short coherence times and constrained gate topologies. Hardware-efficient ans\"atze have been used to find the ground state energies of several small molecules~\cite{kandala2017hardware,kandala2018extending}. However they are unlikely to be suitable for larger systems, as they take into account no details of the chemical system being simulated. \textcite{ParticleHole} attempted to tackle this issue, by proposing hardware efficient ans\"atze which conserve particle number, and so permit the use of a chemically motivated initial state. This proposal has been experimentally demonstrated \cite{ganzhorn2018gateefficient} on a superconducting system.

\textcite{Barren} showed that using hardware efficient ans\"atze with random initial parameters makes the energy gradient essentially zero among most directions of Hilbert space, making classical optimisation extremely difficult. This effect becomes exponentially more prominent as the number of qubits and circuit depth increases. This suggests that randomly initialised hardware efficient ans\"atze are not a scalable solution for problems in quantum computational chemistry. \textcolor{black}{While techniques have been proposed to combat this problem~\cite{grant2019barren}, further work is needed to determine their efficacy beyond small system sizes.}\\

\paragraph{Chemically inspired ans\"atze \\}
\addcontentsline{toc}{subsection}{\hspace{1cm}b. Chemically inspired ans\"atze}
Chemically inspired ans\"atze result from adapting classical computational chemistry algorithms to run efficiently on quantum computers. Most notably, the coupled cluster (CC) method discussed in Sec.~\ref{Subsubsec:CC} can be extended to produce the unitary coupled cluster (UCC) ansatz~\cite{hoffmann1988ucc,bartlett1989ucc}. The UCC method creates a parametrized trial state by considering excitations above the initial reference state, and can be written as 
\begin{equation}\label{UCC}
	\begin{aligned}
		U(\vec{\theta}) &= e^{T - T^\dag}, \\ 
	\end{aligned}
\end{equation}
where $T = \sum_i T_i$, and
\begin{equation}\label{UCC2}
	\begin{aligned}
		T_1 &=  \sum_{i \in virt, \alpha \in occ} t_{i \alpha} a^\dag_i a_\alpha, \\ 
		T_2 &= \sum_{i, j \in virt, \alpha, \beta \in occ} t_{i j \alpha \beta}  a^\dag_i a^\dag_j a_\alpha a_\beta, \\ 
		&...
	\end{aligned}
\end{equation}
and $occ$ are occupied orbitals in the reference state, and $virt$ are orbitals that are initially unoccupied in the reference state. The UCC method is intractable on classical computers, but can be efficiently implemented on a quantum computer. It was originally proposed for quantum computational chemistry by \textcite{yung2014transistor} and \textcite{peruzzo2014variational}. A comprehensive review of the UCC method for quantum computational chemistry is given by \textcite{romero2017strategies}.

The UCC method retains all of the advantages of the CC method, with the added benefits of being fully variational, and able to converge when used with multi-reference initial states. The UCC ansatz is typically truncated at a given excitation level -- usually single and double excitations (known as UCCSD). We show a canonical implementation of the UCCSD ansatz in Sec.~\ref{Subsec:H2Illustrate}. The canonical UCCSD implementation requires $\mathcal{O}(M^3 N^2)$ gates when using the Jordan-Wigner mapping (where $M$ is the number of spin-orbitals, and $N$ is the number of electrons)~\cite{romero2017strategies}. This has been improved to $\mathcal{O}(M^3 N)$ gates, with a depth of $\mathcal{O}(M^2 N)$ using swap networks~\cite{ogorman2019swapnetworks}, or $\mathcal{O}(M^3) - \mathcal{O}(M^4)$ gates using a low-rank decomposition of the UCCSD tensor~\cite{motta2018low}. 

These gate counts assume that a single Trotter step can be used to implement the UCC operator, which has previously been shown to yield accurate results~\cite{romero2017strategies,ParticleHole}. \textcolor{black}{However, this approach has been questioned by \textcite{grimsley2019trotterizeduccsd} who found that depending on how the operators in the Trotterized UCCSD ansatz are ordered, the optimised energies of systems displaying significant electron correlation can vary by amounts larger than chemical accuracy. As a result, determining a suitable (or even optimal) ordering of the operators in the UCCSD ansatz may be an interesting and important area of future research.}

In practice, the gate scaling is typically better than the bounds given above, as many excitations are forbidden by the symmetry point groups of molecular orbitals. For example, the LiH molecule in an STO-3G basis naively has around 200 excitation operators to consider. However, taking into account symmetries and a reduced active space, one can achieve accurate results while considering only around 12 excitation operators~\cite{TrappedIon}. Moreover, we can use classically tractable methods to get initial approximations for the remaining non-zero parameters~\cite{PRXH2, romero2017strategies}, which makes the classical optimisation step of the VQE easier. \\

Alternative variants of the UCC ansatz have also been proposed for solving problems in quantum computational chemistry. These include: the Bogoliubov-UCC ansatz~\cite{dallaire2018low} (a quasiparticle variant of UCC that is suitable for more general Hamiltonians than the UCC ansatz, potentially including the pairing terms present in superconductivity or three body terms present in nuclear physics), the `low-depth circuit ansatz'~\cite{dallaire2018low} (a heuristic which attempts to mimic the aforementioned Bogoliubov-UCC ansatz using a lower depth circuit), orbital optimised-UCCD~\cite{wataru2019ooucc,sokolov2019ooUCC} (which treats single excitations on the classical computer, rather than on the quantum computer, to reduce the depth and number of gates in the circuit), $k$-UpCCGSD~\cite{lee2018gucc} and $k$-uCJ~\cite{matsuzawa2019jastrow} (heuristic ans\"atze comprised of repeated layers of selected UCC operators), adaptive-VQE~\cite{grimsley2019adapt, tang2019adaptqubit} (which creates an adaptive ansatz designed to maximise recovery of the correlation energy), and qubit coupled-cluster~\cite{ryabinkin2018qcc, ryabinkin2019qcc} (a heuristic method which produced lower gate counts than the UCC ansatz when applied to several small molecules).

\paragraph{Hamiltonian variational ansatz \\}
\addcontentsline{toc}{subsection}{\hspace{1cm}c. Hamiltonian variational ansatz}
There are also variational ans\"atze that lie between the \textcolor{black}{extremes of chemically inspired ans\"atze and hardware efficient ans\"atze.} One important example is the Hamiltonian variational ansatz (also commonly referred to as a Trotterized adiabatic state preparation ansatz), proposed by \textcite{PhysRevA.92.042303}. This ansatz was inspired by adiabatic state preparation and the quantum approximate optimisation algorithm (a similar quantum-classical hybrid algorithm for combinatorial optimisation problems~\cite{farhi2014quantum}). The idea is to Trotterize an adiabatic path to the ground state, using a number of Trotter steps that may be insufficient for accurate results. One can then variationally optimise the Trotter evolution times to create a variational ansatz for the ground state. The number of parameters in this ansatz scales linearly with the number of Trotter steps, $S$. \textcolor{black}{Mathematically, we write that 
\begin{equation}
    U = \prod_s^S \prod_i \mathrm{exp}\left(\theta_i^s P_i \right)
\end{equation}
where $P_i$ are the Pauli strings in the Hamiltonian.}

The efficiency of this ansatz is determined by the number of terms in the Hamiltonian. This leads to a scaling of approximately $\mathcal{O}(M^4$) when considering a Gaussian molecular orbital basis~\cite{PhysRevA.92.042303}. \textcite{motta2018low} improved this asymptotic scaling using the low rank decomposition method discussed in Sec.~\ref{Subsec:reduction}. They reduced the number of gates required to implement a Trotter step of the Hamiltonian to $\mathcal{O}(M^2 \mathrm{log}_2(M))$ with increasing molecular size, and $\mathcal{O}(M^3)$ for fixed molecular size and increasing basis size.
The electronic structure Hamiltonian in a plane wave dual basis only contains $\mathcal{O}(M^2)$ terms. \textcite{KivLinearDepth} showed that for Hamiltonians of this form, we can implement Trotter steps in depth $\mathcal{O}(M)$, using $\mathcal{O}(M^2)$ two qubit gates.

The Hamiltonian variational ansatz appears particularly suitable for the Fermi-Hubbard model, which only has $\mathcal{O}(M)$ terms in its Hamiltonian. \textcite{KivLinearDepth} showed that it is possible to implement Trotter steps of the Fermi-Hubbard Hamiltonian with $\mathcal{O}(\sqrt{M})$ depth and $\mathcal{O}(M^{1.5})$ gates on a linear array of qubits with nearest-neighbour connectivity. \textcite{jiang2018hubbard} improved this result for the case of a 2D array of qubits with nearest-neighbour connectivity. They showed that it is possible to prepare initial states of the Fermi-Hubbard model using $\mathcal{O}(M^{1.5})$ gates, and perform Trotter steps of the Hamiltonian using $\mathcal{O}(M)$ gates for each Trotter step. It has also been noted~\cite{KivLinearDepth} that it is possible to use locality preserving mappings~\cite{jiang2018majorana,VerCirac} to perform Trotter steps in constant depth. However, this comes at a cost of requiring additional qubits, and may have a large constant factor gate overhead.

\subsubsection{Measurement}\label{Subsubsec:Measurement}
In Sec.~\ref{Subsec:VQE} we described the Hamiltonian averaging method, whereby the expectation value of each term in the Hamiltonian is estimated through repeated state preparation and measurement. This procedure can be used to calculate the 1-RDM and 2-RDM of the system. \textcite{romero2017strategies,VQETheoryNJP} showed that the number of measurements $N_m$, required to estimate the energy to a precision $\epsilon$, is bounded by
\begin{equation}
    N_m = \frac{\big{(} \sum_i |h_i| \big{)}^2}{\epsilon^2},
\end{equation}
where $h_i$ are the coefficients of each Pauli string in the Hamiltonian. This leads to a scaling of $\mathcal{O}(M^6/\epsilon^2)$ in a Gaussian orbital basis, and $\mathcal{O}(M^4/\epsilon^2)$ for a plane wave dual basis~\cite{cao2019review,Babbush2017low,mcclean2014locality}.

These scalings may be problematic for large molecular calculations. For example, \textcite{PhysRevA.92.042303} found that around $10^{13}$ samples would be required per energy evaluation for a 112 spin-orbital molecule such as Fe$_2$S$_2$. Fortunately, strategies have been proposed to reduce the cost of measurement. Several groups have proposed using heuristics to group together commuting terms~\cite{VQETheoryNJP,kandala2017hardware,izmaylov2018revising,verteletskyi2019vqemeasurements}, which appears to reduce the measurement cost by a constant factor. \textcolor{black}{Some authors have also considered using additional unitary transforms to enable more terms to be grouped, reducing the number of groups of terms from $\mathcal{O}(M^4)$ to $\mathcal{O}(M^3)$~\cite{gokhale2019measurement,gokhale2019measurement2,crawford2019measurement,yen2019measurement,zhao2019measurement,jena2019measurement}. By grouping terms at a fermionic level, rather than a qubit level, \textcite{bonet2019nearlyoptimal} devised a method for measuring the fermionic 2-RDM using $\mathcal{O}(M^2)$ circuits (an additional gate depth of $\mathcal{O}(M)$ is required). Similar results were obtained by \textcite{jiang2019ternary}, by combining Bell basis measurements, and the ternary-tree mapping described in Sec.~\ref{Subsubsec:BKencoding}. Another interesting direction to explore may be to combine locality preserving fermion-to-qubit mappings with efficent results for qubit tomography~\cite{cotler2019overlapping, bonet2019nearlyoptimal,paini2019approximate}.} Although we might expect that  it would be optimal to divide the Hamiltonian into the fewest possible number of groups, this is not always the case. \textcite{VQETheoryNJP} showed that it is important to consider the covariances between Hamiltonian terms within groups, as if these covariances are positive, then it may be better to measure the terms separately. These covariances can be estimated using classical approximations of the target state. 

\textcolor{black}{If we are only interested in measuring the energy expectation value, rather than the 1 and 2-RDMs, \textcite{huggins2019efficientmeasurements} showed that it is possible to use the low rank Hamiltonian decomposition technique (discussed in Sec.~\ref{Subsubsec:LowRank}) and orbital basis rotations to divide the Hamiltonian into $\mathcal{O}(M)$ measurement groups. When taking into account covariances between commuting terms, the total number of measurements required for small molecules was reduced by several orders of magnitude. This measurement scheme also reduces the locality of strings in the qubit Hamiltonian, thus reducing the impact of qubit read-out error.}

\subsubsection{Classical optimisation}\label{Subsubsec:VQEOptimisation}
As discussed above, classical optimisation is a crucial aspect of the VQE. However, it is often difficult to minimise complicated functions in high dimensional parameter spaces. Classical optimisation routines must be both fast and accurate. They also need to be robust to noise, which will be significant in near-term quantum computers. 

Classical optimisation algorithms can be broadly divided into two classes: direct search and gradient based methods. Direct search algorithms are considered more robust to noise than gradient based methods, but may require more function evaluations~\cite{DirectSearch}.

Below, we summarise the results of previous experimental and numerical investigations into classical optimisation algorithms used in conjunction with the VQE. We also discuss methods to assist the classical optimisation procedure. \\

\paragraph{Previous optimisation studies\\}
\addcontentsline{toc}{subsection}{\hspace{1cm}a. Previous optimisation studies}
Prior experimental VQE implementations have been limited to small systems by the number of qubits available. As a result, the parameter space to optimise over is relatively small, so previous results may not be indicative of how these optimisation algorithms will perform for large problems. However, these studies are able to demonstrate which methods can cope with the high noise rates of current hardware. Direct search methods (such as Nelder-Mead simplex~\cite{peruzzo2014variational,PhysRevA.95.020501,TrappedIon}, simulated annealing~\cite{TrappedIon}, particle swarm~\cite{PhysRevX.8.011021,Santagatieaap9646} covariance matrix adaptation~\cite{sagastizabal2019symmetry}, and the dividing rectangles algorithm~\cite{kokail2019latticeschwinger}) were found to exhibit good convergence to the ground state energy of small systems, despite relatively high noise rates.

\textcolor{black}{The simultaneous perturbation stochastic approximation (SPSA) algorithm, has also been used to successfully find the ground state energy of small molecules, despite relatively large uncertainties due to shot noise, and physical error rates~\cite{kandala2017hardware,kandala2018extending,ganzhorn2018gateefficient}. The SPSA algorithm is an approximate gradient based method, which steps `downhill' along a randomly chosen direction in parameter space. In contrast, canonical gradient descent based methods have struggled to find the ground state, due to the high levels of noise present in current quantum devices~\cite{peruzzo2014variational}.}\\

The number of numerical studies comparing a subset of optimisation algorithms for the VQE has grown rapidly -- hence, a full discussion of every study is beyond the scope of this review. Instead, we summarise the major findings, and highlight the different approaches for optimisation.

In comparisons of `out-of-the-box' optimisers~\cite{VQETheoryNJP,romero2017strategies}, the Nelder-Mead simplex algorithm utilised in experimental studies was outperformed by both other direct search approaches, and gradient based approaches, such as L-BFGS-B. All methods were liable to becoming trapped in local minima, even for small systems. This problem was partly mitigated by using a good initialisation strategy, such as using a chemically motivated (M\"oller-Plesset) guess for UCC excitation amplitudes~\cite{romero2017strategies}. Similar results were obtained using heuristic methods of optimisation and initialisation~\cite{PhysRevA.92.042303}. Other investigations have considered optimisation using algorithms related to gradient descent, such as: stochastic gradient descent~\cite{sweke2019gradient,harrow2019gradient}, variational imaginary time evolution~\cite{mcardle2018variational}, and quantum natural gradient descent~\cite{stokes2019natural,yamamoto2019naturalgrad}. A limitation of the majority of these studies is that realistic noise has rarely been considered, so it is still unclear exactly how effective each method may be in practice.

\textcolor{black}{An alternative approach to optimising all of the parameters simultaneously, is to optimise them sequentially ~\cite{nakanishi2019sequential,parrish2019jacobi,StructureLearning19}. One can analytically minimise the energy with respect to a single parameter, and can then sequentially optimise the cost function over all of the parameters. Repeated sequential optimisations are required to find a good estimate of the energy minimum.}

\textcolor{black}{Recent numerical work has explored the possibility of using classical neural networks to optimise VQE circuits for small instances of the Fermi-Hubbard model~\cite{verdon2019learningtolearn, wilson2019metalearning}. This approach (known as `meta-learning') works by training the neural networks on many random instances of the system being studied. The networks were able to successfully `learn' to optimise the VQE ansatz, and showed indications that they may be able to generalise to larger systems than those on which they were trained~\cite{verdon2019learningtolearn}. The approach also demonstrated some resilience to noise included in the simulations~\cite{wilson2019metalearning}.}\\

\paragraph{Related methods of optimisation\\}
\addcontentsline{toc}{subsection}{\hspace{1cm}b. Related methods of optimisation}
Methods which aid classical optimisation, but that are not optimisation algorithms in their own right, have also been proposed.

\begin{figure}[!h]
{\includegraphics{{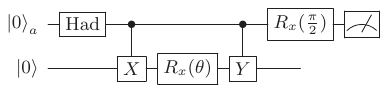}}}
\caption{A quantum circuit to calculate the gradient of a toy VQE simulation. In this toy problem, the ansatz used is $\ket{\Psi} = R_x(\theta)\ket{0}$, and the Hamiltonian is $H=Y$. The energy is given by $E(\theta) = \bra{\Psi(\theta)} H \ket{\Psi(\theta)} = \bra{0} R_x^\dag (\theta) Y R_x(\theta) \ket{0}$. The energy gradient, $\frac{\partial{E}}{\partial{\theta}} = \frac{i}{2}(\bra{0} X R_x^\dag (\theta) Y R_x(\theta) \ket{0} - \bra{0} R_x^\dag (\theta) Y R_x(\theta) X \ket{0}$). This is the expectation value generated by the circuit above.}\label{Qgrad}
\end{figure}

Quantum circuits have been proposed to calculate the analytic gradient of the energy with respect to one of the parameters~\cite{mitarai2018imaginaryindirect,guerreschi2017derivatives,schuld2019gradients}. This avoids the use of finite difference formulae, which restrict the accuracy of gradient evaluation, as the finite difference considered is limited by the noise in the energy evaluation. The quantum gradient method makes use of the differentiability of parametrized unitary operators. parametrized unitaries can be written as exponentials of the parameter and an anti-Hermitian operator, which are simple to differentiate. A circuit to obtain the gradient of a toy VQE simulation is shown in Fig.~\ref{Qgrad}.

Several works have used concepts from adiabatic quantum computing to aid the classical optimisation procedure. \textcite{PhysRevA.92.042303} proposed an `annealed variational' method alongside the Hamiltonian variational ansatz. This technique can be generalised to any ansatz with a repeating, layered structure. We assume that the ansatz is composed of $S$ layers. We first decompose the Hamiltonian of interest into $H_s = H_0 + sH_1$, where $H_0$ is a Hamiltonian that is efficiently solvable with a classical computer, and $H_1$ is a difficult Hamiltonian to solve. When $s=1$, the Hamiltonian is equivalent to the problem Hamiltonian. The annealed variational method works by considering the $S$ layers as separate, distinct ans\"atze. The input state to the first layer of the ansatz is the ground state of $H_{s=0}$. We optimise this first layer to find the ground state of $H_{s=1/S}$. This state is then the input into layer $2$, which is optimised to find the ground state of $H_{s=2/S}$. This process is repeated until the final layer, which takes the ground state of $H_{s=(S-1)/S}$ as its input, and targets the ground state of $H_{s=1}$. All of these steps are then combined, and used as the starting point for a standard VQE optimisation. A similar technique was proposed by \textcite{saez2018adiabatic}.

\textcite{ParticleHole} introduced a transformation of the Hamiltonian, such that it only measures the energy of excitations above the Hartree-Fock state (the correlation energy). Because only the correlation energy is calculated, fewer measurements are required and classical optimisation becomes easier. Overall, simulated VQE calculations on small molecules were sped up by a factor of 2 to 4~\cite{ParticleHole}.\\

\subsection{Evaluation of excited states}\label{Subsec:excited}
In this section, we discuss methods used to evaluate the excited states of chemical Hamiltonians. \textcolor{Black}{These are of interest for calculations of spectral properties, such photodissociation rates and absorption bands. We are sometimes interested in obtaining excited states with specific properties, such as a certain spin, or charge~\cite{ryabinkin2019constrainedvqe, yen2019symmetryprojectors,ryabinkin2018symmetry}. The techniques used to obtain these special states are similar in nature to the methods we discuss for finding general excited states.} There is still no clear consensus as to which method may perform best for finding excited states. As such, we describe each of the leading methods in turn, and discuss their advantages and limitations. \textcolor{black}{We note that this area has seen rapid recent development. Consequently, we do not discuss initial proposals to calculate excited states, such as the folded spectrum method~\cite{VQETheoryNJP} and WAVES~\cite{Santagatieaap9646}, which are comparatively less efficient than more recent proposals.}

\subsubsection{Quantum subspace expansion}\label{Subsubsec:SubspaceExpansionExcited}

The quantum subspace expansion (QSE) method was originally introduced to find excited states~\cite{PhysRevA.95.042308}, but has proven to be one of the most useful techniques introduced to near-term quantum computational chemistry, with benefits including better ground state estimation (discussed below) and mitigation of hardware errors (discussed in Sec.~\ref{Subsec:SubspaceMitigation}). 

The original formulation of the QSE used a polynomial number of additional measurements to find the excited states of a quantum system~\cite{PhysRevA.95.042308}. The motivation for this was that the higher order reduced density matrices (RDMs) can be obtained by expanding the wavefunction in a subspace around the ground state. These RDMs can then be used to find the linear response excited eigenstates. \textcite{PhysRevA.95.042308} considered the single excitation linear response subspace around the fermionic ground state. This subspace is spanned by the states $a_i^\dag a_j \ket{E_0}$ for all possible $i, j$, which corresponds to measuring the 3- and 4-RDMs. This is designed to target the low-lying excited states, which are assumed to differ from the ground state by a small number of excitations.

The excited states can be found by solving a generalised eigenvalue problem in fermionic Fock space
\begin{equation}
	H^\mathrm{QSE}C = S^\mathrm{QSE}CE, 
\end{equation}
with a matrix of eigenvectors $C$, and a diagonal matrix of eigenvalues $E$. The Hamiltonian projected into the subspace is given by
\begin{equation}
	H_{ij,kl}^\mathrm{QSE} = \bra{E_0}a_i a_j^\dag H a_k^\dag a_l \ket{E_0}.
\end{equation}
The overlap matrix, required because the subspace states are not orthogonal to each other, is given by 
\begin{equation}
	S_{ij,kl}^\mathrm{QSE} = \bra{E_0}a_i a_j^\dag a_k^\dag a_l \ket{E_0}.
\end{equation}
The QSE was experimentally demonstrated in a two qubit superconducting system to find the ground and excited states of H$_2$~\cite{PhysRevX.8.011021}. \textcolor{black}{\textcite{ollitrault2019eom} showed that the QSE with single and double excitations can be understood as an approximation to the `equation of motion' (EoM) method for finding excited states, which was originally introduced as a classical method in quantum chemistry by \textcite{rowe1968eom}. \textcite{ollitrault2019eom} showed how the EoM method can be implemented on a quantum computer. They found the EoM method to be robust to noise when demonstrated experimentally, and more accurate for finding excitation energies than the QSE, in numerical tests.}\\

As discussed above, the QSE has also been extended to provide better estimates of the ground state energy. \textcite{takeshita2019virtualorbs} showed how the QSE can be used to recover the energy contribution of virtual orbitals, without requiring additional qubits to represent the virtual orbitals. This was achieved by considering a QSE with single and double excitations, where the double excitations take two electrons from an active space, to the virtual space. Naively, this would require the measurement of matrix elements like $H^{ijkl}_{\alpha \beta \gamma \delta , \epsilon \zeta \eta \theta} = \bra{E_0} a_\alpha a^\dag_\beta a_\gamma a^\dag_\delta \left ( a^\dag_i a^\dag_j a_k a_l \right ) a^\dag_\epsilon a_\zeta a^\dag_\eta a_\theta \ket{E_0}$, of which there are $\mathcal{O}(M^{12})$. However, because the double excitations are restricted to exciting two electrons into the virtual space, it is possible to use Wick's theorem, and contract over the operators in the virtual space, to express the matrix element above in terms of matrix elements involving only 8 creation/annihilation operators (which defines the 4-RDM). These operators only act on the active space orbitals, $M_A$, which reduces the number of measurements required to $\mathcal{O}(M_A^8)$, and means that no additional qubits are required. Those authors demonstrate this method by numerically simulating a cc-PVDZ calculation on H$_2$ (which normally requires 20 qubits), using just 4 qubits, and the additional measurements described above.

\textcolor{black}{The QSE was also extended by \textcite{huggins2019qse}, who devised a low cost method to extend the subspace from being excitations above a reference state, to being any state that is efficiently preparable on the quantum computer. The method thus allows for creation of flexible ansatz states, without a dramatic increase in the circuit depth. This approach was further refined by \textcite{stair2019multirefkrylov}, who provided a more efficient approach to realising this technique, when the ansatz circuit used is composed of products of exponentiated Pauli operators. \\}

\textcolor{black}{The main drawback of the QSE is the large number of measurements that may be required to obtain the 4-RDM of the system. In general there are $\mathcal{O}(M^8)$ elements to measure in the 4-RDM, compared to $\mathcal{O}(M^4)$ elements for the Hamiltonian. The cost can be somewhat reduced by approximating the 4-RDM using products of lower order RDMs, and perturbative corrections. In addition, using the linear response excitation operators described above, we are limited in our description of excited states. This can be problematic for systems which need higher order excitations to accurately describe the excited states~\cite{watson2012butadiene, lee2018gucc}.} 

We provide more information on the QSE method in Sec.~\ref{Sec:errorMitigation}, where we discuss how it can also be used to mitigate the effects of errors.

\subsubsection{Overlap-based methods}\label{Subsubsec:OverlapMethods}
Overlap based methods exploit the orthogonality of energy eigenstates. Once an eigenstate is found, we can find other eigenstates by ensuring that they are orthogonal to the original state~\cite{higgott2018variational, endo2018discovering}. After finding the ground state $\ket{E_0}$ of a Hamiltonian $H$ with a VQE calculation, we replace the Hamiltonian with 
\begin{align}
H^\prime = H + \alpha \ket{E_0}\bra{E_0},
\end{align}
where $\alpha$ is chosen to be sufficiently large compared to the energy scale of the system. The ground state of the updated Hamiltonian $H^\prime$ is no longer $\ket{E_0}$, but the first excited state $\ket{E_1}$ of the original Hamiltonian $H$. This process can be repeated to obtain higher energy eigenstates. The energy of the updated Hamiltonian, $\bra{\Psi (\vec{\theta})} H^\prime \ket{\Psi (\vec{\theta})}=\bra{\Psi (\vec{\theta})} H \ket{\Psi (\vec{\theta})} +\alpha \braket{\Psi (\vec{\theta})| E_0}\braket{E_0|\Psi (\vec{\theta})} $ can be obtained by measuring each term separately. We can measure the first term using the Hamiltonian averaging procedure described in Sec.~\ref{Subsec:VQE}. The second term can be obtained from circuits which calculate the overlap between the states, such as the SWAP test. The SWAP test approach requires a circuit that has twice as many gates as the ansatz used (but is of the same depth), and has twice as many qubits. We can also use a method which requires twice as many gates and is twice as deep as the original ansatz, but does not require additional qubits~\cite{higgott2018variational}. The additional resources required are the main limitation of this approach to calculating excited states. Overlap-based techniques were numerically investigated by \textcite{lee2018gucc}, who also considered the propagation of errors resulting from only obtaining an approximation of lower lying eigenstates, rather than the true eigenstates.

\textcolor{black}{ \subsubsection{Contraction VQE methods}\label{Subsubsec:ContractionVQE}
The Subspace-search variational quantum eigensolver (SSVQE)~\cite{nakanishi2018subspace} and the multistate  contracted  variational quantum eigensolver (MCVQE)~\cite{parrish2019quantum} methods are related to the overlap based methods described above, in that they are driven by the orthogonality of energy eigenstates. The differences lie in how they enforce this orthogonality, as well as how the ordering of the eigenstates is determined. In the overlap based methods, orthogonality is enforced by including the overlap between states in the cost function. In contrast, the contraction VQE methods exploit conservation of orthogonality between states under a unitary transform. To be more specific, a set of orthonormal approximate eigenstates $\{ \ket{\phi_i}_{i=0}^k \}$ are chosen using an efficient classical approach. The quantum computer must be able to be initialised in any of these states. The orthogonality of these states is invariant under the application of a unitary ansatz circuit. The aim of the ansatz circuit is to generate linear combinations of these initial states, to form good approximations of the low energy subspace of the system. This is achieved by optimising the ansatz over an ensemble cost function, 
\begin{align}\label{Eq:costfunctionSS}
\mathcal{C}(\vec{\theta})= \sum_{j=0}^k \bra{\phi_j}U^\dag(\vec{\theta})H U(\vec{\theta})\ket{\phi_j}.
\end{align}}
\textcolor{black}{Both of the contraction VQE methods then use further processing to find the ordering of the energy eigenstates. The SSVQE method uses a similar quantum-classical hybrid approach to the ordinary VQE. In contrast, the MCVQE method uses classical diagonalisation of a Hamiltonian matrix obtained in the low energy subspace (similar to the way in which the quantum subspace expansion works). 
These contraction VQE methods differ from the overlap based methods, in that they obtain all of the eigenstates at the same time, rather than sequentially. Consequently, all eigenstates should be obtained to a similar accuracy. However, it may be difficult to find a unitary circuit that simultaneously prepares many eigenstates on an equal footing, which may make these contraction methods difficult to realise in practice.}

\section{Error mitigation for chemistry}
\label{Sec:errorMitigation}
As discussed in Sec.~\ref{Subsec:QuantumComputing}, physical qubits accumulate errors during computation, due to their interaction with the environment, and our imperfect control. While the effects of these errors can be suppressed using quantum error correction, this requires a considerable increase in the number of qubits used for the computation. If these errors are not dealt with, they will corrupt the results of our algorithms, rendering the calculations meaningless. This was confirmed for the case of chemistry calculations by \textcite{sawaya2016error}, who used numerical simulations to show the impact that noise has on phase estimation based chemistry calculations. Phase estimation based approaches (discussed in Sec.~\ref{Subsec:qpe}) require long circuit depths, and so implicitly assume the use of quantum error correction. We will discuss the resources required to carry out these calculations in Sec.~\ref{Subsec:LongTermQuantumResources} -- however, it suffices to say here that they are considerably greater than the resources we have available at time of writing. We claimed in Sec.~\ref{Subsec:VQE} that the reduced coherence time requirements of the VQE make it more suitable for noisy quantum hardware, and may enable the extraction of accurate results without the use of quantum error correction. It has been shown both theoretically~\cite{VQETheoryNJP} and experimentally~\cite{PRXH2} that the VQE can be inherently robust to some coherent errors, such as qubit over-rotation. However, small experimental demonstrations of the VQE have shown that noise can still prevent us from reaching the desired levels of accuracy~\cite{kandala2017hardware,TrappedIon}. Consequently, it appears that additional techniques are required, if we are to perform classically intractable chemistry calculations, without error correction.

The methods described in this section have been developed to mitigate errors, rather than correct them. These techniques are only effective when used with low depth circuits, such that the total error rate in the circuit is low. However, the additional resources required are much more modest than for full error correction. In general, these techniques only require a multiplicative overhead in the number of measurements required, if the error rate is sufficiently low. Many of these techniques were introduced for use in general near-term algorithms, and so can be applied to problems beyond chemistry simulation. \\

As we are dealing with errors, it becomes necessary to consider mixed states, rather than just pure states. As such, we now switch to the density matrix formalism of quantum mechanics.

We consider a quantum circuit that consists of $G$ unitary gates applied to an initial reference state $\ket{\bar{0}}$. The output state if errors do not occur is given by
\begin{equation}
		\rho_0 = \mathcal{U}_G \circ \dots \circ \mathcal{U}_2 \circ \mathcal{U}_1(\ket{\bar{0}}\bra{\bar{0}}),
\end{equation}
where for a density matrix $\rho$, $\mathcal{U}(\rho) = U \rho U^\dag$. We extract information from the circuit by measuring a Hermitian observable, $O$
\begin{equation}
		\bar{O}_0 = \textrm{Tr}[\rho_0 O].
\end{equation}

If each gate is affected by a noise channel $\mathcal{N}_i$, the prepared state becomes 
\begin{equation}
		\rho = \prod_i \mathcal{N}_i \circ \mathcal{U}_i(\ket{\bar{0}}\bra{\bar{0}}),
\end{equation}
and the measurement result becomes $\bar{O} = \textrm{Tr}[\rho O]$.
In general, we cannot recover the noiseless state $\rho_0$ from the noisy state $\rho$ without error correction. However, the error mitigation methods discussed below can approximate the noiseless measurement result $\bar{O}_0$ from the noisy measurement result, $\bar{O}$ when the error rate is sufficiently low. \textcolor{black}{It is important to note that error mitigation schemes are not a scalable solution to the problem of noise in quantum hardware. In order to scale up computations to arbitrarily large sizes, fault-tolerant, error corrected quantum computers are required.}

\begin{figure}[t]
\includegraphics[width=7cm]{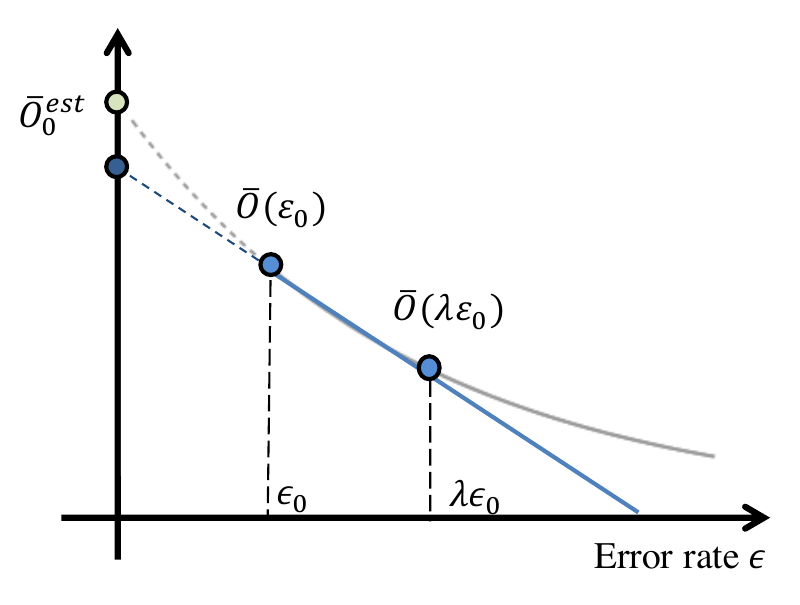}
\caption{A comparison of linear (blue, lower) and exponential (grey, upper) extrapolation. The horizontal axis is the error rate of each gate and the vertical axis is the expectation value of the measured observable, $\bar{O}$. This figure was reproduced from \textcite{endo2017practical}, with permission.}
\label{extrap}
\end{figure}

\subsection{Extrapolation}\label{Subsec:Extrapolation}
The extrapolation method~\cite{Li2017,PhysRevLett.119.180509} works by intentionally increasing the dominant error rate, $\epsilon_0$, by a factor $\lambda$, and inferring the error free result by extrapolation. We can increase the error rate using the techniques described by \textcite{Li2017, kandala2018extending}. The technique is based on Richardson extrapolation~\cite{Richardson299}, which to first order corresponds to linear extrapolation using two points. We could also take a linear or higher order fit with several data points. For the former case, the estimated value of the observable is given by 
\begin{equation}
		\bar{O}_0^{\mathrm{est}} = \frac{\lambda \bar{O}(\epsilon_0) - \bar{O}(\lambda \epsilon_0)}{\lambda - 1}.
\end{equation}
While this method can improve the accuracy of calculations, it requires additional measurements in order to keep the variance of the measured observable the same as the non-extrapolated case. The extrapolation method has been demonstrated for VQE experiments in both molecular chemistry~\cite{kandala2018extending,ollitrault2019eom}, and nuclear physics~\cite{shehab2019deuteronextrap}.

Exponential extrapolation was introduced by \textcite{endo2017practical} as a more appropriate extrapolation technique for large quantum circuits. A comparison between the two extrapolation methods is shown in Fig.~\ref{extrap}. 
\textcite{otten2018recovering} have also extended the extrapolation method to the scenario where the error rates of different gates are increased by different factors.

\subsection{Probabilistic error cancellation}\label{Subsec:Quasiprob}
\begin{figure}[t]
\includegraphics[width=8cm]{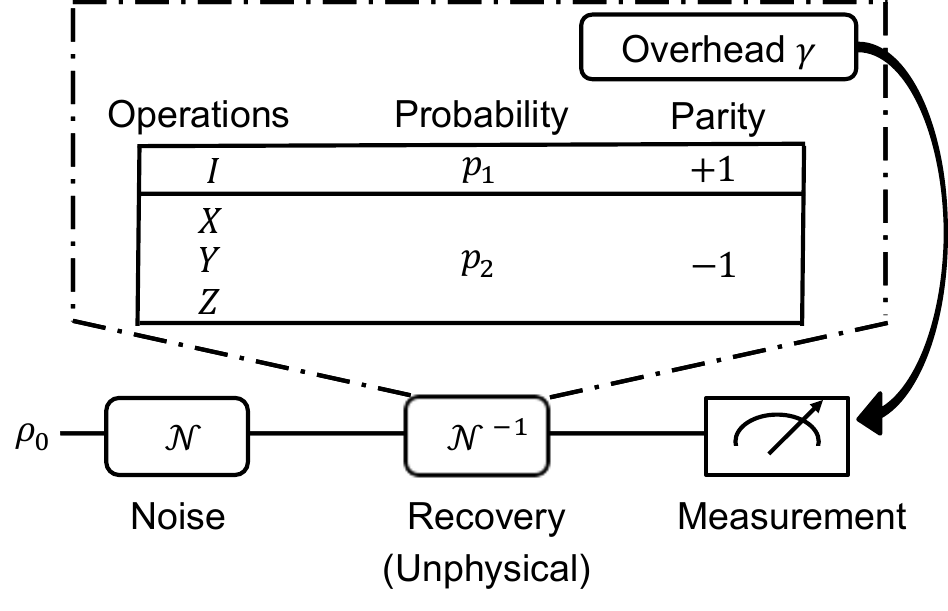}
\caption{A schematic of the probabilistic error cancellation method for a depolarising error resulting from a single qubit gate. After the gate is applied, there is a noise channel $\mathcal{N}$. The method works by effectively realising the inverse channel $\mathcal{N}^{-1}$. This is achieved by randomly applying one of the $X$, $Y$ or $Z$ operators with probability $p_2$, or the identity gate with $p_1$. The expectation values resulting from the circuits are combined. If one of the Pauli matrices was applied to realise the inverse channel, the resulting expectation value is subtracted, rather than added (parity $-1$). The overhead $\gamma$ determines the number of additional measurements required to keep the variance of the error mitigated result equal to the variance of the noisy result. This can be generalised to multi-qubit gates as described in the main text.}
\label{quasiprob}
\end{figure}

The probabilistic error cancellation method introduced by \textcite{PhysRevLett.119.180509} works by effectively realising the inverse of an error channel, $\mathcal{N}^{-1}$, such that $\mathcal{N}^{-1}(\mathcal{N}(\rho_0)) = \rho_0$. Because realising the inverse channel is in general an unphysical process, we use the scheme depicted in Fig.~\ref{quasiprob} to effectively realise the inverse channel by focusing only on measurement results.

As an example, we consider the case of a depolarising error channel,
\begin{equation}
	\rho = \mathcal{D}(\rho_0) = \left(1-\frac{3}{4}p\right)\rho_0+\frac{p}{4}(X\rho_0 X+Y\rho_0 Y+Z\rho_0 Z).
\end{equation}
The unphysical inverse channel is 
\begin{equation}
\begin{aligned}
		\rho_0 = \mathcal{D}^{-1}(\rho) 
		=\gamma[p_1\rho-p_2(X\rho X+Y\rho Y+Z\rho Z)],
		\end{aligned}
\end{equation}
where the coefficient $\gamma = (p+2)/(2-2p)\ge1$, $p_1 = (4-p)/(2p+4)$, and $p_2=p/(2p+4)$ in this case.

We cannot directly realise $\mathcal{D}^{-1}$ due to the minus sign before $p_2$. However, we can consider and correct its effect on the expectation value $\bar{O}_0$
\begin{equation}
\begin{aligned}
	\bar{O}_0 &= \tr[O\mathcal{D}^{-1}(\rho)],\\
	 &= \gamma[p_1\braket{O}_{\rho}-p_2(\braket{O}_{X\rho X}+\braket{O}_{Y\rho Y}+\braket{O}_{Z\rho Z})],\\
	 &=\gamma[p_1\braket{O}_{\rho}-p_2(\braket{XOX}_{\rho}+\braket{YOY}_{\rho }+\braket{ZOZ}_{\rho })],
\end{aligned}
\end{equation}
where $\braket{O}_{\rho} = \tr[O\rho]$. We can therefore measure $O$, $XOX$, $YOY$, $ZOZ$, and linearly combine the measurement results to effectively realise the inverse channel, thus obtaining the noiseless measurement result $\bar{O}_0$. The variance in our estimate of $\bar{O}_0$ is increased by a factor of $\gamma^G$, where $\gamma$ is the overhead coefficient, and $G$ is the number of gates in the circuit.

In practice, it is not possible to exactly measure all of the possible terms resulting from errors if there are many gates in the circuit. Instead, we can consider only the most important terms, which result from a small number of errors occurring. If the error rate is low, then the other terms can be considered negligibly small. After each single qubit gate, we can apply $X$, $Y$ or $Z$ operators with probability $p_2$, or the identity gate with $p_1$. We repeat that circuit variant many times to extract the expectation value, and multiply the expectation value by $(-1)^{G_p}$, where $G_p$ is the number of additional $X$, $Y$ or $Z$ gates that were applied in that circuit iteration. We then sum up the values for several circuit variants and multiply by $\gamma$ to obtain the error mitigated result. This method can also be extended to multi-qubit gates. For example, for two qubit gates in the depolarising noise model, after each two qubit gate we insert one of the pairs: $XI, IX, YI, IY, ZI, IZ$ (parity $-1$) with probability $p_2$, one of the pairs $XX, YY, ZZ, XY, YX, XZ, ZX, YZ, ZY$ (parity $+1$) with probability $p_2$ and $II$ (parity $+1$) with probability $p_1$. \\

The probabilistic error cancellation method described above has been shown to work for general Markovian noise~\cite{endo2017practical}, and has also been extended to work with temporally correlated errors and low frequency noise~\cite{huo2018temporally}. Probabilistic error cancellation requires full knowledge of the noise model associated with each gate. This can be obtained from either process tomography, or a combination of process and gate set tomography. The latter approach reduces the effect of errors due to state preparation and measurement~\cite{endo2017practical}. The probabilistic error cancellation method has been experimentally demonstrated on both superconducting~\cite{song2018quasiprobability} and trapped ion~\cite{zhang2019quasiprob} systems.

\subsection{Quantum subspace expansion}\label{Subsec:SubspaceMitigation}
The quantum subspace expansion (QSE)~\cite{PhysRevA.95.042308} described in Sec.~\ref{Subsubsec:SubspaceExpansionExcited} can mitigate errors in the VQE, in addition to calculating the excited energy eigenstates. This method is most effective at correcting systematic errors, but can also suppress some stochastic errors. Suppose that we use the VQE to find an approximate ground state $\ket{\tilde{E}_0}$. Noise may cause this state to deviate from the true ground state $\ket{E_0}$. For example, if $\ket{\tilde{E}_0} = X_1\ket{E_0}$, we can simply apply an $X_1$ gate to recover the correct ground state.

However, as we do not know which errors have occurred, we can instead consider an expansion in the subspace $\{\ket{P_i \tilde{E}_0}\}$, where $P_i$ are matrices belonging to the Pauli group.
Then, one can measure the matrix representation of the Hamiltonian in the subspace,
\begin{equation}
	H^\mathrm{QSE}_{ij} = \bra{\tilde{E}_0}P_i H P_j\ket{\tilde{E}_0}.
\end{equation}
As the subspace states are not orthogonal to each other, we should also measure the overlap matrix 
\begin{equation}
	S^\mathrm{QSE}_{ij} = \bra{\tilde{E}_0}P_i P_j\ket{\tilde{E}_0}.
\end{equation}
By solving the generalised eigenvalue problem 
\begin{equation}
	H^\mathrm{QSE}C = S^\mathrm{QSE}CE, 
\end{equation}
with a matrix of eigenvectors $C$ and diagonal matrix of eigenvalues $E$, we can get the error mitigated spectrum of the Hamiltonian. A small number of Pauli group operators are typically considered, in order to minimise the required number of measurements. The QSE has been experimentally demonstrated, using a two qubit superconducting system to measure the ground and excited state energies of H$_2$~\cite{PhysRevX.8.011021}.

\subsection{Symmetry based methods}\label{Subsec:StabiliserMitigation}
It is also possible to mitigate some errors by using symmetry checks on a suitably constructed ansatz state~\cite{mcardle2018mitigation,bonet2018mitigation}. A key concern for the VQE is preserving particle number, as states with electron number far from the true value appear to have a larger energy variance than those with smaller particle number errors~\cite{sawaya2016error}. Consequently, we can perform `checks' on quantities which should be conserved (such as number of electrons, or $s_z$ value), discarding the results if the measured value is not as expected. 

This can be achieved in a number of ways: by using stabiliser checks with additional ancilla qubits~\cite{mcardle2018mitigation,bonet2018mitigation}, by taking additional measurements and performing post-processing~\cite{bonet2018mitigation} (this has been experimentally demonstrated in superconducting qubits~\cite{sagastizabal2019symmetry}), by enforcing physically derived constraints on the form of the measured 1 and 2-RDMs (known as the $n$-representability constraints)~\cite{rubin2018nrep}, or by using the low-rank + orbital rotations measurement technique of \textcite{huggins2019efficientmeasurements} (discussed in Sec.~\ref{Subsubsec:Measurement}). The latter method appears to be the most powerful of these, requiring a modest additional circuit depth, limited connectivity, and enabling effective post-selection on the electron number and $s_z$ value, rather than just the parities of these quantities. These methods of error mitigation can be combined with some of the other techniques discussed above, such as extrapolation.

A related extension of the quantum subspace expansion was developed by \textcite{mcclean2019decoding}, who effectively engineered additional symmetries in the system using an error correcting code. By post-processing measurements from multiple iterations, they could detect some errors, or effectively realise a limited form of error correction. In order to maintain a polynomial cost for the procedure, the authors introduced a stochastic sampling scheme for the code stabilisers. 

\textcolor{black}{The choice of fermion-to-qubit mapping can also introduce additional symmetries. For example, both the generalised Bravyi-Kitaev superfast transform~\cite{setia2018superfast}, and the majorana loop stabiliser code~\cite{jiang2018majorana} (discussed in Sec.~\ref{Subsubsec:OtherEncodings}) introduce additional qubits, whose values are constrained by the mappings. By performing suitable stabiliser checks on these qubits, single qubit errors can be detected and corrected.}

\subsection{Other methods of error mitigation}\label{Subsec:MoreMitigation}
Other methods of error mitigation have been proposed, but require further research in order to assess how they can be best incorporated into chemistry calculations. One such method is the quantum variational error corrector~\cite{johnson2017qvector}, which uses a variational algorithm to construct noise-tailored quantum memories. Another example is individual error reduction~\cite{otten2018individual}. This method uses error correction to protect a single qubit, while leaving the rest of the physical qubits subject to noise. The process is repeated several times, with each physical qubit being protected in turn. Post-processing the results produces a more accurate expectation value than would be obtained without the mitigation technique.

\section{Illustrative examples}\label{Sec:illustrate}

In this section we illustrate many of the techniques described in the previous sections of this review, by explicitly demonstrating how to map electronic structure problems onto a quantum computer. We do this in second quantisation for the Hydrogen molecule (H$_2$) in the STO-3G, 6-31G and cc-PVDZ bases, and Lithium Hydride (LiH) in the STO-3G basis (as described in Sec.~\ref{Subsec:basis}). Across these examples, we showcase the Jordan--Wigner (JW), Bravyi--Kitaev (BK) and BK tree mappings (as described in Sec.~\ref{Subsec:2ndencoding}), reduction of active orbitals using the Natural Molecular Orbital (NMO) basis (as described in Sec.~\ref{Subsec:OrbitalReduction}), reduction of qubits using symmetry conservation (as described in Sec.~\ref{Subsubsec:HamiltonianReduction}) and the unitary coupled cluster ansatz (as described in Sec.~\ref{Subsubsec:VQEAnsatz}). These examples are designed to familiarise the reader with the key steps of formulating a quantum computational chemistry problem. Many of these techniques are applicable to both ground state and general chemical problems.

\subsection{Hydrogen}\label{Subsec:H2Illustrate}

\begin{figure}[t]
{\includegraphics[width=9cm]{{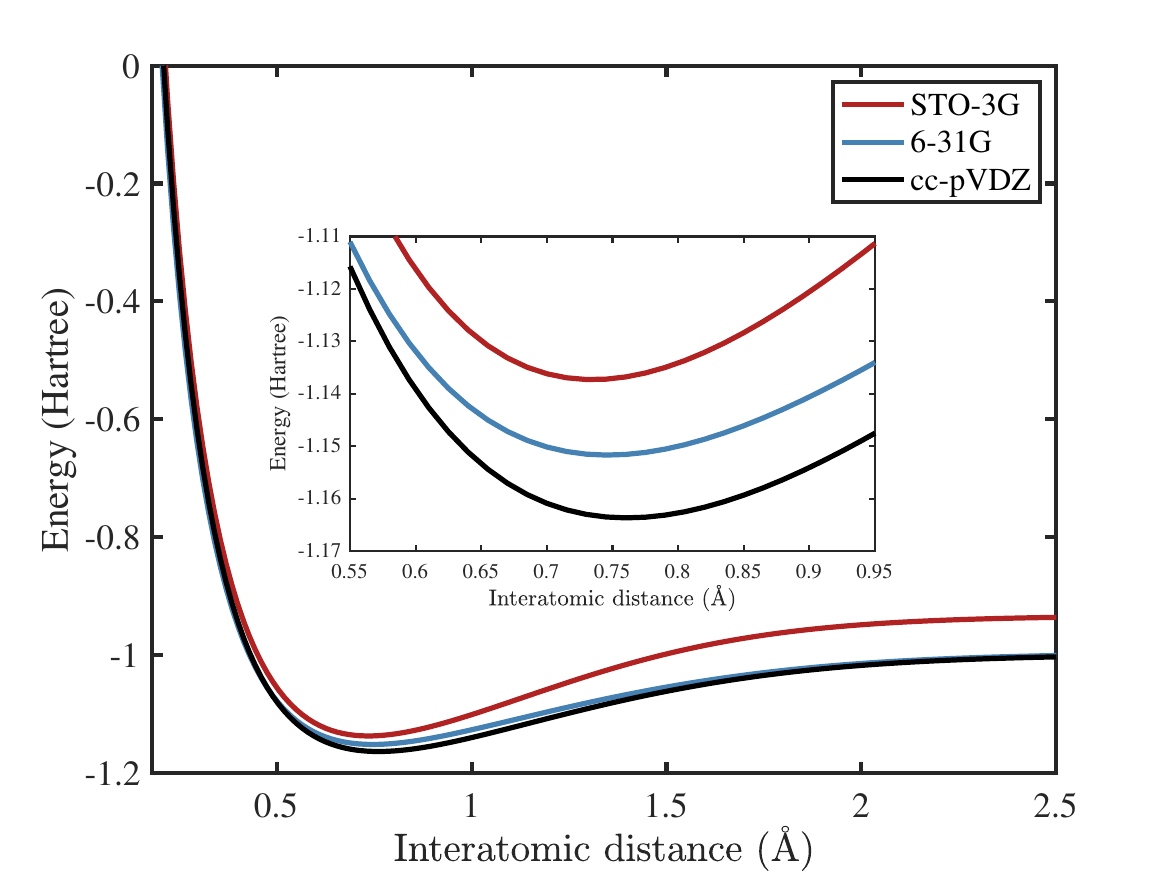}}}
\caption{Comparing the ground state dissociation curves of H$_2$ for a range of basis sets.}\label{h2energies}
\end{figure}

The continuous space molecular Hamiltonian for H$_2$ is given by Eq.~\eqref{ElectronicStructureH} (which we reproduce here), with two electrons and two nuclei 
\begin{equation}\label{H2ContinuousHamil}
	H_{\mathrm{H_2}} = -\sum_i\frac{\nabla^2_i}{2} -\sum_{i,I}\frac{Z_I}{|\mathbf{r}_i-\mathbf{R}_I|}+\frac{1}{2}\sum_{i\neq j}\frac{1}{|\mathbf{r}_i-\mathbf{r}_j|}.
\end{equation}
To convert this Hamiltonian into the second quantised representation,
\begin{equation}\label{2ndQHamilIllustrate}
	H = \sum_{p,q}h_{pq}a^\dag_p a_q + \frac{1}{2}\sum_{p,q,r,s}h_{pqrs}a^\dag_p a^\dag_q a_ra_s,
\end{equation}
with 
\begin{equation}
	\begin{aligned}\label{IntegralsIllustrate}
		h_{pq}&=\int \mathrm{d}\textbf{x} \phi_p^*(\textbf{x}) \left(-\frac{\nabla^2}{2} -\sum_{I}\frac{Z_I}{|\mathbf{r}-\mathbf{R}_I|}\right) \phi_q(\mathbf{x}),\\
		 h_{pqrs}&=\int \mathrm{d}\mathbf{x}_1 \mathrm{d}\mathbf{x}_2\frac{\phi_p^*(\mathbf{x}_1) \phi_q^*(\mathbf{x}_2)\phi_r(\mathbf{x}_2) \phi_s(\mathbf{x}_1)}{|\mathbf{x}_1-\mathbf{x}_2|},
	\end{aligned}
\end{equation}
we need to select a basis set, $\phi_p(\mathbf{x})$. As discussed in Sec.~\ref{Subsec:basis}, this is a discrete set of functions which are used to approximate the spin-orbitals of the molecule. By considering a larger number of orbitals (and the Slater determinants that they can generate), we are able to recover a larger proportion of the correlation energy in a molecule, resulting in a more accurate estimate of the true ground state energy. Fig.~\ref{h2energies} shows the H$_2$ ground state dissociation curves in the STO-3G, 6-31G and cc-PVDZ bases. We can see that the differences in energy between the three minima are considerably larger than chemical accuracy ($1.6$~mHartree). This highlights that working in a suitably large basis set is crucial for obtaining accurate results.

\subsubsection{STO-3G basis}
The STO-3G basis for H$_2$ includes only the $\{1s\}$ orbital for each hydrogen atom. The $1s$ orbital is represented by a linear combination of three Gaussian type orbitals. Each hydrogen atom contributes one orbital, and there are two possible spins for each orbital -- resulting in a total of 4 spin-orbitals for STO-3G H$_2$. We denote these orbitals as 
\begin{equation}
	| 1s_{A \uparrow} \rangle, | 1s_{A \downarrow} \rangle, | 1s_{B \uparrow} \rangle, | 1s_{B \downarrow} \rangle,
\end{equation}
where the subscript $A$ or $B$ denotes which of the two atoms the spin-orbital is centred on, and the $\uparrow / \downarrow$ denotes the $s_z$ value of the electron in the spin-orbital. For convenience, we work in the molecular orbital basis for H$_2$, which is simple to construct manually. These single electron molecular spin-orbitals are given by 
\begin{equation}
	\begin{aligned}
		|\sigma_{g \uparrow} \rangle = \frac{1}{\sqrt{N_g}}(| 1s_{A \uparrow} \rangle + | 1s_{B \uparrow} \rangle), \\
		|\sigma_{g \downarrow} \rangle = \frac{1}{\sqrt{N_g}}(| 1s_{A \downarrow} \rangle + | 1s_{B \downarrow} \rangle), \\
		|\sigma_{u \uparrow} \rangle = \frac{1}{\sqrt{N_u}}(| 1s_{A \uparrow} \rangle - | 1s_{B \uparrow} \rangle), \\
		|\sigma_{u \downarrow} \rangle = \frac{1}{\sqrt{N_u}}(| 1s_{A \downarrow} \rangle - | 1s_{B \downarrow} \rangle),
	\end{aligned}
\end{equation}
where $N_{g/u}$ are normalisation factors that depend on the overlap between the atomic orbitals, $N_g = 2(1+\langle 1s_A | 1s_B \rangle) $, $N_u = 2(1-\langle 1s_A | 1s_B \rangle) $. We can write a Slater determinant in the occupation number basis as
\begin{equation}
	\begin{aligned}
		\ket{\psi} = 
		\ket{f_{\sigma_{u \downarrow}}, f_{\sigma_{u \uparrow}}, f_{\sigma_{g \downarrow}}, f_{\sigma_{g \uparrow}}},
	\end{aligned}
\end{equation}
where $f_i=1$ if spin-orbital $i$ is occupied, and $f_i=0$ if spin-orbital $i$ is unoccupied. We can now calculate the integrals given in Eq.~(\ref{IntegralsIllustrate}) using these molecular orbitals. These integrals have been calculated for a large number of basis sets, and the results can be obtained by using a computational chemistry package~\cite{frish2016Gaussian, parrish2017psi4, pyquante, qiming2017pyscf}. We must then map the problem Hamiltonian from being written in terms of creation and annihilation operators, to being written in terms of qubit operators. Using the JW encoding, we can obtain the 4 qubit Hamiltonian for H$_2$
\begin{equation}
	\begin{aligned}
	H &=h_0 I + h_1 Z_0 + h_2 Z_1 + h_3 Z_2 + h_4 Z_3  \\
		&+h_5 Z_0 Z_1 + h_6 Z_0 Z_2 +
		h_7 Z_1 Z_2 + h_8 Z_0 Z_3+h_9 Z_1 Z_3  \\
		& + h_{10} Z_2 Z_3 +
		h_{11} Y_0 Y_1 X_2 X_3 + h_{12} X_0 Y_1 Y_2 X_3 \\
		&+h_{13} Y_0 X_1 X_2 Y_3 + h_{14} X_0 X_1 Y_2 Y_3.
	\end{aligned}
\end{equation}
While it is important to understand this procedure, every step from selecting a basis to producing an encoded qubit Hamiltonian can be carried out using a quantum computational chemistry package such as OpenFermion~\cite{mcclean2017openfermion}, Qiskit Aqua~\cite{ibm2018qiskit}, or QDK-NWChem~\cite{low2019nwchem}. \\

In the JW encoding, it is simple to construct the Hartree-Fock (HF) state for the H$_2$ molecule. The HF state for H$_2$ in the occupation number basis is given by
\begin{equation}\label{H2HF}
	\ket{\Psi_{\mathrm{HF}}^{\mathrm{H_2}}} = \ket{0011}. 
\end{equation}
This represents the Slater determinant
\begin{equation}
	\Psi_{\mathrm{HF}}^{\mathrm{H_2}}(\mathbf{r_1}, \mathbf{r_2}) = \frac{1}{\sqrt{2}} (\sigma_{g \uparrow}(\mathbf{r_1}) \sigma_{g \downarrow}(\mathbf{r_2}) - \sigma_{g \uparrow}(\mathbf{r_2}) \sigma_{g \downarrow}(\mathbf{r_1}) ), 
\end{equation}
where $\mathbf{r_i}$ is the position of electron $i$. The most general state for H$_2$ (with the same $s_z$ value and electron number as the HF state) is given by
\begin{equation}
	\ket{\Psi^{\mathrm{H_2}}} = \alpha \ket{0011} + \beta \ket{1100} + \gamma \ket{1001} + \delta \ket{0110},
\end{equation}
and the ground state of the H$_2$ molecule at its equilibrium bond distance is given by~\cite{helgaker2014molecular}
\begin{equation}\label{H2ground}
	\ket{\Psi_{g}^{\mathrm{H_2}}} = 0.9939 \ket{0011} - 0.1106 \ket{1100}.
\end{equation}
The first determinant in the ground state wavefunction is the HF state for H$_2$, showing that a mean-field solution is a good approximation for this molecule at this interatomic distance. The second determinant represents the antibonding state, and accounts for dynamical correlation between the electrons due to their electrostatic repulsion. While the HF determinant dominates at the equilibrium separation, at large separation the two determinants contribute equally to the wavefunction. This is because the bonding and antibonding configurations become degenerate. We require both determinants to accurately describe the state, ensuring that only one electron locates around each atom. This is an example of static correlation, which can also be dealt with using multiconfigurational self-consistent field methods, as described in Sec.~\ref{Subsubsec:HFsec}. \\

\begin{figure}[t]
{\includegraphics{{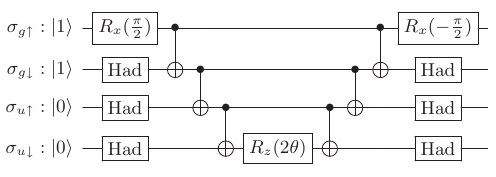}}}
\caption{The circuit for implementing the UCCSD operator for H$_2$ in the STO-3G basis, as given by Eq.~\eqref{h2uccsd}. The $R_x(\frac{\pi}{2})$ and $\mathrm{Had}$ gates rotate the basis such that the exponentiated operator applied to the corresponding qubit is either $Y$ or $X$, respectively. Single excitation terms are implemented with a change of basis~\cite{TrappedIon}.} \label{UCCH2}
\end{figure}

As discussed previously, in order to find the ground state of the H$_2$ molecule (using either the VQE or phase estimation), we need to construct the state on the quantum computer. Here we explicitly derive the unitary coupled cluster (UCC) ansatz with single and double excitations (UCCSD) for H$_2$. As discussed in Sec.~\ref{Subsubsec:VQEAnsatz}, the UCCSD operator we seek to realise is given by
\begin{equation}
	\begin{aligned}
		U &= e^{(T_1 - T_1^\dag) + (T_2 - T_2^\dag)}, \\ 
		T_1 &=  \sum_{i \in virt, \alpha \in occ} t_{i \alpha} a^\dag_i a_\alpha, \\ 
		T_2 &= \sum_{i, j \in virt, \alpha, \beta \in occ} t_{i j \alpha \beta}  a^\dag_i a^\dag_j a_\alpha a_\beta, \\ 
	\end{aligned}
\end{equation}
where $occ$ are initially occupied orbitals in the HF state, $virt$ are initially unoccupied orbitals in the HF state, and $t_{i \alpha}$ and $t_{ij\alpha \beta}$ are variational parameters to be optimised. For H$_2$, the only operators which do not change the $s_z$ value of the molecule when acting upon the HF state are: $a_2^\dag a_0,~a_3^\dag a_1,~a_3^\dag a_2^\dag a_1 a_0$. Other valid operators are equivalent to these operators, and can be combined with them, such as $ a_3^\dag a_0^\dag a_1 a_0 = -a_3^\dag a_1 $. As a result, the UCCSD operator takes the form
\begin{equation}
	\begin{aligned}
		U = e^{t_{02}(a_2^\dag a_0 - a_0^\dag a_2) + t_{13}(a_3^\dag a_1 - a_1^\dag a_3) + t_{0123}(a_3^\dag a_2^\dag a_1 a_0 - a_0^\dag a_1^\dag a_2 a_3)}. 
	\end{aligned}
\end{equation}
We can split this operator using Trotterization with a single Trotter step
\begin{equation}\label{TrotterUCC}
	\begin{aligned}
		U =~&e^{t_{02}(a_2^\dag a_0 - a_0^\dag a_2)} \times e^{t_{13}(a_3^\dag a_1 - a_1^\dag a_3)} \\
			&\times e^{t_{0123}(a_3^\dag a_2^\dag a_1 a_0 - a_0^\dag a_1^\dag a_2 a_3)}.
	\end{aligned}
\end{equation}
Using the JW encoding, we find that
\begin{equation}
	\begin{aligned}
		(a_2^\dag a_0& - a_0^\dag a_2) = \frac{i}{2}(X_2 Z_1 Y_0 - Y_2 Z_1 X_0) \\
		(a_3^\dag a_1& - a_1^\dag a_3) = \frac{i}{2}(X_3 Z_2 Y_1 - Y_3 Z_2 X_1) \\
		(a_3^\dag a_2^\dag & a_1 a_0 - a_0^\dag a_1^\dag a_2 a_3) = \\
		&\frac{i}{8}(X_3 Y_2 X_1 X_0 + Y_3 X_2 X_1 X_0 + Y_3 Y_2 Y_1 X_0 + Y_3 Y_2 X_1 Y_0 \\
		&- X_3 X_2 Y_1 X_0 - X_3 X_2 X_1 Y_0 - Y_3 X_2 Y_1 Y_0 - X_3 Y_2 Y_1 Y_0).
	\end{aligned}
\end{equation}
It was shown by \textcite{romero2017strategies} that all Pauli terms arising from the same excitation operators commute. As a result, each of the exponentials in Eq.~(\ref{TrotterUCC}) can be separated into a product of exponentials of a single Pauli string. For example
\begin{equation}
	\begin{aligned}
		e^{t_{02}(a_2^\dag a_0 - a_0^\dag a_2)} = e^{\frac{i t_{02}}{2}X_2 Z_1 Y_0} \times e^{\frac{-i t_{02}}{2}Y_2 Z_1 X_0}.
	\end{aligned}
\end{equation}
\textcite{TrappedIon} simplified the UCCSD operator for H$_2$ by implementing the single excitation terms as basis rotations, and combining terms in the double excitation operator (by considering the effect of each term on the HF state). This latter technique is only possible because there is only one double excitation operator for this molecule, and so is not a scalable technique in general. The UCCSD operator is simplified to 
\begin{equation}\label{h2uccsd}
	\begin{aligned}
		U = e^{-i \theta X_3 X_2 X_1 Y_0} .
	\end{aligned}
\end{equation}
This can be implemented using the circuit ~\cite{whitfield2011simulation} shown in Fig.~\ref{UCCH2}.

Applying the simplified UCCSD operator to the HF state in Eq.~(\ref{H2HF}) gives 
\begin{equation}
    \begin{aligned}
    U\ket{0011} &=
    \left ( \mathrm{cos}(\theta)I - i\mathrm{sin}(\theta)X_3X_2X_1Y_0 \right )\ket{0011},  \\&= \mathrm{cos}(\theta) \ket{0011} - \mathrm{sin}(\theta) \ket{1100},
    \end{aligned}
\end{equation}
which can reproduce the ground state given by Eq.~(\ref{H2ground}).

\subsubsection{6-31G basis}
As discussed in Sec.~\ref{Subsec:basis}, H$_2$ in the 6-31G basis has a double-zeta representation of the valence electrons. This means we have 8 spin-orbitals to consider in total; \{$1s_\uparrow, 1s_\downarrow, 1s'_\uparrow, 1s'_\downarrow$\} from each atom. Working in the canonical orbital basis, (obtained by performing a Hartree--Fock calculation) we show how to construct Bravyi-Kitaev encoded states of 6-31G H$_2$. The BK transform matrix for an 8 spin-orbital system is given by
\begin{equation}
\begin{pmatrix}
    1 & 0 & 0 & 0 & 0 & 0 & 0 & 0\\
    1 & 1 & 0 & 0 & 0 & 0 & 0 & 0\\
    0 & 0 & 1 & 0 & 0 & 0 & 0 & 0\\
    1 & 1 & 1 & 1 & 0 & 0 & 0 & 0\\
    0 & 0 & 0 & 0 & 1 & 0 & 0 & 0\\
 	0 & 0 & 0 & 0 & 1 & 1 & 0 & 0\\
 	0 & 0 & 0 & 0 & 0 & 0 & 1 & 0\\
 	1 & 1 & 1 & 1 & 1 & 1 & 1 & 1\\
 \end{pmatrix}.
\end{equation}
We order the spin-orbitals such that the first $M/2$ spin-orbitals are spin up, and the final $M/2$ spin-orbitals are spin down. When the spin-orbitals are ordered in this way, the 4\textsuperscript{th} entry in the BK encoded vector is the sum (mod 2) of the spin up occupancies, which sums to the number of spin up electrons. Moreover, the 8\textsuperscript{th} entry is the sum (mod 2) of all of the orbital occupancies, which sums to the number of electrons. As these quantities are conserved, we can remove these two qubits from the simulation, following the procedure of Sec.~\ref{Subsec:reduction}. If the spin-orbitals are arranged `up-down, up-down', then while the 8\textsuperscript{th} entry is still equal to the number of electrons, the 4\textsuperscript{th} entry is no longer necessarily equal to a conserved quantity. The JW mapped HF state (8 qubits) is given by $\ket{00010001}$. Using the matrix given above, the BK mapped HF state (8 qubits) is $\ket{00111011}$. When the two conserved qubits are removed, the BK mapped HF state (6 qubits) is $\ket{011011}$.

\subsubsection{cc-PVDZ basis}

As discussed in Sec.~\ref{Subsec:basis}, the cc-PVDZ basis for H$_2$ includes a double-zeta representation of the valence shell, and additional polarisation orbitals. Each atom contributes \{$1s, 1s', 2p_x, 2p_y, 2p_z$\} orbitals, resulting in 20 spin-orbitals in total. In order to reduce our active space, we first change to the natural molecular orbital (NMO) basis, using the single particle reduced density matrix (1-RDM), as discussed in Sec.~\ref{Subsec:OrbitalReduction}. We first obtain the 1-RDM for H$_2$ in the cc-PVDZ basis with a classically tractable configuration interaction singles and double (CISD) calculation. 

We perform a unitary diagonalisation of this matrix, and rotate the orbitals by the same unitary matrix. This constitutes a change of basis to the NMOs of the molecule. 
The diagonalised 1-RDM is given by \textit{Diag}(1.96588, 0.00611, 0.02104, 0.0002, 0.00001, 0.00314, 0.00314, 0.00016, 0.00016, 0.00016). There are only 10 diagonal entries in this 1-RDM because the spin-up and spin-down entries for the same spatial orbitals have been combined. As discussed in Sec.~\ref{Subsec:OrbitalReduction}, the diagonal entries are the natural orbital occupation numbers (NOONs). We can see that the 5\textsuperscript{th} orbital has a NOON that is 20 times smaller than the next smallest NOON. As a result, we consider this spatial orbital to always be empty, and so remove all terms involving it from the Hamiltonian. This leaves a Hamiltonian acting on $M=18$ spin-orbitals. We now map these into qubits using the BK-tree method. 

To map fermionic orbitals to qubits we follow a similar procedure to that shown in Fig.~\ref{Fig:BKLiH} for the LiH molecule. The reader will find that the 9\textsuperscript{th} and 18\textsuperscript{th} orbitals store the number of spin up electrons and total number of electrons, respectively. As a result, they can be removed. This reduces the problem to one of 16 qubits. 

The lowest energy computational basis state of cc-PVDZ H$_2$ in the Jordan-Wigner encoding (18 qubits) is $\ket{000000001000000001}$. The corresponding BK-tree mapped state (16 qubits) is given by $\ket{0001011100010111}$. \textcolor{black}{We stress that while this procedure may seem complex, it can in fact be easily implemented using the functions available in the aforementioned quantum computational chemistry packages.}

\subsection{Lithium Hydride STO-3G basis}\label{Subsec:LiHIllustrate}

\begin{figure*}[t]
{\includegraphics[width=11cm]{{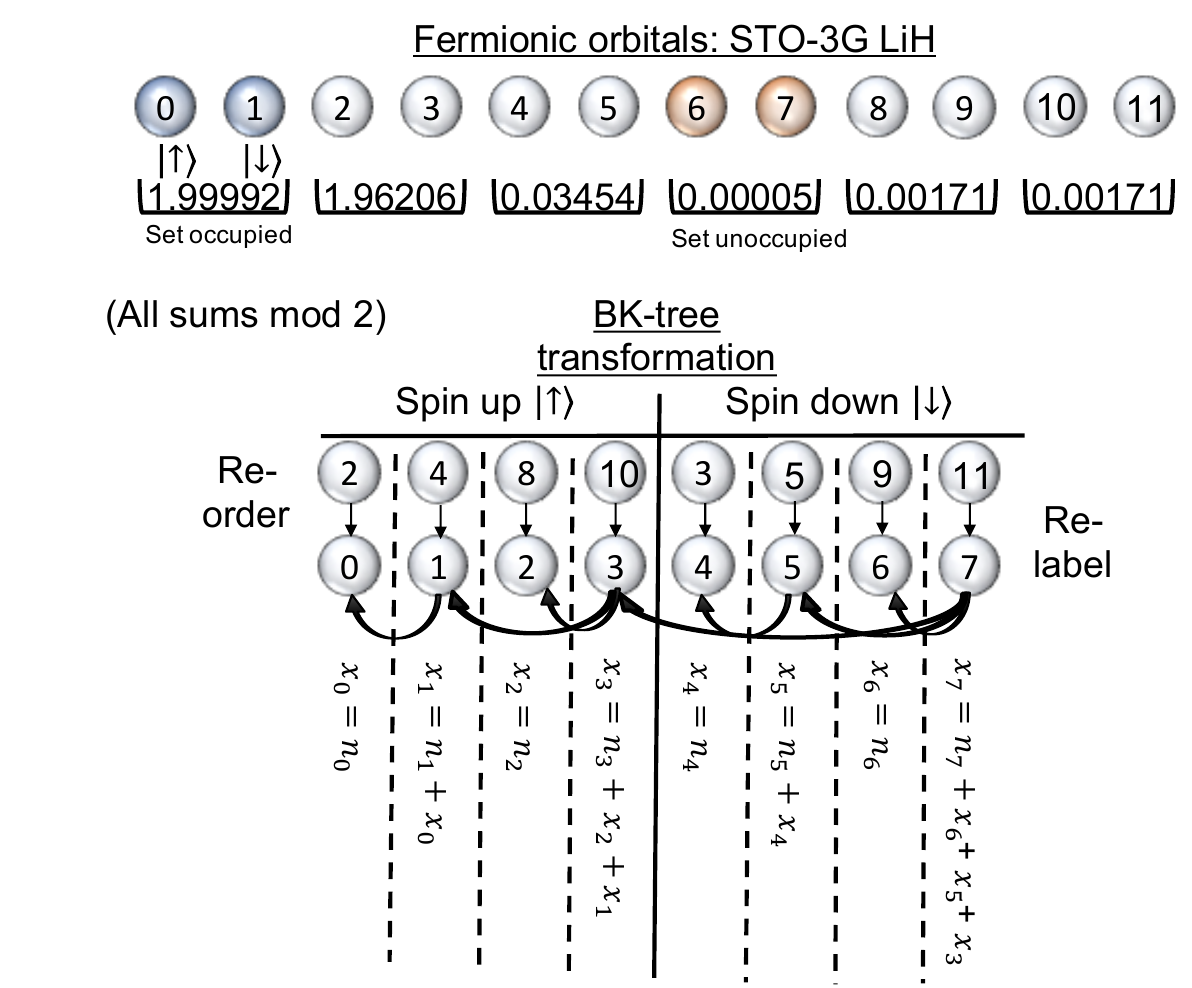}}}

\caption{A pictorial representation of the fermion-to-qubit mapping procedure for LiH in the STO-3G basis. The fermionic natural molecular orbitals (NMO) are initially arranged `spin up, spin down, spin up, spin down, ...', and have their corresponding natural orbital occupation number (NOON) below. As the NOON of orbitals 6 and 7 is so small, they can be assumed unfilled, and removed from the Hamiltonian. As the combined NOON of orbitals 0 and 1 is close to 2, they can be assumed filled, and removed from the Hamiltonian. We then rearrange the remaining orbitals to be `all spin up, all spin down', and re-label them from 0 to 7. We then perform the BK-tree mapping by constructing the Fenwick tree, Fen(0,7), as described in Fig.~\ref{Fig:FenTreeFig}. The value $x_i$ is the value of the $i$\textsuperscript{th} qubit under the BK-tree mapping, while $n_i$ is the value of the $i$\textsuperscript{th} qubit under the JW mapping. We see that qubit 3 stores the sum $\sum_{i=0}^3 n_i$, and qubit 7 stores the sum $\sum_{i=0}^{7} n_i$. As these sums are conserved quantities, these qubits do not flip throughout the simulation, and so can be removed from the Hamiltonian as described in Sec.~\ref{Subsec:reduction}.}\label{Fig:BKLiH}
\end{figure*}

\begin{figure*}[hbtp]
{\includegraphics[width=17cm]{{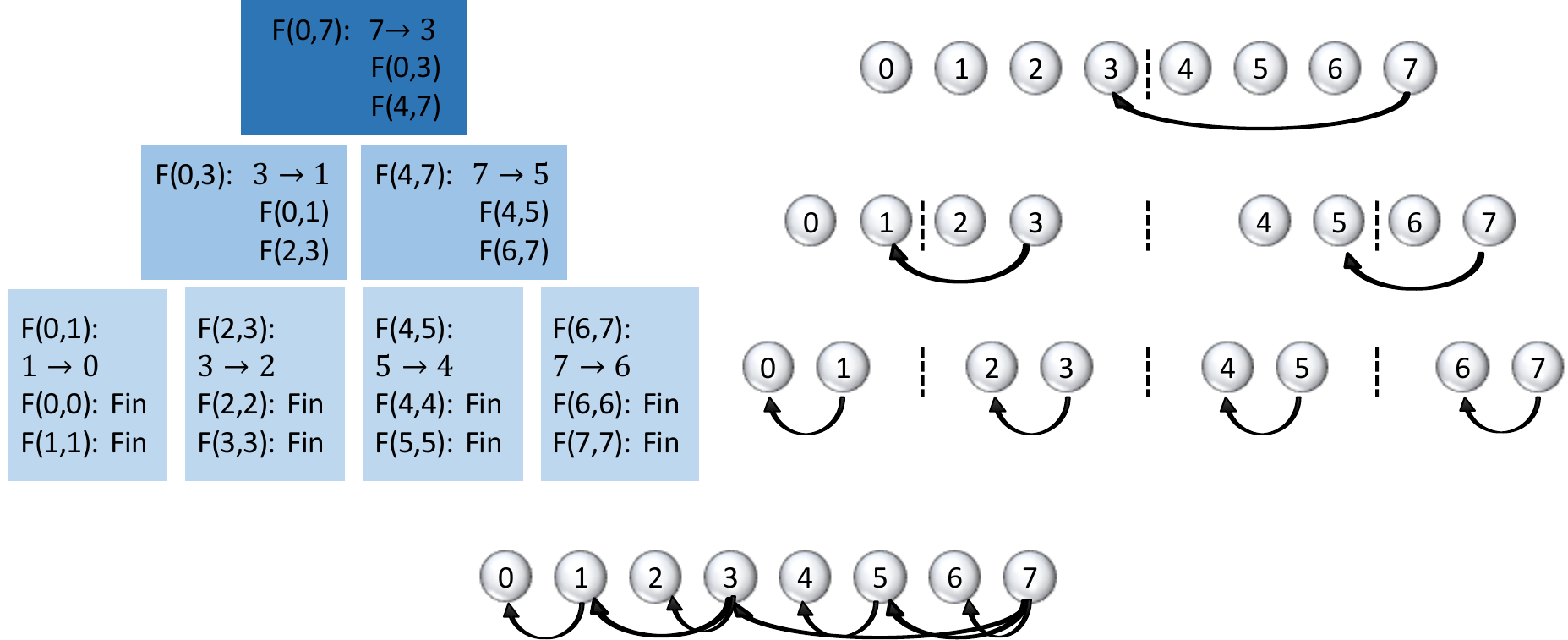}}}

\caption{A pictorial representation of the Fenwick tree construction for LiH, shown in Fig.~\ref{Fig:BKLiH}. We carry out the BK-tree mapping by constructing the Fenwick tree, Fen(0,7), as described in Algorithm.~\ref{tcolorbox:Gradient}. The algorithmic steps are shown on the left hand side of the figure, while the corresponding actions are shown on the right hand side. The notation $X \rightarrow Y$ means connect spin-orbital $X$ to spin-orbital $Y$ with an arrow. `Fin' means that the corresponding branch of the Fenwick tree is finished. The finished Fenwick tree Fen(0,7) is shown at the bottom of the figure.}\label{Fig:FenTreeFig}
\end{figure*}

For LiH in the STO-3G basis, we consider \{$1s, 2s, 2p_x, 2p_y, 2p_z$\} functions for lithium, and a single \{$1s$\} orbital for hydrogen. This gives a total of 12 spin-orbitals. We can reduce this problem to one of six qubits, as illustrated in Fig.~\ref{Fig:BKLiH}. The 1-RDM in the canonical orbital basis from a CISD calculation on LiH (at an internuclear separation of 1.45~$\dot{\mathrm{A}}$) is given by\\

\scalebox{0.8}{$
\begin{pmatrix}
\mathbf{1.9999} &	\mathbf{-0.0005} &	\mathbf{0.0006} &	0.0000&	0.0000&	\mathbf{-0.0010}	\\
\mathbf{-0.0005}&	\mathbf{1.9598}&	\mathbf{0.0668}&	0.0000&	0.0000&	\mathbf{0.0084}	\\
\mathbf{0.0006}&	\mathbf{0.0668}&	\mathbf{0.0097}&	0.0000&	0.0000&	\mathbf{-0.0138}	\\
0.0000&	0.0000&	0.0000&	\mathbf{0.0017}&	0.0000&	0.0000	\\
0.0000&	0.0000&	0.0000&	0.0000&	\mathbf{0.0017}	&	0.0000	\\
\mathbf{-0.0010}&	\mathbf{0.0084}&	\mathbf{-0.0138}&	0.0000&	0.0000&	\mathbf{0.0273}
 \end{pmatrix}.$}

~\\ 
There are only 6 rows/columns in this 1-RDM because the spin-up and spin-down entries for the same spatial orbitals have been combined. We diagonalise this 1-RDM, moving to the NMO basis. The NMO 1-RDM is given by\\

\scalebox{0.8}{$
\begin{pmatrix}
\mathbf{1.99992} &	0.00000&	0.00000&	0.00000&	0.00000&	0.00000	\\
0.00000&	\mathbf{1.96206} &	0.00000&	0.00000&	0.00000&	0.00000	\\
0.00000&	0.00000&	\mathbf{0.03454} &	0.00000&	0.00000&	0.00000	\\
0.00000&	0.00000&	0.00000&	\mathbf{0.00005} &	0.00000&	0.00000	\\
0.00000&	0.00000&	0.00000&	0.00000&	\mathbf{0.00171}	&	0.00000	\\
0.00000&	0.00000&	0.00000&	0.00000&	0.00000&	\mathbf{0.00171}	
 \end{pmatrix}.$
}

~\\
The first orbital has a NOON close to two, and so we consider it to always be doubly occupied. We can then remove any terms containing $a_0^\dag, a_0, a_1^\dag, a_1$ (the spin-orbitals corresponding to the first spatial orbital in the 1-RDM) from the Hamiltonian. In contrast, the fourth orbital has a very small NOON. As a result, we assume that this orbital is never occupied, and so remove the two corresponding fermion operators from the Hamiltonian. This leaves a Hamiltonian acting on 8 spin-orbitals. As the number of orbitals is now a power of 2, we can use either the BK or BK-tree mappings to remove the 2 qubits associated with conservation symmetries. We use the BK-tree mapping in order to provide an explicit example of Fenwick tree construction. The Fenwick tree tells us which qubits store which orbitals in the BK-tree mapping. We denote the Fenwick tree for the $M$ orbitals as $\textrm{Fen}(0,M-1)$. We can obtain this data structure using an iterative algorithm, which we reproduce from \textcite{havlicek2017operator} below. The generation of the Fenwick tree for the LiH molecule using this algorithm is shown in Fig.~\ref{Fig:FenTreeFig}.

\begin{tbox}[label=tcolorbox:Gradient]{Algorithm.1 : Fenwick tree generation}
{\fontfamily{qcr}\selectfont
	\small{
	Define $\textrm{Fen}(L, R)$ \\ \\
	If $L \neq R$:
	\begin{itemize}
			\item Connect $R$ to $\textrm{Floor}\left(\frac{R + L}{2}\right)$;
			\item $\textrm{Fen}\left(L,~\textrm{Floor}\left(\frac{R + L}{2}\right)\right)$;
		       \item $\textrm{Fen}\left(\textrm{Floor}\left(\frac{R + L}{2}\right) + 1, R\right)$.
 	\end{itemize}

	Else:
	\begin{itemize}
			\item End the current Fenwick tree.
 	\end{itemize}
	}}
\end{tbox}
 Our final Hamiltonian acts on 6 qubits, but differs in energy from the full 12 qubit Hamiltonian by only $0.2$~mHartree. A similar procedure is described by \textcite{TrappedIon}.

\section{Discussion and Conclusions}\label{Sec:conclude}
We have now reviewed the key concepts in both classical, and quantum, computational chemistry. In particular, we have discussed and shown how to map chemical problems onto quantum computers, and how to solve them to obtain both the ground and excited states. We now turn our attention to how these techniques compare to the established classical methods discussed in Sec.~\ref{Sec:classicalChemistry}. We will first review the applicability and limitations of the various classical methods, in Sec.~\ref{Subsec:ClassicalLimits}. This will highlight the problems for which quantum computers may one day be useful. We discuss the resources required for such calculations in Sec.~\ref{Subsec:LongTermQuantumResources}. We will see that the resources required are considerably greater than what we currently have available, at time of writing. Consequently, in Sec.~\ref{Subsec:NearFutureOutlook} we consider routes towards these calculations. This will include both heuristic calculations on classically intractable system sizes, and exact calculations on smaller system sizes. We conclude this review in Sec.~\ref{Sec:conclude} with a blueprint for future investigations.

\subsection{Classical limits}\label{Subsec:ClassicalLimits}
As discussed in Sec.~\ref{Sec:classicalChemistry}, there are many different methods used in classical computational chemistry, which all seek to approximate the true ground state energy of the system of interest, to varying degrees of accuracy. In general, the cost of applying these methods grows with the accuracy of the results that they produce, and the size of the system simulated. In this section, we discuss the system sizes that one can typically accurately treat using some of the commonly used classical methods. This will help us to elucidate where, and when, quantum computers may become useful for chemistry simulation. As in previous sections, $M$ denotes the number of spin-orbitals considered in basis set approaches, and $N$ denotes the number of electrons in the system.

At one end of the spectrum are density functional theory (DFT), and the Hartree-Fock (HF) method. These calculations are often very efficient to run, and can treat large systems. As a result, they are used widely in chemistry and materials science. However, these techniques can struggle to achieve highly accurate results for strongly correlated systems, and are not systematically improvable. Consequently, they are often used for obtaining qualitative results for large system sizes. We do not expect these calculations to be replaced by those on quantum computers, given the large system sizes that are simulated. 

At the other end of the spectrum, are exact calculations -- by which we mean the exact energy that can be obtained from the model of the system. It is important to note that these `exact' calculations are rarely performed. Moreover, the degree of accuracy depends on the details included in the calculation, such as: the inclusion of relativistic corrections, whether the Born-Oppenheimer approximation is used, or whether nuclear vibrational and rotational contributions are included. Grid-based simulations (as described in Sec.~\ref{Subsubsec:firstqclassical}) provide the most accurate results -- but can only be carried out for a very small number of particles, because of the large number of grid points required. Exact results (albeit with a less accurate model of the system) can also be obtained by carrying out basis set, full configuration interaction (FCI) calculations (Sec.~\ref{Subsubsec:2ndqclassical}), which are less computationally expensive than grid based approaches. However, the cost of these calculations still scales exponentially with the system size, so they are only applicable to small systems, like the dinitrogen molecule~\cite{rossi1999fci}. In the context of condensed matter physics, these calculations are referred to as `exact diagonalisation' and are possible up to system sizes of around 20 to 30 lattice sites, in the case of Fermi-Hubbard models~\cite{jiang2018hubbard,yamada2005hubbard}.\\

The vast majority of calculations carried out by the computational chemistry community do not achieve this level of accuracy. Instead, approximate, less costly methods are used, such as: configuration interaction (Sec.~\ref{Subsubsec:CI}), coupled cluster (Sec.~\ref{Subsubsec:CC}) multiconfigurational self-consistent field (Sec.~\ref{Subsubsec:MCSCF}), tensor network methods, or quantum Monte Carlo. 
An exhaustive comparison of these methods is beyond the scope of this review, as is attempting to catalogue the ever-evolving list of the largest calculations performed. However, we will briefly highlight some of the system sizes where these methods have been successfully applied. 

Coupled cluster methods (often CCSD) are some of the most widely used high-accuracy methods. They are applicable to large systems (hundreds of spin-orbitals~\cite{bartlett2007coupledclusterrmp}) which do not display strong static correlation. Examples include: the DNA base guanine (C$_5$H$_5$N$_5$O) in a cc-PVTZ basis~\cite{hobza2002dna}, or the hydrocarbon octane (C$_8$H$_{18}$) also in a cc-PVTZ basis~\cite{takeshi2018decomposition}. While CC methods can also be applied to strongly correlated systems~\cite{watson2012butadiene,simonscollab2015hubbard}, higher excitation degrees are often required, making the method more costly to implement.

Quantum Monte Carlo has many variants, and has been used to obtain results comparable to FCI in relatively small systems (the Cr$_2$ molecule with 24 active electrons in 30 spin-orbitals~\cite{tubman2016qfci}, or the fluorine atom in a cc-PV5Z basis with additional basis functions~\cite{booth2010qfci}), as well as state of the art results in Fermi-Hubbard systems that are accurate to around 100 sites~\cite{simonscollab2015hubbard}, and less accurate results in other, larger systems~\cite{austin2012quantummontecarlo}. However, Monte Carlo methods are not without their own shortcomings, including the infamous `sign problem', that affects fermionic simulations~\cite{ortiz2001fermionic, austin2012quantummontecarlo}.

Tensor network methods, such as density matrix renormalisation group (DMRG), have proven effective for dealing with systems displaying strong static correlation. They provide an alternative approach to CASSCF (see Sec.~\ref{Subsubsec:MCSCF}) approaches~\cite{knecht2016dmrgchemistry}, allowing the treatment of larger active spaces, including those of metalloenzyme complexes with active spaces of over 70 spin-orbitals~\cite{sharma2014low,kurashige2013entangled}. This is larger than the roughly 30 to 40 spin-orbitals that can be treated with a CASSCF approach~\cite{lischka2018excitedstatesmultiref}. Tensor network methods are also useful for treating systems in condensed matter physics, including Fermi-Hubbard models that are accurate to around 100 sites~\cite{simonscollab2015hubbard}. While tensor network methods are best suited to dealing with systems with strong static correlation, recent work has investigated post-DMRG methods to recover dynamic correlation~\cite{knecht2016dmrgchemistry,yanai2015dmrgchemistry}. For more information on the use of DMRG in quantum chemistry, we refer the reader to the reviews by \textcite{olivares2015dmrg,szalay2015dmrgchemistry}. 

We can see from the above discussion that there appears to be an untreated `sweet spot', of systems with around 100 to 200 spin-orbitals, that require high accuracy calculations. These systems are too strongly correlated to be tackled with methods like HF, DFT or even CCSD. They are also too large to be reliably dealt with using DMRG or quantum Monte Carlo, and much too large for classical FCI methods. Interestingly, many problems of scientific interest fit this description, including: transition metal catalysts~\cite{vogiatzis2019catalysis,podewitz2011bioinorganic} and the Fermi-Hubbard model~\cite{simonscollab2015hubbard}. As we have discussed throughout this review, a small quantum computer, with around 100 perfect qubits, would be able to calculate the FCI energy of a system with around 100 spin-orbitals in polynomial time. This would imply that these problems are among the best targets for quantum computers.

It is important to note that being able to accurately predict the ground state energy of 100 spin-orbital systems still leaves us far from our long-term goal of designing new medicines and materials with simulations. For example, \textcite{takeshi2018decomposition} noted that over 95~\% of the approved drug molecules in DrugBank~5.0 are larger than these 100-200 spin-orbital systems that we might aim to simulate with a small, error-corrected quantum computer. However, in practice it is not always necessary to perform highly accurate calculations on the entirety of a large molecule or enzyme. Instead, problem decomposition approaches can be utilised, whereby the most important part of the system is accurately simulated, and then integrated with a potentially less accurate simulation of the less challenging parts of the system. This approach has been investigated in the context of quantum computing by \textcite{takeshi2018decomposition, rubin2016dmet, Reiher201619152,kreula2016dmft,bauer2016dmft}.\\

\subsection{Quantum resources: medium to long-term}\label{Subsec:LongTermQuantumResources}

As discussed in Sec.~\ref{Subsec:2ndencoding}, quantum computers can store the FCI wavefunction of $M$ spin-orbitals using only $M$ qubits in second quantisation. However, as discussed in Sec.~\ref{Sec:QCS}, we must also take into consideration the qubit overhead of error correction. Initial work suggested that around $10^{18}$ gates would be necessary to perform phase estimation on a system of around 100 spin-orbitals (excluding the overhead of error correction)~\cite{wecker2014gates}. This estimate was subsequently reduced through a series of algorithmic optimisations, as described in Sec~\ref{Para:HamSimGaussians}.

These initial estimates did not focus on specific problems of interest, and neglected the overhead of quantum error correction, which is necessitated by the large number of gates needed. Fault-tolerant resource estimations have since been carried out for two main systems: small transition metal molecules, and condensed phase materials (including 2D Fermi-Hubbard models, 2D and 3D electron gases, and solid materials, such as lithium or diamond). When performing fault-tolerant resource estimates, one must specify details of the problem, the hardware considered, and the error correcting code used. All resource estimates to date have focused on the 2D surface code, due to its high threshold, and suitability for architectures with a 2D nearest-neighbour connectivity. In the standard model of surface code resource estimation, Clifford gates (such as Pauli gates and the CNOT gate) are considered to be of negligible cost, while non-Clifford gates (such as the T gate, or Toffoli gate) are more costly. This is because these non-Clifford gates cannot be natively implemented in a fault-tolerant way in the surface code, but instead are typically implemented using magic state distillation and teleportation~\cite{bravyi2005magicstateoriginal}, which is often expensive~\cite{campbell2017roads}. As a result, algorithm complexities are measured in terms of the number of T and/or Toffoli gates that they contain, as these are often the dominant contribution to the cost of the algorithm. There has been considerable work to reduce the cost of operations in the surface code (including magic state distillation), which has reduced the overhead of error correction by several orders of magnitude~\cite{Litinski2019gameofsurfacecodes,litinski2019notascostly,fowler2018lowoverheadlatticesurgery,Gidney2019efficientmagicstate,trout2015magicstateforchemistry}. These improvements, combined with the algorithmic advances described throughout this review, have contributed to a significant reduction in the resources required for chemistry calculations, compared to the initial estimates. In order to distinguish algorithmic advances from fault-tolerance improvements, we list both the number of T and/or Toffoli gates required for the simulation, as well as the corresponding number of physical qubits, obtained using the best fault-tolerance procedures available at that work's time of writing. \\

\textcite{Reiher201619152} carried out a fault-tolerant resource estimation for the problem of biological dinitrogen fixation, as described in Sec.~\ref{Subsubsec:ProblemsOfInterest}. Those authors calculated the resources required to perform an FCI calculation on an active space of 54 electrons in 108 spin-orbitals for FeMo-co, using a Trotter-based approach to phase estimation. They found that this would require around $10^{14}$ T gates. Assuming the best physical error rates ($10^{-3}$) at our time of writing, this would require around 200 million physical qubits, and take on the order of weeks (10~ns to implement a T gate, including surface code decoding) or months (100~ns per T gate)~\cite{Reiher201619152}. We note that those authors were considering a targeted majorana fermion-based quantum computer, with physical error rates $10^3$ times lower than has been demonstrated in trapped ion or superconducting qubits, at our time of writing. 

\textcolor{black}{A similar resource estimation (although using the more accurate FeMo-co active space of \textcite{li2018nitrogenase}, with 113 electrons in 152 spin-orbitals) was carried out by \textcite{berry2019qubitizationlowrank}, who used the algorithm based on qubitization and low-rank decompositions of the Hamiltonian, described in Sec.~\ref{Subsubsec:ChemistryTimeEvolution}. This approach reduced the resources required to around $10^{11}$ Toffoli gates (which are currently the bottleneck for this approach). While a complete fault-tolerant resource analysis for this new approach has not yet been performed, those authors showed that the cost of Toffoli gate distillation is equivalent to around 1 million physical qubits, working for 2 months (assuming $10^{-3}$ error rates and surface code cycle times of \SI{1}{\micro\second}), at our time of writing.\\}

Other works have conducted similar resource estimations for equivalently sized problems (100-200 spin-orbitals), but have focused on matter in the condensed phase. This enables the use of the plane wave dual basis (Sec.~\ref{Subsubsec:PlaneWave}), which as discussed in Sec.~\ref{Para:HamSimPlane} can reduce the costs of simulations. \textcite{babbush2018encoding} used the algorithm based on qubitization in a plane dual wave basis (discussed in Sec.~\ref{Para:HamSimPlane}) to obtain resource estimates of around $2 \times 10^8$ T gates for a 128 spin-orbital 3D homogeneous electron gas (with similar results for other 3D materials), and around $7.1 \times 10^8$ T gates for the 2D Fermi-Hubbard model with 100 lattice sites (200 spin-orbitals). Assuming error rates of
$10^{-3}$, \SI{1}{\micro\second} to implement a T gate (including surface code decoding), and \SI{10}{\micro\second} feedforward, this led to resources of around 2-3 million qubits, running for tens of hours (using the best fault-tolerance protocols available when that work was completed). Subsequent improvements in fault-tolerance protocols have further reduced these physical resources~\cite{Litinski2019gameofsurfacecodes,litinski2019notascostly,fowler2018lowoverheadlatticesurgery,Gidney2019efficientmagicstate}.

\textcolor{black}{\textcite{kivlichan2019condensedtrotter} also performed resource estimations for condensed matter and solid state problems, using the Trotter based algorithm discussed in Sec.~\ref{Para:HamSimPlane}. They considered simulations targeting a size-extensive error in the energy (which is appropriate when considering scaling to the thermodynamic limit). They found that simulations of a 100 site Fermi-Hubbard model, and a 128 spin-orbital homogeneous electron gas would both require on the order of $10^6$ Toffoli gates and $10^7-10^8$ T gates. Assuming error rates of $10^{-3}$, surface code error detection times of \SI{1}{\micro\second}, and surface code error decoding times of \SI{10}{\micro\second}, they require around 400,000 -- 600,000 physical qubits, running for a couple of hours. As these simulations consider a loose-but-rigorous bound on the energy error (as discussed in Sec.~\ref{Subsubsec:OutstandingQPEproblems}), these resource estimates may be overly pessimistic. \textcite{kivlichan2019condensedtrotter} found that if an intensive (e.g. absolute) error in the energy is targeted, then their algorithm was less efficient than the qubitization algorithm of \textcite{babbush2018encoding}.\\ }

All of the resource estimation papers above focus on the cost of phase estimation, and assume that the system is prepared in an initial state that has a sufficiently large overlap with the true ground state. As discussed in Sec.~\ref{Subsubsec:QPEstateprep}, there are many techniques for state preparation, including adiabatic state preparation, or variational approaches. \textcite{tubman2018orthogonality} presented an algorithm which can prepare a suitable initial state by leveraging a classical adaptive configuration interaction method. This algorithm was numerically shown to provide good estimates for the ground states of many systems of interest in chemistry, physics and materials science. This stems from the fact that the dominant Slater determinants in the wavefunction typically converge much more quickly than the correlation energy. \textcite{tubman2018orthogonality} showed that the cost of this state preparation algorithm is considerably lower than the cost for phase estimation detailed above, and hence, may be neglected for many systems of interest. This technique can also be combined with that of \textcite{berry2018improved}, which used classically obtained knowledge of the energy eigenvalues to reduce the number of times that phase estimation must be repeated. This reduced the necessity of having a large overlap with the ground state. \\

One might assume that grid based methods will require considerably larger quantum computers than basis set approaches, given the former's non-compact description of the wavefunction. However, as described above, the figure of merit for fault-tolerant calculations is often the circuit depth -- in particular the number of T or Toffoli gates. While an initial investigation into fault-tolerant grid based simulation was performed~\cite{jones2012faster}, it did not calculate the total number of T gates required, or the number of qubits, for systems of interest. As such, it is not directly comparable to the methods described above. The algorithm investigated in that work (the algorithm of \textcite{kassal2008polynomial}) has since been surpassed by the algorithm of \textcite{KivReal}. Moreover, there have been many improvements in fault-tolerant circuit design and magic state distillation since the work of \textcite{jones2012faster}. As classical computers are limited to small grid based calculations, it would be an interesting direction of future research to establish the quantum resources required to surpass these small, high accuracy calculations.\\

These results suggest that certain calculations with around $100$ spin-orbitals may be better suited to early quantum computers than others. In particular, materials in the condensed phase and Fermi-Hubbard simulations so far require considerably fewer resources than simulations of individual molecules. \textcolor{black}{Despite these promising results, and recent, rapid improvements, we see that it still requires on the order of $100,000$ physical qubits to surpass classical techniques. Current quantum computers possess only around $100$ physical qubits, and we have yet to demonstrate a fully protected logical qubit. It may be many years before we possess a quantum computer with the resources required to implement these algorithms, given the challenges in scaling up hardware and performing quantum error correction~\cite{gambetta2017errorcorrected}.} In order to attempt to solve classically intractable chemistry problems before that time, different approaches are required. We discuss the potential paths towards these classically intractable simulations below.

\subsection{Quantum resources: near to medium-term}\label{Subsec:NearFutureOutlook}
As discussed above, existing estimates for surpassing classical calculations require on the order of 100,000 physical qubits, in order to implement quantum error correction. The first generations of quantum computers will be significantly smaller than this. Nevertheless, there are many interesting avenues to pursue with these first machines.

As discussed in Sec.~\ref{Subsec:VQE}, the variational quantum eigensolver (VQE) has received significant attention in recent years, due to its short required circuit depth (compared to phase estimation). It has been speculated that the VQE may enable small quantum computers with 100 to 200 physical qubits to surpass classical methods. Considerable further work is required to demonstrate that this will be possible. The VQE is a heuristic approach, which attempts to generate an approximation to the ground state wavefunction that is better than classical methods, using a short circuit. It is difficult to prove that a given circuit will be able to obtain a good estimate for the ground state, especially when the difficulty of classical optimisation is considered.

In general, the longer the circuit is, the better it can approximate the ground state wavefunction. However, the length of circuit that we can implement without error correction is heavily limited by noise. A simple calculation demonstrates the limited number of gates that we have available. If we assume a discrete error model for our circuit, such that error events happen probabilistically and independently following each gate in the circuit, then even with an optimistic two qubit gate error rate of 0.01~\% (10 times better than the error rates achieved to date), we could only carry out around 10,000 gates before we expect one or more errors to occur in the circuit with high probability. While the error mitigation techniques discussed in Sec.~\ref{Sec:errorMitigation} may enable us to recover accurate results from a circuit deeper than 10,000 gates, it seems unlikely that these methods alone would enable more than a small multiplicative increase in the circuit depth.\\

It is unclear whether we would be able to surpass classical methods for any chemistry problems with this number of gates. The Fermi-Hubbard model is one of the leading candidates for such a simulation. As described in Sec.~\ref{Subsubsec:VQEAnsatz}, the Hamiltonian variational ansatz is particularly efficient for this problem. We can prepare initial states of the Fermi-Hubbard model using $\mathcal{O}(M^{1.5})$ gates, and perform Trotter steps of the Hamiltonian using $\mathcal{O}(M)$ gates for each Trotter step~\cite{jiang2018hubbard}. Previous work has shown that the Hamiltonian variational ansatz performs well for the Fermi-Hubbard model, although it is not yet known how many Trotter steps may be required for accurate results. \textcite{PhysRevA.92.042303} achieved good convergence for a 12 site problem with 20 Trotter steps. Further work has shown promising results for both the ground state problem~\cite{reiner2018hubbardground}, and dynamics simulation~\cite{reiner2018hubbarddynamics}, both in the presence of realistic noise rates. These results were obtained using less efficiently scaling circuits than those described above. Nevertheless, even if the number of Trotter steps required to find the ground state scaled only linearly with the number of spin-orbitals, the total algorithm scaling would be $\mathcal{O}(M^2)$, which is around 40,000 two-qubit gates for a $10 \times 10$ site Fermi-Hubbard model (which requires 200 qubits). This rough estimate neglects constant factors, and so the number of gates required would likely be higher. More thorough resource estimates for the same problem on a silicon quantum architecture were performed by \textcite{cai2019hubbard}. Our estimate of 40,000 two-qubit gates is approximately equal to the number of gates that we were limited to by noise in our back-of-the-envelope calculation above. There are a number of routes to try and overcome the issue of noise in such a calculation. We may be able to combine existing error mitigation techniques, or try to develop new ones. We could also utilise ans\"atze which appear robust to noise~\cite{kim2017robust,kim2017noise,borregaard2019noiserobust}.\\

A less widely discussed approach is to perform error corrected VQE simulations. The aim would be to suppress the error rate to a value low enough to obtain chemically accurate energies from the simulation. For example, we could use fermion-to-qubit mappings which enable the detection and/or correction of single qubit errors~\cite{setia2018superfast, jiang2018majorana} (discussed in Sec.~\ref{Subsubsec:OtherEncodings} and Sec.~\ref{Subsec:StabiliserMitigation}). Alternatively, we could explore using small error correcting codes. \textcolor{black}{An initial foray into this area was conducted by \textcite{urbanek2019detectionchemistry}, who experimentally implemented a VQE calculation on the H$_2$ molecule encoded in the [[4,2,2]] error detecting code. This calculation showed improved accuracy over an un-encoded calculation, due to the use of post-selection.} Nevertheless, the use of error correcting codes is complicated by the difficulty of producing error protected T gates. As such, it is important to ask if variational algorithms can be implemented with fewer T gates than their phase estimation based counterparts. As an example, the phase estimation based algorithm of \textcite{kivlichan2019condensedtrotter} is already only a constant factor less efficient (in terms of T/Toffoli gates) than just implementing time evolution under the Fermi-Hubbard Hamiltonian directly (as is required for a Hamiltonian variational ansatz). Moreover, synthesizing an arbitrary angle rotation gate can require at least 10 to 100 T gates. For the hypothetical $\mathcal{O}(M^2)$ scaling variational algorithm described above, we may still need around $4 \times 10^5$ to $4 \times 10^6$ T gates, neglecting constant factors. This is comparable to the T gate counts required for phase estimation based approaches described in Sec.~\ref{Subsec:LongTermQuantumResources}. In addition, the long duration of such a computation could be problematic, given the potentially large number of measurements required by the VQE. \\

An alternative approach to doggedly pursuing classically intractable problems is to use chemistry calculations whose results we do know as a benchmark of our technology. This effort has arguably already begun, following the publication of many VQE experiments on small molecules like H$_2$ and LiH, in a range of different hardware systems. This proposal has recently been formalised by \textcite{mccaskey2019benchmark,nam2019watervqe}. We cannot expect to surpass classical methods without first reproducing classically known results. In a similar vein, once we are able to experimentally demonstrate error corrected logical qubits, the next step will be to perform small, error corrected demonstrations of the algorithms described throughout this review. This approach was recommended by \textcite{love2012backtothefuture}, who charted the evolution of classical computational chemistry milestones since the 1930's, and selected target problems to emulate. Equivalent targets would be small Fermi-Hubbard models, or the G1 set of molecules~\cite{pople1989g1}. This is a small set of molecules whose energy is known extremely accurately. For many of the molecules in the G1 set, an FCI calculation on a sufficiently accurate basis set would be classically intractable (although they are typically not necessary due to highly accurate approximate methods and experimental results). As a result, this may be an excellent test case for future quantum computers.

\subsection{Summary and outlook}\label{Subsec:TheEnd}
This review has sought to be accessible to both scientists working on quantum computing, and those working on computational chemistry. We have discussed the key methods used in classical computational chemistry, and how these have been incorporated into quantum algorithms. 
Emphasis has been placed on the key differences between quantum and classical methods of chemistry simulation, and the resulting benefits that quantum computing is widely predicted to bring to the field of computational chemistry. 

However, we have also shown that quantum methods still face many challenges, not least the high error rates and low qubit counts of existing hardware. Ultimately, the success of quantum computational chemistry will depend on our ability to construct larger and better controlled quantum computers. The question of how large, and how well controlled these machines must be will be determined by the quality of the procedures that we have developed to carry out calculations of interest. It is therefore crucial that we continue to develop and optimise new: algorithms, mappings, error correction codes and procedures, basis sets, and error mitigation techniques. Below, we highlight potential research directions to aid in this goal. \\

In the realm of variational algorithms, a wide range of ans\"atze, chemistry mappings, classical optimisation routines, and error mitigation techniques have been proposed in recent years. However, the vast majority of these proposals: were tested on small system sizes, performed a limited number of comparisons to other techniques, and were not optimised for maximum efficacy. We suggest that future work should begin to collate the existing proposals, and determine which look most promising. This review is a first step in this direction, as is the growing availability of quantum computational chemistry packages, and software libraries to emulate quantum computers. Fast numerical simulations may enable us to test variational algorithms on systems with up to around 30 to 40 qubits. This may begin to show which methods are most suitable for near-term hardware. These calculations should be performed both with, and without noise, in order to ascertain the all-round performance of the various techniques.
As quantum hardware develops, this effort can be migrated onto real systems, in order to test whether our algorithms are as effective as we expect them to be. We expect that this more focused research program will lead to new developments, as well as the optimisation of existing methods.

It is more difficult to construct a road-map for phase estimation based approaches to solving chemistry problems, given the higher degree of sophistication of these methods, and the fact that there is less variation between the different approaches. It is also difficult to anticipate breakthroughs that can lead to a large reduction in required resources, such as the introduction of the qubitization technique or tightening of Trotter error bounds. One potential route to new developments is to investigate areas that appear to be less well explored. One example may be error correction procedures that are tailored specifically for chemistry problems. Alternatively, one could import ideas that are well established in classical computational chemistry. This is how transformative ideas like the plane wave basis sets and low rank Hamiltonian decomposition entered the field. \\

Successful exploration of these future directions may only prove possible through close collaboration between chemists and quantum information scientists. We hope that this review helps to develop a common language for these two groups, facilitating this important collaboration.

\begin{acknowledgments}
This work was supported by BP plc and by the EPSRC National Quantum Technology Hub in Networked Quantum Information Technology (EP/M013243/1). A.A.G. acknowledges Anders G Froseth for his generous support, as well as the Vannevar Bush Faculty Fellowship program of the US Department of Defense. We thank R. Babbush for insightful comments. S.M. and X.Y. thank L. Lindoy for initial discussions on basis sets. SE is supported
by Japan Student Services Organization (JASSO)
Student Exchange Support Program (Graduate Scholarship
for Degree Seeking Students).
\end{acknowledgments}

\bibliography{ChemistryReviewBib}

\end{document}